\documentclass[12pt]{article}
\pdfoutput=1
\usepackage{latexsym}
\usepackage{amssymb, amsfonts, amsmath}
\usepackage{graphicx} 
\usepackage{indentfirst}
\usepackage{bbm}
\usepackage{amssymb}
\usepackage{verbatim}
\usepackage{amsmath, amsthm, amssymb}
\usepackage{mathrsfs}
\usepackage{amsfonts}
\usepackage{dsfont}
\usepackage{csquotes}
\usepackage{hyperref}
\usepackage{url}
\usepackage{xcolor}
\usepackage{xpatch}
\usepackage[english]{babel}

\usepackage[style=phys, articletitle=true, biblabel=brackets, backend=bibtex,
chaptertitle=false, pageranges=false, eprint=true,%
natbib=true, maxbibnames=10, giveninits=true, sorting=none]{biblatex} % Minimalistic styling 

\DeclareFieldFormat[article]{journaltitle}{\mkbibitalic{#1\isdot}} % Use \mkbibemph or \mkbibitalic
\makeatletter
\newrobustcmd{\mkbibfixedbrackets}[1]{%
	\begingroup
	\blx@blxinit
	\blx@setsfcodes
	\bibleftbracket#1\bibrightbracket
	\endgroup}

\renewbibmacro*{doi+eprint+url}{%
	\iftoggle{bbx:doi}
	{\printfield{doi}}
	{}%
	\ifboolexpr{test {\iffieldequalstr{eprinttype}{arxiv}} or test {\iffieldequalstr{eprinttype}{arXiv}}}
	{\setunit{\addspace}\newblock}
	{\newunit\newblock}%
	\iftoggle{bbx:eprint}
	{\usebibmacro{eprint}}
	{}%
	\newunit\newblock
	\iftoggle{bbx:url}
	{\usebibmacro{url+urldate}}
	{}}

\xpatchbibdriver{online}
{\newunit\newblock
	\iftoggle{bbx:eprint}
	{\usebibmacro{eprint}}
	{}}
{\ifboolexpr{test {\iffieldequalstr{eprinttype}{arxiv}} or test {\iffieldequalstr{eprinttype}{arXiv}}}
	{\setunit{\addspace}\newblock}
	{\newunit\newblock}%
	\iftoggle{bbx:eprint}
	{\usebibmacro{eprint}}
	{}}
{}{}

\DeclareFieldFormat{eprint:arxiv}{%
	\mkbibbrackets{%
		\ifhyperref
		{\href{http://arxiv.org/\abx@arxivpath/#1}{%
				arXiv\addcolon
				\nolinkurl{#1}%
				\iffieldundef{eprintclass}
				{}
				{\addspace\UrlFont{\mkbibfixedbrackets{\thefield{eprintclass}}}}}}
		{arXiv\addcolon
			\nolinkurl{#1}%
			\iffieldundef{eprintclass}
			{}
			{\addspace\UrlFont{\mkbibfixedbrackets{\thefield{eprintclass}}}}}}}
\makeatother

\parskip=\medskipamount

\newcommand{\cA}{{\cal A}}
\newcommand{\cB}{{\cal B}}

\newcommand{\cD}{{\cal D}}
\newcommand{\cE}{{\cal E}}
\newcommand{\cF}{{\cal F}}

\newcommand{\cH}{{\cal H}}
\newcommand{\cI}{{\cal I}}
\newcommand{\cJ}{{\cal J}}

\newcommand{\cN}{{\cal N}}
\newcommand{\cO}{{\cal O}}

\newcommand{\cQ}{{\cal Q}}

\newcommand{\cT}{{\cal T}}

\newcommand{\cZ}{{\cal Z}}

\def\a{\alpha}

\def\b{\beta}

\def\d{\delta}
\def\e{\epsilon}
\def\f{\phi}
\def\g{\gamma}

\def\l{\lambda}

\def\q{\theta}
\def\r{\rho}
\def\s{\sigma}
\def\t{\tau}

\def\x{\xi}

\def\D{\Delta}
\def\F{\Phi}
\def\J{\Psi}

\def\P{\Pi}
\def\Q{\Theta}

\def\U{\Upsilon}
\def\X{\Xi}

\newcommand{\ve}{\varepsilon}                            %new
\newcommand{\pa}{\partial}                           %new

\newcommand{\abar}{\bar{a}}
\newcommand{\bbar}{\bar{b}}

\newcommand{\be}{\begin{equation}}
\newcommand{\ee}{\end{equation}}
\newcommand{\bea}{\begin{eqnarray}}
\newcommand{\eea}{\end{eqnarray}}

\newcommand{\ba}{\begin{array}}
\newcommand{\ea}{\end{array}}

\def\double #1{#1{\hbox{\kern-2pt $#1$}}}

\newcommand{\bsubeq}{\begin{subequations}}
\newcommand{\esubeq}{\end{subequations}}

\numberwithin{equation}{section}
\begin{document}

%%%%%%%%%%%%%%%%%%%%%%%%%%%%%%%%%%%%%%%%%%%%%%%%%%%%%%%%%%%%%%%%%%%%%%%%%
\begin{titlepage}
\begin{flushright}
November, 2021
\end{flushright}
\vspace{2mm}

\begin{center}
\Large \bf Three-point functions of a superspin--2 current multiplet in 3D, $\cN=1$
superconformal theory \\
\end{center}

\begin{center}
{\bf
Evgeny I. Buchbinder and Benjamin J. Stone}

{\footnotesize{
{\it School of Physics M013, The University of Western Australia\\
35 Stirling Highway, Crawley W.A. 6009, Australia}} ~\\
}
\end{center}
\begin{center}
\texttt{Email: evgeny.buchbinder@uwa.edu.au, \\ benjamin.stone@research.uwa.edu.au}
\end{center}

\vspace{4mm}

\begin{abstract}
\baselineskip=14pt

%%%%%%%%%%%%%%%%%%%%%%%%%%%%%%%%%%%%%%%%%%%%%%%%%%%%%%%%%%%%%%%%%%%%%%%%%%%%%

We consider $\cN=1$ superconformal field theories in three dimensions possessing a conserved current multiplet $\cF_{ (\a_{1} \a_{2} \a_{3} \a_{4}) }$ 
which we refer to as the superspin--2 current multiplet. At the component level it contains a conserved spin--2 current different from the energy-momentum tensor
and a conserved fermionic higher spin current of spin 5/2. 
Using a superspace formulation, we calculate correlation functions involving $\cF$, focusing particularly on the three-point function $ \langle \cF \cF \cF \rangle $. 
After imposing the constraints arising from conservation equations and invariance under permutation of superspace points, we find that the parity-even and parity-odd sectors of this three-point function are each fixed up to a single coefficient. The presence of the parity-odd contribution is rather non-trivial, as there is an apparent tension between supersymmetry and the existence of parity-odd structures. 

%%%%%%%%%%%%%%%%%%%%%%%%%%%%%%%%%%%%%%%%%%%%%%%%%%%%%%%%%%%%%%%%%%%%%%%%%%%%%%%%%%%%

\end{abstract}
\end{titlepage}

\newpage
\renewcommand{\thefootnote}{\arabic{footnote}}
\setcounter{footnote}{0}

\tableofcontents
\vspace{1cm}
\bigskip\hrule
%%%%%%%%%%%%%%%%%%%%%%%%%%%%%%%%%%%%%%%%%%%%%%%%%%%%%%%%%%%%%%%%%%%%%%%%%

%%%%%%%%%%%%%%%%%%%%%%%%%%%%%%%%%%%%%%%%%%%%%%%%%%%%%%%%%%%%%%%%%%%%%%%%%
\section{Introduction}\label{section1}

A peculiar feature of three-dimensional conformal field theories is the presence of parity violating, or parity-odd, structures in the three-point functions of conserved currents such as the energy-momentum tensor and vector currents~\cite{Giombi:2011rz}. 
These structures were not considered in the systematic studies of~\cite{Osborn:1993cr, Erdmenger:1996yc}, which utilised a group-theoretic approach to solve for the correlation functions of conserved currents in a generic number of space-time dimensions.\footnote{For earlier work concerning correlation functions of conserved currents in conformal field theory, the reader may consult
	refs.~\cite{Polyakov:1970xd, Migdal:1971xh, Migdal:1971fof, Schreier:1971um, Ferrara:1972cq,  Ferrara:1973yt, Koller:1974ut, Mack:1976pa, Fradkin:1978pp, Stanev:1988ft}.} 
Parity-odd structures are not present in free theories but have been shown to arise in Chern--Simons theories interacting with parity violating matter.
In various approaches and contexts they were studied 
in~\cite{Giombi:2011kc, Costa:2011mg, Aharony:2011jz, Maldacena:2011jn, Maldacena:2012sf, 
	Aharony:2012nh, GurAri:2012is, Nizami:2013tpa, Giombi:2016zwa, Aharony:2018pjn, Skvortsov:2018uru, Jain:2021wyn, Chowdhury:2017vel}.

In general, besides the energy-momentum tensor and vector currents, conformal field theories also possess currents of higher spin. 
In~\cite{Maldacena:2011jn} Maldacena and Zhiboedov proved under certain assumptions (see below) that all correlation functions of 
higher-spin currents in three-dimensional conformal field theory
are equal to that of a free theory. In particular, it implies that they do not have parity-odd contributions. 
This theorem was later generalised to higher-dimensional cases in~\cite{Stanev:2013qra, Alba:2013yda, Alba:2015upa}. 
These results can be viewed as the analogue of the Coleman-Mandula theorem~\cite{Coleman:1967ad} for conformal field theories. 

In this paper we will be interested in ${\mathcal N}=1$ superconformal field theories in three dimensions. 
The general formalism to construct the two- and three-point functions of conserved currents in three-dimensional superconformal field theories was developed 
in~\cite{Park:1999cw, Buchbinder:2015qsa, Buchbinder:2015wia, Kuzenko:2016cmf} (a similar formalism
in four dimensions was developed in~\cite{Park:1997bq, Osborn:1998qu, Kuzenko:1999pi} and in six dimensions in~\cite{Park:1998nra}). 
In supersymmeric theories, conserved currents are contained within supermultiplets. The energy-momentum tensor lies in the supercurrent multiplet~\cite{Ferrara:1974pz}, 
which in three dimensions also contains a fermionic supersymmetry current. On the other hand, a vector current becomes a component of the flavour current multiplet. 
As was pointed out in~\cite{Buchbinder:2015qsa,Buchbinder:2021gwu} there is an apparent tension between supersymmetry and the existence of parity violating structures in the three-point functions of conserved currents. In particular, three-point functions containing the supercurrent and flavour current multiplets admit only parity-even contributions. Combining this with the Maldacena-Zhiboedov theorem, it follows that supersymmetric conformal field theories do not admit parity-odd contributions to the three-point functions of conserved currents for any spin unless the assumptions of the theorem are violated. 

The strongest assumption of the Maldacena-Zhiboedov theorem is that the conformal field theory under consideration possesses a unique
conserved spin--2 current --- the energy-momentum tensor. However, in the same article~\cite{Maldacena:2011jn} Maldacena and Zhiboedov
showed that the existence of a conserved fermionic higher-spin current implies that there is more than one conserved current of spin 2. 
In supersymmetric theories conserved currents belong to supermultiplets which contain both bosonic and fermionic currents. This implies that a supersymmetric conformal field theory possessing a bosonic higher spin current also possesses a fermionic higher spin current (and vice-versa), 
thus it is conceivable that there exists another conserved current of spin 2. This, in turn, implies that in three-dimensional superconformal field theories
the assumptions of~\cite{Maldacena:2011jn} might be violated and the properties of correlation functions of higher spin currents might be more subtle. 

In this paper, we will assume that the ${\mathcal N}=1$ superconformal field theory under consideration possesses a spin--2 conserved current different from the energy-momentum tensor. It naturally sits in the supermultiplet 
\begin{equation}
{\mathcal F}_{\a_1 \a_2\a_3\a_4}={\mathcal F}_{(\a_1 \a_2\a_3\a_4)}= {\mathcal F}_{\a(4)}\,, 
\label{0.1}
\end{equation}
and satisfies the conservation equation
\begin{equation}
D^{\a_1} {\mathcal F}_{\a_1 \a_2\a_3\a_4}=0\,. 
\label{0.2}
\end{equation}
The superfield ${\mathcal F}_{\a(4)}$ contains two independent conserved currents (see Section \ref{section3})
\begin{equation} \label{0.3}
J_{ \a_{1} \a_{2} \a_{3} \a_{4} }(x) = \cF_{ \a_{1} \a_{2} \a_{3} \a_{4} }(z) | \, , \hspace{5mm} Q_{\a_{1} \a_{2} \a_{3} \a_{4}, \a }(x) = D_{\a} \cF_{ \a_{1} \a_{2} \a_{3} \a_{4} }(z) | \, ,
\end{equation}
where, as usual, bar-projection means setting all Grassmann odd variables to zero. We will refer to ${\mathcal F}_{\a(4)}$ as to the 
``superspin--2 current multiplet". The component current $J_{\a(4)}$ is a conserved spin--2 current different from the energy momentum tensor, though it satisfies similar properties 
(the latter belongs to the supercurrent multiplet $\cJ_{\a(3)}$), while $Q_{\a_{1} \a_{2} \a_{3} \a_{4}, \a_5 }= Q_{\a(5)}$ is a conserved fermionic current of spin 5/2. 
We will not discuss here particular realisations of superconformal theories possessing a conserved superspin--2 multiplet, 
our interest here is to explore how the ${\mathcal N}=1$ superconformal symmetry 
constraints the three-point correlation functions involving ${\mathcal F}_{\a(4)}$. 

Our main result is that the three point function, 
\begin{equation}
\langle \cF_{\a(4)} (z_1) \cF_{\b(4)} (z_2) \cF_{\g(4)} (z_3) \rangle \, ,
\label{0.4}
\end{equation}
is fixed by the ${\mathcal N}=1$ superconformal symmetry up to one parity-even and one parity-odd structure. 
Our analysis is technically quite involved; the analytic superfield consideration turns out to be quite intractable and we were required to complete both superfield and component analysis with the aid of the \textit{xAct} package \cite{MARTINGARCIA2008597} for \textit{Mathematica}, which contains an advanced suite of tools designed for tensor analysis. The three-point function~\eqref{0.4} contains two independent component correlators (all others can found in terms of these two by virtue of the 
conservation law~\eqref{0.2})
\begin{align}
&\langle J_{\a(4)} (x_1) \, J_{\b(4)} (x_2) \,  J_{\g(4)}(x_{3}) \rangle \, , \hspace{10mm} 
\langle Q_{\a(5)}(x_1) \, J_{\b(4)}(x_2) \, Q_{\g(5)}(x_{3}) \rangle \, .  \label{0.5}
\end{align} 
These two correlators were analysed analytically, however, to provide a complete check that all the necessary conditions are satisfied we had to also perform some numerical analysis. 
We also discuss some basic mixed three-point functions involving $\cF_{\a(4)}$.\footnote{A more detailed study of the mixed three-point functions involving the superspin-2 multiplet, the supercurrent and the flavour current multiplet will be presented elsewhere.}
In particular we analyse the three-point function $\langle \cO(z_{1}) \, \cF_{\a(4)} (z_{2}) \, \cO(z_{3}) \rangle$, where $\cO(z)$ is a scalar superfield of dimension $\D$. We found that it is fixed up to a single parity-even tensor structure. We also compute the three-point function
\begin{equation}
\langle \cF_{\a(4)} (z_1) \, L^{\abar}_{\b} (z_2) \, L^{\bbar}_{\g} (z_3) \rangle \, ,
\end{equation}
where $L^{\abar}_{\a}(z)$ is the non-abelian flavour current multiplet. We found that this three-point function is also fixed up to a single parity-even tensor structure, which is in disagreement with the result previously reported in~\cite{Nizami:2013tpa}, which used a different approach (see Subsection \ref{subsection5.2} for details).
In our approach the analysis of this correlation function is relatively straightforward as it can be studied analytically, so we are confident in our result.

The paper is organised as follows. In Section \ref{section2} we introduce the superconformal building blocks that are essential to the construction of two- and three-point correlation functions of primary operators.
In Section \ref{section3} we analyse the structure of the supermultiplet $\cF$; in particular we define the component fields in the multiplet and determine the constraints on them resulting from the superfield conservation equations.
Section \ref{section4} is devoted to studying the three-point function $\langle \cF \cF \cF \rangle$. 
First we impose the constraints resulting from the conservation of $\cF$ and invariance under permutation of superspace points $z_{1}$ and $z_{2}$;
we show that these constraints are sufficient to fix the parity-even and parity-odd sectors each up to a single coefficient. 
Next we check invariance under permutation of superspace points $z_{1}$ and $z_{3}$ which requires a combination of both analytic and numerical methods. 
As a result, we show that the three-point function is fixed by the $\cN=1$ superconformal symmetry up to two independent tensor structures, 
one is parity-even while the other is parity-odd.
Section \ref{section5} is devoted to the study of mixed correlation functions involving the superfield $\cF$. We compute the three-point function of $\cF$ with two scalar superfield insertions, and the three-point function of $\cF$ with two non-abelian flavour current multiplets. In Section \ref{section6} we provide a brief summary of the work 
and some future directions. The appendices \ref{AppA}, \ref{AppB} and \ref{AppC} are devoted to our conventions, technical details and some consistency checks.

%%%%%%%%%%%%%%%%%%%%%%%%%%%%%%%%%%%%%%%%%%%%%%%%%%%%%%%%%%%%%%%%%%%%%%%%%

%%%%%%%%%%%%%%%%%%%%%%%%%%%%%%%%%%%%%%%%%%%%%%%%%%%%%%%%%%%%%%%%%%%%%%%%%
\section{Superconformal building blocks}\label{section2}

In this section we will review the pertinent details of the group theoretic formalism used to compute correlation functions of primary superfields. For a more detailed review of our conventions the reader may consult \cite{Buchbinder:2015qsa, Buchbinder:2021gwu}.

\subsection{Superconformal transformations}
Consider 3D, $\cN=1$ Minkowski superspace $\mathbb{M}^{3 | 2}$, parameterised by coordinates $z^{A} = (x^{a} , \q^{\a})$, where $a = 0,1,2$, $\a = 1,2$ are Lorentz and spinor indices respectively. Under infinitesimal superconformal transformations, the superspace coordinates transform as
\begin{equation}
\d z^{A} = \x z^{A}  \hspace{3mm} \Longleftrightarrow \hspace{3mm} \d x^{a} = \x^{a}(z) + \text{i} (\g^{a})_{\a \b} \, \x^{\a}(z) \, \q^{\b} \, , 
\hspace{8mm} \d \q^{\a} = \x^{\a}(z) \,, 
\label{new1}	
\end{equation}
where $\x^{a}(z)$ is a conformal Killing supervector
\begin{equation}
\x = \x^{A}(z) \, \partial_{A} = \x^{a}(z) \, \partial_{a} + \x^{\a}(z) D_{\a} \, , \label{Superconformal Killing vector field}
\end{equation}
which satisfies the master equation $[\x , D_{\a} ] \propto D_{\b}$. From the master equation we find
\begin{equation}
\x^{\a} = \frac{\text{i}}{6} D_{\b} \x^{\a \b} \, ,
\end{equation}
which, in particular, implies the conformal Killing equation
\begin{equation}
\partial_{a} \x_{b} + \partial_{b} \x_{a} = \frac{2}{3} \eta_{a b} \partial_{c} \x^{c} \, .
\label{new2}	
\end{equation}
The solutions to the master equation are called the conformal Killing supervector fields of Minkowski 
superspace \cite{Buchbinder:1998qv,Kuzenko:2010rp}, which span a Lie algebra isomorphic to the superconformal algebra $\mathfrak{osp}(1 | 2 ; \mathbb{R})$. 

Now consider a generic tensor superfield $\F_{\cA}(z)$ transforming in a representation $T$ of the Lorentz group with respect to the 
index $\cA$.\footnote{We assume that the representations $T$ is irreducible.} Such a superfield is called primary with dimension $q$ if its superconformal transformation law is 
\begin{equation}
\d \F_{\cA} = - \x \F_{\cA} - q \s(z) \F_{\cA} + \l^{\a \b}(z) (M_{\a \b})_{\cA}{}^{\cB} \F_{\cB} \,,
\label{new6}
\end{equation}
where $\x$ is the superconformal Killing vector, and the matrix $M_{\a \b}$ is the Lorentz generator. The $z$-dependent parameters $\s(z)$, $\l^{\a \b}(z)$ associated with $\x$ are defined as follows
\begin{equation}
\l_{\a \b}(z) = - D_{(\a} \x_{\b)} \, , \hspace{5mm} \s(z) = D_{\a} \x^{\a} \, . 
\label{new4}
\end{equation}
%

%%%%%%%%%%%%%%%%%%%%%%%%%%%%%%%%%%%%%%%%%%%%%%%%%%%%%%%%%%%%%%%%%%%%%%%%
\subsection{Two-point and three-point building blocks}

\noindent \textbf{Two-point building blocks:}

Given two superspace points $z_{1}$ and $z_{2}$, we can define the two-point functions
\begin{equation}
\boldsymbol{x}_{12}^{\alpha \beta} = (x_{1} - x_{2})^{\alpha \beta} + 2 \text{i} \theta^{(\alpha}_{1} \theta^{\beta)}_{2} - \text{i} \theta^{\a}_{12} \theta^{\b}_{12} \, ,  \hspace{10mm} \theta^{\alpha}_{12} = \theta_{1}^{\alpha} - \theta_{2}^{\alpha} \,. \label{Two-point building blocks 1}
\end{equation}
Note that $\boldsymbol{x}_{21}^{\a \b} = - \boldsymbol{x}_{12}^{\b \a}$.
It is convenient to split the two-point function \eqref{Two-point building blocks 1} into symmetric and antisymmetric parts as follows
\begin{equation}
\boldsymbol{x}_{12}^{\a \b} = y_{12}^{\a \b} + \frac{\text{i}}{2} \ve^{\alpha \beta} \theta^{2}_{12} \, , \hspace{10mm} \q_{12}^{2} = \q_{12}^{\a} \q_{12 \a} \, ,
\label{Two-point building blocks 1 - properties 1}
\end{equation}
where $y_{12}^{\a \b}$ is the symmetric part of $\boldsymbol{x}_{12}^{\alpha \beta}$, 
\begin{equation}
y_{12}^{\a \b}= (x_{1} - x_{2})^{\alpha \beta}+ 2 \text{i} \theta^{(\alpha}_{1} \theta^{\beta)}_{2}\,. 
\label{newnew1}
\end{equation}
It can also be represented by the three-vector $y_{12}^m = - \frac{1}{2} (\g^{m})_{\a \b} y_{12}^{\a \b}$.
Next we introduce the two-point objects
\begin{subequations}
	\begin{gather}
	\boldsymbol{x}_{12}^{2} = -\frac{1}{2} \boldsymbol{x}_{12}^{\a \b} \boldsymbol{x}_{12 \a \b} \, , \hspace{5mm} \label{Two-point building blocks 2} \\
	\hat{\boldsymbol{x}}_{12}^{\a \b} = \frac{\boldsymbol{x}_{12}^{\a \b}}{\sqrt{ \boldsymbol{x}_{12}^{2}}} \, , \hspace{10mm} \hat{\boldsymbol{x}}_{12 \a}{}^{\g} \hat{\boldsymbol{x}}_{12 \g}{}^{\b} = \d_{\a}{}^{\b} \, . \label{Two-point building blocks 3}
	\end{gather}
\end{subequations}
Hence, we find
\begin{equation}
(\boldsymbol{x}_{12}^{-1})^{\a \b} = - \frac{\boldsymbol{x}_{12}^{\b \a}}{\boldsymbol{x}_{12}^{2}} \, . \label{Two-point building blocks 4}
\end{equation}
These objects are essential in the construction of correlation functions of primary superfields. We also have the useful differential identities
\begin{equation}
D_{(1) \g} \boldsymbol{x}_{12}^{\a \b} = - 2 \text{i} \q^{\b}_{12} \d_{\g}^{\a} \, , \hspace{10mm} D_{(1) \a} \boldsymbol{x}_{12}^{\a \b} = - 4 \text{i} \q^{\b}_{12} \, , \label{Two-point building blocks 1 - differential identities}
\end{equation}
where $D_{(i) \a}$ is the standard covariant spinor derivative \eqref{Covariant spinor derivatives} acting on the superspace point $z_{i}$. 

%%%%%%%%%%%%%%%%%%%%%%%%%%%%%%%%%%%%%%%%%%%%%%%%%%%%%%%%%%%%%%%%%%%%%%%

\noindent \textbf{Three-point building blocks:}

Given three superspace points $z_{i}$, $i = 1,2,3$, one can define the following three-point building blocks
\begin{subequations} \label{Three-point building blocks 1}
	\begin{align}
	\boldsymbol{X}_{1 \, \a \b} &= -(\boldsymbol{x}_{21}^{-1})_{\a \g}  \boldsymbol{x}_{23}^{\g \d} (\boldsymbol{x}_{13}^{-1})_{\d \b} \, , 
	\label{Three-point building blocks 1a} \\[2mm]
	\Q_{1 \, \a} &= (\boldsymbol{x}_{21}^{-1})_{\a \b} \q_{12}^{\b} - (\boldsymbol{x}_{31}^{-1})_{\a \b} \q_{13}^{\b} \, , \label{Three-point building blocks 1b}
	\end{align}
\end{subequations}
and, similarly, $(\boldsymbol{X}_{2}, \Q_{2}), \ (\boldsymbol{X}_{3}, \Q_{3})$ which can be found from~\eqref{Three-point building blocks 1}
by cyclic permutation. 
Next we define 
\begin{equation}
\boldsymbol{X}_{1}^{2} = - \frac{1}{2} \boldsymbol{X}_{1}^{\a \b}  \boldsymbol{X}_{1 \, \a \b} = \frac{\boldsymbol{x}_{23}^{2}}{\boldsymbol{x}_{13}^{2} \boldsymbol{x}_{12}^{2}} \, , 
\hspace{10mm}  \Q_{1}^{2} = \Q^{\a}_{1} \Q_{1 \, \a} \, . \label{Three-point building blocks 2}
\end{equation}
We also define the normalised building block, $\hat{\boldsymbol{X}}_{1}$, and the inverse of $\boldsymbol{X}_{1}$,
\begin{equation}
\hat{\boldsymbol{X}}_{1 \, \a \b} = \frac{\boldsymbol{X}_{1 \,\a \b}}{\sqrt{ \boldsymbol{X}_{1}^{2}}} \, , \hspace{10mm}	(\boldsymbol{X}_{1}^{-1})^{\a \b} = - \frac{\boldsymbol{X}_{1}^{\b \a}}{\boldsymbol{X}_{1}^{2}} \, .
\end{equation}  
There are also useful identities involving $\boldsymbol{X}_{i}$ and $\Q_{i}$ at different superspace points, e.g.,
\begin{subequations}
	\begin{align}
	\boldsymbol{x}_{13}^{\a \a'} \boldsymbol{X}_{3 \, \a' \b'} \boldsymbol{x}_{31}^{\b' \b} &= - (\boldsymbol{X}_{1}^{-1})^{\b \a} \, , \label{Three-point building blocks 1a - property} \\[2mm]
	\Q_{1 \, \g} \boldsymbol{x}_{13}^{\g \d} \boldsymbol{X}_{3 \, \d \b} &= \Q_{3 \, \b} \, . \label{Three-point building blocks 1b - property}
	\end{align}
\end{subequations}
The three-point objects~\eqref{Three-point building blocks 1a} and \eqref{Three-point building blocks 1b} have many properties similar to those of the two-point building blocks. Now if we decompose $\boldsymbol{X}_1$ into symmetric and antisymmetric parts similar to \eqref{Two-point building blocks 1 - properties 1} we have
\begin{equation}
\boldsymbol{X}_{1 \, \a \b} = X_{1 \, \a \b} - \frac{\text{i}}{2} \ve_{\a \b} \Q_{1}^{2} \, , \hspace{10mm} X_{1 \, \a \b} = X_{1 \, \b \a} \, , \label{Three-point building blocks 1a - properties 3}
\end{equation}
where the symmetric spinor $X_{1 \, \a \b}$ can be equivalently represented by the three-vector $X_{1 \, m} = - \frac{1}{2} (\g_{m})^{\a \b} X_{1 \, \a \b}$. 
%Here it is useful to note the similarities between our formalism and that of Osborn \& Petkou; the three-vectors $X_{(i) \, m}$ are precisely the same building blocks defined in (2.15) of \cite{Osborn:1998qu}, this allows for seamless comparison between the supersymmetric and non-supersymmetric results. 
Now let us introduce analogues of the covariant spinor derivative and supercharge operators involving the three-point objects,
\begin{equation}
\cD_{(1) \a} = \frac{\partial}{\partial \Q^{\a}_{1}} + \text{i} (\g^{m})_{\a \b} \Q^{\b}_{1} \frac{\partial}{\partial X^{m}_{1}} \, , \hspace{5mm} \cQ_{(1) \a} = \text{i} \frac{\partial}{\partial \Q^{\a}_{1}} + (\g^{m})_{\a \b} \Q^{\b}_{1} \frac{\partial}{\partial X^{m}_{1}} \, , \label{Supercharge and spinor derivative analogues}
\end{equation}
which obey the standard anti-commutation relations
\begin{equation}
\big\{ \cD_{(i) \a} , \cD_{(i) \b} \big\} = \big\{ \cQ_{(i) \a} , \cQ_{(i) \b} \big\} = 2 \text{i} \, (\g^{m})_{\a \b} \frac{\partial}{\partial X^{m}_{i}} \, .
\end{equation}
Some useful identities involving~\eqref{Supercharge and spinor derivative analogues} are
\begin{equation}
\cD_{(1) \g} \boldsymbol{X}_{1 \, \a \b} = - 2 \text{i} \ve_{\g \b} \Q_{1 \, \a} \, , \hspace{5mm} \cQ_{(1) \g} \boldsymbol{X}_{1 \, \a \b} = - 2 \ve_{\g \a} \Q_{1 \, \b} \, . \label{Three-point building blocks 1a - differential identities 1}
\end{equation}
We must also account for the fact that various primary superfields obey certain differential equations. Using \eqref{Two-point building blocks 1 - differential identities} we arrive at the following
\begin{subequations}
	\begin{align}
	D_{(1) \g} \boldsymbol{X}_{3 \, \a \b} &= 2 \text{i} (\boldsymbol{x}^{-1}_{13})_{\a \g} \Q_{3 \, \b} \, , &  D_{(1) \a} \Q_{3 \, \b} &= - (\boldsymbol{x}_{13}^{-1})_{\b \a} \, , \label{Three-point building blocks 1c - differential identities 1}\\[2mm]
	D_{(2) \g} \boldsymbol{X}_{3 \, \a \b} &= 2 \text{i} (\boldsymbol{x}^{-1}_{23})_{\b \g} \Q_{3 \, \b} \, , &  D_{(2) \a} \Q_{3 \, \b} &= (\boldsymbol{x}_{23}^{-1})_{\b \a} \, . \label{Three-point building blocks 1c - differential identities 2}
	\end{align}
\end{subequations}
Now given a function $f(\boldsymbol{X}_{3} , \Q_{3})$, there are the following differential identities which arise as a consequence of \eqref{Three-point building blocks 1a - differential identities 1}, \eqref{Three-point building blocks 1c - differential identities 1} and \eqref{Three-point building blocks 1c - differential identities 2}:
\begin{subequations}
	\begin{align}
	D_{(1) \g} f(\boldsymbol{X}_{3} , \Q_{3}) &= (\boldsymbol{x}_{13}^{-1})_{\a \g} \cD_{(3)}^{\a} f(\boldsymbol{X}_{3} , \Q_{3}) \, ,  \label{Three-point building blocks 1c - differential identities 3} \\[2mm]
	D_{(2) \g} f(\boldsymbol{X}_{3} , \Q_{3}) &= \text{i} (\boldsymbol{x}_{23}^{-1})_{\a \g} \cQ_{(3)}^{\a} f(\boldsymbol{X}_{3} , \Q_{3}) \, .  \label{Three-point building blocks 1c - differential identities 4}
	\end{align}
\end{subequations}
These identities are essential for imposing differential constraints on correlation functions.

\noindent \textbf{Building blocks in components:}

For future reference
we will also review the non-supersymmetric conformal blocks detailed in \cite{Osborn:1993cr}. 
These objects will appear in component reduction of superspace correlation functions. The two-point and three-point structures are defined as follows
\begin{align}
x_{ij} &= x_{i} - x_{j} \, , & X_{ij} &= \frac{x_{ik}}{x_{ik}^{2}} - \frac{x_{jk}}{x_{jk}^{2}} \, , & i, j, k &= 1, 2 , 3 \, .
\end{align}
These objects may be obtained by bar-projection of the superspace variables defined in Section \ref{section2} as follows
\begin{align}
(x_{ij})_{m} = -\frac{1}{2} (\g_{m})^{\a \b} (\boldsymbol{x}_{ij})_{\a \b} \big| \, ,  \hspace{10mm} (X_{ij})_{m} = -\frac{1}{2} (\g_{m})^{\a \b} (\boldsymbol{X}_{k})_{\a \b} \big| \, .
\label{a00}
\end{align}
Here $(i, j, k)$ is a cyclic permutation of $(1, 2, 3)$. That is, 
\begin{equation}
X_{12} = \frac{x_{13}}{x_{13}^{2}} - \frac{x_{23}}{x_{23}^{2}}\,, \qquad 
(X_{12})_{m} = -\frac{1}{2} (\g_{m})^{\a \b} (\boldsymbol{X}_{3})_{\a \b} \big|\,, \  {\rm etc.} \,. 
\end{equation}
In addition, we introduce the inversion tensor, $I_{a_{1} a_{2}}$, and its representation acting on rank--2 symmetric traceless tensors, $\cI_{a_{1} a_{2}, m_{1} m_{2}}$
\begin{subequations}
	\begin{align}
	I_{a_{1} a_{2}}(X) &= \eta_{a_{1} a_{2}} - \frac{2 X_{a_{1}} X_{a_{2}}}{X^{2}} \, , \\[2mm]
	\cI_{a_{1} a_{2}, m_{1} m_{2}}(X) &= I_{a_{1} n_{1}}(X) \, I_{a_{2} n_{2}}(X) \, \cE^{n_{1} n_{2}}{}_{m_{1} m_{2} } \, , \label{Conformal inversion tensors}
	\end{align}
\end{subequations}
where we have introduced the projection operator 
\begin{equation}
\cE_{m_{1} m_{2} , n_{1} n_{2} } = 
\frac{1}{2} \, \big( \eta_{m_{1} n_{1}} \eta_{m_{2} n_{2}} + \eta_{m_{1} n_{2}} \eta_{m_{2} n_{1}} \big) - \frac{1}{3} \, \eta_{m_{1} m_{2}} \eta_{n_{1} n_{2}} \, .
\end{equation}
%

%%%%%%%%%%%%%%%%%%%%%%%%%%%%%%%%%%%%%%%%%%%%%%%%%%%%%%%%%%%%%%%%%%%%%%%

\subsection{Correlation functions of primary superfields}

%%%%%%%%%%%%%%%%%%%%%%%%%%%%%%%%%%%%%%%%%%%%%%%%%%%%%%%%%%%%%%%%%%%%%%%

The two-point correlation function of a primary superfield $\F_{\cA}$ and its conjugate $\bar{\F}^{\cB}$ is fixed by the superconformal symmetry as follows
\begin{equation}
\langle \F_{\cA}(z_{1}) \, \bar{\F}^{\cB}(z_{2}) \rangle = c \, \frac{T_{\cA}{}^{\cB}(\hat{\boldsymbol{x}}_{12})}{(\boldsymbol{x}_{12}^{2})^{q}} \, , 
\end{equation} 
where $c$ is a constant coefficient. The denominator of the two-point function is determined by the conformal dimension $q$ of $\F_{\cA}$, which guarantees that the correlation function transforms with the appropriate weight under scale transformations.

Concerning the three-point functions; let $\F$, $\J$, $\P$ be primary superfields with conformal dimensions $q_{1}$, $q_{2}$ and $q_{3}$ respectively. The three-point function may be constructed using the general expression
\begin{align}
\langle \F_{\cA_{1}}(z_{1}) \, \J_{\cA_{2}}(z_{2}) \, \P_{\cA_{3}}(z_{3}) \rangle =& \frac{ T^{(1)}{}_{\cA_{1}}{}^{\cA'_{1}}(\hat{\boldsymbol{x}}_{13}) \,  T^{(2)}{}_{\cA_{2}}{}^{\cA'_{2}}(\hat{\boldsymbol{x}}_{23}) }{(\boldsymbol{x}_{13}^{2})^{q_{1}} (\boldsymbol{x}_{23}^{2})^{q_{2}}}
\; \cH_{\cA'_{1} \cA'_{2} \cA_{3}}(\boldsymbol{X}_{3}, \Q_{3}, U_{3}) \, , \label{Three-point function - general ansatz}
\end{align} 
where the tensor $\cH_{\cA_{1} \cA_{2} \cA_{3}}$ is highly constrained by the superconformal symmetry as follows:
\begin{enumerate}
	\item[\textbf{(i)}] Under scale transformations of superspace $z^{A} = (x^{a},\q^{\a}) \mapsto z'^{A} = (\l^{-2} x^{a} , \l^{-1} \q^{\a})$, the three-point building blocks transform as $\cZ = (\boldsymbol{X},\Q) \mapsto \cZ' = (\l^{2} \boldsymbol{X}, \l \Q)$. As a consequence, the correlation function transforms as 
	\begin{equation}
	\langle \F_{\cA_{1}}(z_{1}') \, \J_{\cA_{2}}(z_{2}') \, \P_{\cA_{3}}(z_{3}') \rangle = (\l^{2})^{q_{1} + q_{2} + q_{3}} \langle \F_{\cA_{1}}(z_{1}) \, \J_{\cA_{2}}(z_{2}) \,  \P_{\cA_{3}}(z_{3}) \rangle \, ,
	\end{equation}
	which implies that $\cH$ obeys the scaling property
	\begin{equation}
	\cH_{\cA_{1} \cA_{2} \cA_{3}}(\l^{2} \boldsymbol{X}, \l \Q, U) = (\l^{2})^{q_{3} - q_{2} - q_{1}} \, \cH_{\cA_{1} \cA_{2} \cA_{3}}(\boldsymbol{X}, \Q, U) \, , \hspace{5mm} \forall \l \in \mathbb{R} \, \backslash \, \{ 0 \} \, .
	\end{equation}
	This guarantees that the correlation function transforms correctly under conformal transformations.
	
	\item[\textbf{(ii)}] If any of the fields $\F$, $\J$, $\P$ obey differential equations, such as conservation laws in the case of conserved current multiplets, then the tensor $\cH$ is also constrained by differential equations. Such constraints may be derived with the aid of identities \eqref{Three-point building blocks 1c - differential identities 3}, \eqref{Three-point building blocks 1c - differential identities 4}.
	
	\item[\textbf{(iii)}] If any (or all) of the superfields $\F$, $\J$, $\P$ coincide, the correlation function possesses symmetries under permutations of superspace points, e.g.
	\begin{equation}
	\langle \F_{\cA_{1}}(z_{1}) \, \F_{\cA_{2}}(z_{2}) \, \P_{\cA_{3}}(z_{3}) \rangle = (-1)^{\e(\F)} \langle \F_{\cA_{2}}(z_{2}) \, \F_{\cA_{1}}(z_{1}) \, \P_{\cA_{3}}(z_{3}) \rangle \, ,
	\end{equation}
	where $\e(\F)$ is the Grassmann parity of $\F$. As a consequence, the tensor $\cH$ obeys constraints which will be referred to as ``point-switch identities". 
\end{enumerate}

The constraints above fix the functional form of $\cH$ (and therefore the correlation function) up to finitely many parameters. Hence the procedure described above reduces the problem of computing three-point correlation functions to deriving the tensor $\cH$ subject to the above constraints.	

%%%%%%%%%%%%%%%%%%%%%%%%%%%%%%%%%%%%%%%%%%%%%%%%%%%%%%%%%%%%%%%%%%%%%%%%%
\section{Component structure of a superspin-2 current multiplet }\label{section3}
%%%%%%%%%%%%%%%%%%%%%%%%%%%%%%%%%%%%%%%%%%%%%%%%%%%%%%%%%%%%%%%%%%%%%%%%%

In this paper we will be interested in three-point functions of a superspin-2 current multiplet described by 
the totally symmetric superfield $\cF_{\a(4)} :=   \cF_{ \a_{1} \a_{2} \a_{3} \a_{4} }(z)$, satisfying the conservation equation 
\begin{equation}
D^{\a_{1}} \cF_{ \a_{1} \a_{2} \a_{3} \a_{4} }(z) = 0 \, . \label{F - conservation equation}
\end{equation}
In three dimensions this superfield admits the follwing Taylor expansion
\begin{align}\label{eeee}
\begin{split}
\cF_{ \a_{1} \a_{2} \a_{3} \a_{4} }(z) &= J_{\a_{1} \a_{2} \a_{3} \a_{4} }(x) + Q_{\a_{1} \a_{2} \a_{3} \a_{4}, \a }(x) \, \q^{\a} \\
& \hspace{25mm} + \q_{( \a_{1}} S_{\a_{2} \a_{3} \a_{4} ) }(x) + \q^{2} B_{\a_{1} \a_{2} \a_{3} \a_{4} }(x) \, .
\end{split}
\end{align}
It can be convenient to express some of these fields in vector notation as follows
\begin{equation}
J_{ \a_{1} \a_{2} \a_{3} \a_{4} }(x) = ( \g^{a_{1}} )_{\a_{1} \a_{2}} ( \g^{a_{2}} )_{\a_{3} \a_{4}} J_{a_{1} a_{2} }(x) \, , 
\end{equation}
where $J_{a_{1} a_{2}}$ is symmetric and traceless; a similar treatment follows for the other fields in the multiplet. Imposing the conservation equation is then tantamount to the following constraints on the component fields
\begin{subequations}
	\begin{gather}
	\pa^{a_{1}} J_{a_{1} a_{2}} = 0 \, , \hspace{5mm} \pa^{a_{1}} Q_{a_{1} a_{2} , \a} = 0 \, , \hspace{5mm}  (\g^{a_{1}})_{\d}{}^{\a} Q_{a_{1} a_{2} , \a} = 0 \, , 
	\label{a0.1}\\[2mm]
	B_{a_{1} a_{2}} = \tfrac{\text{i}}{2} \e_{ (a_{1}}{}^{ m n } \pa_{m} J_{a_{2}) n} \, , \hspace{5mm} S_{\a_{1} \a_{2} \a_{3} } = 0 \label{F multiplet - component constraints} \, .
	\end{gather}
\end{subequations}
Hence we see this multiplet contains only two independent component currents: 
a conserved spin--2 field $J_{a_{1} a_{2}}$ satisfying the same properties as the energy momentum tensor, and a conserved spin--5/2 field $Q_{a_{1} a_{2} , \a}$ 
which is conserved and gamma-traceless (the latter guarantees that $Q$ is totally symmetric in spinor notation). 
Let us stress that $J_{a_{1} a_{2}}$ is different from the energy-momentum tensor $T_{a_1 a_2}$, the latter is a component of the supercurrent multiplet ${\cal J}_{\a(3)}$.
The independent components of $\cF_{ \a (4)}$
may be extracted by bar-projection:
\begin{equation}
J_{ \a_{1} \a_{2} \a_{3} \a_{4} }(x) = \cF_{ \a_{1} \a_{2} \a_{3} \a_{4} }(z) | \, , \hspace{5mm} Q_{\a_{1} \a_{2} \a_{3} \a_{4}, \a }(x) = D_{\a} \cF_{ \a_{1} \a_{2} \a_{3} \a_{4} }(z) | \, .
\end{equation}
In addition, under infinitesimal superconformal transformations, the superfield $\cF$ transforms as 
\begin{equation}
\d \cF_{ \a_{1} \a_{2} \a_{3} \a_{4} }(z) = - \xi \cF_{ \a_{1} \a_{2} \a_{3} \a_{4} }(z) - q \s(z) \cF_{ \a_{1} \a_{2} \a_{3} \a_{4} }(z) + 4 \l_{(\a_{1}}{}^{\d}(z) \, \cF_{ \a_{2} \a_{3} \a_{4} ) \d }(z) \, .
\end{equation}
where $q$ is the scaling dimension of $\cF$. The conservation equation \eqref{F - conservation equation} then uniquely fixes the dimension of the field as follows: if we compute $\d D^{\a_{1}} F_{\a_{1} \a_{2} \a_{3} \a_{4}}(z)$, and use the definitions \eqref{new6}, \eqref{new4} we obtain
\begin{equation}
\d \Big(D^{\a_{1}} \cF_{\a_{1} \a_{2} \a_{3} \a_{4}}(z) \Big)= \tfrac{1}{2} ( q - 3 ) D^{2} \x^{\d} \cF_{\d \a_{2}\a_{3} \a_{4} }(z) \, .
\end{equation}
Hence we see that we require $q = 3$ for covariant conservation of $\cF$.

%%%%%%%%%%%%%%%%%%%%%%%%%%%%%%%%%%%%%%%%%%%%%%%%%%%%%%%%%%%%%%%%%%%%%%%%%
\section{Correlation function \texorpdfstring{$\langle \cF \cF \cF \rangle$}{< F F F >}}\label{section4}
%%%%%%%%%%%%%%%%%%%%%%%%%%%%%%%%%%%%%%%%%%%%%%%%%%%%%%%%%%%%%%%%%%%%%%%%%

In this section we will derive an explicit solution for the three-point function $\langle \cF \cF \cF \rangle$. In Subsection \ref{subsection4.1} we impose the constraints which arise due to the superfield conservation equations and invariance under permutation of superspace points $z_{1}$ and $z_{2}$. 
This is already to sufficient to fix the three-point function up to one
parity-even and one parity-odd structure. 
In Subsection \ref{subsection4.2} we computationally analyse the constraints arising from invariance of the three-point function under permutation of superspace points $z_{1}$ and $z_{3}$; this is done by considering the independent component correlators contained within $\langle \cF \cF \cF \rangle$: $\langle J J J \rangle$ and $\langle Q J Q \rangle$. This is followed by a numerical analysis of the point-switch identity for consistency. Most of the tensor expressions are too large to manipulated efficiently by hand, so we make use of \textit{Mathematica} to do most of the lengthy calculations. 

\subsection{Superfield analysis}\label{subsection4.1}
The ansatz for the correlation function $\langle \cF \cF \cF \rangle$ is
%\footnote{To simplify the presentation of expressions involving many indices, in subsequent sections of this paper we will use the notation $\F_{(s_{1} ... s_{k})} = \F_{s(k)}$.}
%
\begin{equation}
\langle \cF_{\a(4)}(z_{1}) \, \cF_{\b(4)}(z_{2}) \, \cF_{\g(4)}(z_{3}) \rangle = \frac{ \prod_{i=1}^{4} \hat{\boldsymbol{x}}_{13 \, \a_{i}}{}^{\a'_{i}} \, \hat{\boldsymbol{x}}_{23 \, \b_{i}}{}^{\b'_{i}} }{(\boldsymbol{x}_{13}^{2})^{3} (\boldsymbol{x}_{23}^{2})^{3}} \, \cH_{\a'(4) \b'(4) \g(4) }( \boldsymbol{X}_{3}, \Q_{3} ) \, , \label{FFF - ansatz}
\end{equation}
where the tensor $\cH$ is independently totally symmetric in the $\a_{i}$, $\b_{i}$ and $\g_{i}$, and is required to satisfy covariant constraints which arise due to conservation equations and invariance under permutations of superspace points. The constraints are summarised below:
\begin{enumerate}
	\item[\textbf{(i)}] \textbf{Homogeneity constraint}
	
	Covariance of the correlation function under scale transformations of superspace results in the following constraint on $\cH$
	\begin{align}
	\cH_{\a(4) \b(4) \g(4) }( \l^{2} \boldsymbol{X} , \l \Q ) = (\l^{2})^{-3} \cH_{\a(4) \b(4) \g(4) }( \boldsymbol{X} , \Q ) \, , \label{FFF - homogeneity constraint}
	\end{align}
	which implies that $\cH$ is a homogeneous tensor field of degree $-3$. This constraint ensures conformal covariance of the three-point function.
	
	\item[\textbf{(ii)}] \textbf{Differential constraints}
	
	The conservation equation \eqref{F - conservation equation} implies that the correlation function must satisfy the following constraint
	\begin{equation}
	D_{(1)}^{\s} \langle \cF_{\s \a(3)}(z_{1}) \, \cF_{\b(4)}(z_{2}) \, \cF_{\g(4)}(z_{3}) \rangle = 0 \, . \label{H - differential constraint}
	\end{equation}
	Application of the identities \eqref{Three-point building blocks 1c - differential identities 3} results in the following differential constraint on $\cH$
	\begin{equation}
	\cD^{\s} \cH_{\s \a(3) \b(4) \g(4) }( \boldsymbol{X} , \Q ) = 0 \, .
	\label{newnew2}
	\end{equation}
	
	\item[\textbf{(iii)}] \textbf{Point-switch identities}
	
	Invariance under permutation of the superspace points $z_{1}$ and $z_{2}$ results in the following constraint on the correlation function
	\begin{equation}
	\langle \cF_{\a(4)}(z_{1}) \, \cF_{\b(4)}(z_{2}) \, \cF_{\g(4)}(z_{3}) \rangle = \langle \cF_{\b(4)}(z_{2}) \, \cF_{\a(4)}(z_{1})  \, \cF_{\g(4)}(z_{3}) \rangle \, ,
	\end{equation}
	which results in the condition
	\begin{equation}
	\cH_{\a(4) \b(4) \g(4) }( \boldsymbol{X} , \Q ) = \cH_{\b(4) \a(4) \g(4) }( -\boldsymbol{X}^{\text{T}} , -\Q ) \, .
	\label{e1}
	\end{equation}
	There is an additional point-switch identity obtained from imposing invariance under permutation of the points $z_{1}$ and $z_{3}$, however it is considerably more complicated so we will discuss it in detail later.
\end{enumerate}
To make subsequent calculations more tractable, it is often convenient to express $\cH$ in terms of its vector equivalent by factoring out gamma matrices as follows
\begin{align}
\cH_{\a_{1} \a_{2} \a_{3} \a_{4} \b_{1} \b_{2} \b_{3} \b_{4} \g_{1} \g_{2} \g_{3} \g_{4} }( \boldsymbol{X} , \Q ) &= (\g^{a_{1}})_{\a_{1} \a_{2}} (\g^{a_{2}})_{\a_{3} \a_{4}} (\g^{b_{1}})_{\b_{1} \b_{2}} (\g^{b_{2}})_{\b_{3} \b_{4}} \nonumber \\
& \hspace{10mm} \times (\g^{c_{1}})_{\g_{1} \g_{2}} (\g^{c_{2}})_{\g_{3} \g_{4}} \cH_{a_{1} a_{2} b_{1} b_{2} c_{1} c_{2} }( \boldsymbol{X} , \Q ) \, . \label{H - vector notation}
\end{align}
This equality holds provided that $\cH$ (in vector notation) is symmetric and traceless in the pairs $a_{i}$, $b_{i}$ and $c_{i}$ respectively. This is seen by requiring that the components antisymmetric in $\a_{2}, \a_{3}$ (and other combinations involving $\b_{i}$ and $\g_{i}$) vanish. Further, since $\cH$ is Grassmann even it admits the Taylor expansion
\begin{equation}
\cH_{a_{1} a_{2} b_{1} b_{2} c_{1} c_{2} }(\boldsymbol{X}, \Q) = F_{a_{1} a_{2} b_{1} b_{2} c_{1} c_{2} }(X) + \Q^{2} G_{a_{1} a_{2} b_{1} b_{2} c_{1} c_{2} }(X) \, . \label{H - expansion}
\end{equation}
At this step it is more convenient to view $F$ and $G$ as functions of the three-vector $X^m$ rather than of $\boldsymbol{X}^{\a \b}$. The point-switch identity \eqref{e1} then implies the following constraints on $F$ and $G$
\begin{subequations}
	\begin{align}
	F_{a_{1} a_{2} b_{1} b_{2} c_{1} c_{2} }(X) &= F_{b_{1} b_{2} a_{1} a_{2} c_{1} c_{2} }(-X) \, , \label{F - x1 to x2} \\
	G_{a_{1} a_{2} b_{1} b_{2} c_{1} c_{2} }(X) &= G_{b_{1} b_{2} a_{1} a_{2} c_{1} c_{2} }(-X) \, . \label{G - x1 to x2}
	\end{align}
\end{subequations}
On the other hand, the differential constraint \eqref{newnew2} results in\footnote{The underlined indices are excluded from the symmetrisation.}
\begin{align}
\pa^{a_{1}} F_{a_{1} a_{2} b_{1} b_{2} c_{1} c_{2} } = 0 \, , \hspace{5mm} 
G_{a_{1} a_{2} b_{1} b_{2} c_{1} c_{2} } =
\tfrac{\text{i}}{2} \e_{(a_{1}}{}^{m n} \pa_{\underline{m}} F_{\underline{n} a_{2}) b_{1} b_{2} c_{1} c_{2} } 
%\tfrac{\text{i}}{4} \e_{a_{1}}{}^{m n} \pa_{m} F_{n a_{2} b_{1} b_{2} c_{1} c_{2} } +\tfrac{\text{i}}{4} \e_{a_{2}}{}^{m n} \pa_{m} F_{n a_{1} b_{1} b_{2} c_{1} c_{2} }% 
\, . \label{F&G constraints}
\end{align}
In the next subsections we will computationally solve for the tensor $\cH$ subject to the constraints listed above using the \textit{xAct} package.

\subsubsection{Parity-even sector}
In the parity-even sector, we begin by constructing a solution for $F$ that is an even function of $X$, hence \eqref{F - x1 to x2} implies
\begin{equation}
F_{a_{1} a_{2} b_{1} b_{2} c_{1} c_{2} }(X) = F_{b_{1} b_{2} a_{1} a_{2} c_{1} c_{2} }(X) \, . \label{F - parity even x1 to x2}
\end{equation}
A general expansion for $F$ consistent with the symmetry property \eqref{F - parity even x1 to x2} may be obtained by introducing the symmetric and traceless basis tensors found in \cite{Osborn:1993cr}. Explicit expressions for the elements of the tensor basis $\{ \U^{i} \}$, $i = 1, ... , 8$, are as follows
\begin{subequations}
	\begin{align}
	\U^{1}_{a_{1} a_{2}}(\hat{X}) &= \hat{X}_{a_{1}} \hat{X}_{a_{2}} - \frac{1}{3} \eta_{a_{1} a_{2}} \, , \hspace{15mm} \hat{X}_{a} = \frac{X_{a}}{\sqrt{X^{2}}} \, , \\[2mm]
	\begin{split}
	\U^{2}_{ a_{1} a_{2} b_{1} b_{2} }(\hat{X}) &= \hat{X}_{a_{1}} \hat{X}_{b_{1}} \eta_{a_{2} b_{2} } + ( a_{1} \leftrightarrow a_{2} , b_{1} \leftrightarrow b_{2} ) \\
	& \hspace{15mm} - \tfrac{4}{3} \, \hat{X}_{a_{1}} \hat{X}_{a_{2}} \eta_{b_{1} b_{2} } - \tfrac{4}{3} \, \hat{X}_{b_{1}} \hat{X}_{b_{2}} \eta_{a_{1} a_{2} } + \tfrac{4}{9} \, \eta_{a_{1} a_{2} } \eta_{b_{1} b_{2} } \, , 
	\end{split} \\[2mm]
	\U^{3}_{ a_{1} a_{2} b_{1} b_{2} } &= \eta_{a_{1} b_{1}} \eta_{a_{2} b_{2} } + \eta_{a_{1} b_{2}} \eta_{a_{2} b_{1} } - \tfrac{2}{3} \, \eta_{a_{1} a_{2}} \eta_{b_{1} b_{2}} \, , \\[2mm]
	%	\end{align}
	%%\end{subequations}
	%%
	%%\begin{subequations}
	%	\begin{align}
	\begin{split}
	\U^{4}_{ a_{1} a_{2} b_{1} b_{2} c_{1} c_{2} }(\hat{X}) &= \U^{3}_{a_{1} a_{2} b_{1} c_{1} } \hat{X}_{b_{2}} \hat{X}_{c_{2}} + ( b_{1} \leftrightarrow b_{2} , c_{1} \leftrightarrow c_{2} ) \\
	& \hspace{15mm} - \tfrac{2}{3} \, \eta_{b_{1} b_{2} } \U^{2}_{a_{1} a_{2} c_{1} c_{2} }(\hat{X}) - \tfrac{2}{3} \, \eta_{c_{1} c_{2} } \U^{2}_{a_{1} a_{2} b_{1} b_{2} }(\hat{X}) \\
	& \hspace{30mm} - \tfrac{8}{9} \, \eta_{b_{1} b_{2} } \eta_{c_{1} c_{2} } \U^{1}_{a_{1} a_{2}}(\hat{X}) \, ,
	\end{split} \\[2mm]
	\begin{split}
	\U^{5}_{ a_{1} a_{2} b_{1} b_{2} c_{1} c_{2} } &= \eta_{a_{1} b_{1}} \eta_{a_{2} c_{1} } \eta_{b_{2} c_{2}} + (a_{1} \leftrightarrow a_{2} , b_{1} \leftrightarrow b_{2} , c_{1} \leftrightarrow c_{2} ) \\
	& \hspace{15mm} - \tfrac{4}{3} \, \eta_{a_{1} a_{2}} \U^{3}_{b_{1} b_{2} c_{1} c_{2} } - \tfrac{4}{3} \, \eta_{b_{1} b_{2}} \U^{3}_{a_{1} a_{2} c_{1} c_{2} } \\
	& \hspace{30mm} - \tfrac{4}{3} \, \eta_{c_{1} c_{2}} \U^{3}_{a_{1} a_{2} b_{1} b_{2} } - \tfrac{8}{9} \, \eta_{a_{1} a_{2}} \eta_{b_{1} b_{2}} \eta_{c_{1} c_{2}} \, .
	\end{split}
	\end{align}
	\label{newnew10}
\end{subequations}
These tensors each possess a variety of symmetry properties, in particular they are symmetric and traceless in pairs of indices. Using this basis we can construct the following set of rank 6 tensors
\begin{subequations}
	\begin{align}
	t^{1}_{a_{1} a_{2} b_{1} b_{2} c_{1} c_{2} }(\hat{X}) &= \U^{5}_{a_{1} a_{2} b_{1} b_{2} c_{1} c_{2} } \, , \\
	t^{2}_{a_{1} a_{2} b_{1} b_{2} c_{1} c_{2} }(\hat{X}) &= \U^{4}_{c_{1} c_{2} a_{1} a_{2} b_{1} b_{2} }(\hat{X}) \, ,\\
	t^{3}_{a_{1} a_{2} b_{1} b_{2} c_{1} c_{2} }(\hat{X}) &= \U^{4}_{a_{1} a_{2} b_{1} b_{2} c_{1} c_{2}}(\hat{X}) + \U^{4}_{b_{1} b_{2} a_{1} a_{2} c_{1} c_{2}}(\hat{X}) \, ,\\
	t^{4}_{a_{1} a_{2} b_{1} b_{2} c_{1} c_{2} }(\hat{X}) &= \U^{3}_{a_{1} a_{2} b_{1} b_{2} } \U^{1}_{ c_{1} c_{2} }(\hat{X}) \, , \\
	t^{5}_{a_{1} a_{2} b_{1} b_{2} c_{1} c_{2} }(\hat{X}) &= \U^{3}_{b_{1} b_{2} c_{1} c_{2} } \U^{1}_{ a_{1} a_{2} }(\hat{X}) + \U^{3}_{ a_{1} a_{2} c_{1} c_{2} } \U^{1}_{ b_{1} b_{2} }(\hat{X}) \, ,\\
	t^{6}_{a_{1} a_{2} b_{1} b_{2} c_{1} c_{2} }(\hat{X}) &= \U^{2}_{ a_{1} a_{2} b_{1} b_{2} }(\hat{X}) \, \U^{1}_{c_{1} c_{2}}(\hat{X}) \, , \\
	t^{7}_{a_{1} a_{2} b_{1} b_{2} c_{1} c_{2} }(\hat{X}) &= \U^{2}_{ a_{1} a_{2} c_{1} c_{2} }(\hat{X}) \, \U^{1}_{b_{1} b_{2}}(\hat{X}) + \U^{2}_{ b_{1} b_{2} c_{1} c_{2} }(\hat{X}) \, \U^{1}_{a_{1} a_{2} }(\hat{X}) \, ,\\
	t^{8}_{a_{1} a_{2} b_{1} b_{2} c_{1} c_{2} }(\hat{X}) &= \U^{1}_{a_{1} a_{2} }(\hat{X}) \, \U^{1}_{b_{1} b_{2} }(\hat{X}) \, \U^{1}_{c_{1} c_{2} }(\hat{X}) \, .
	\end{align}
	\label{newnew11}
\end{subequations}
The $t^{i}_{a_{1} a_{2} b_{1} b_{2} c_{1} c_{2} }$ each possess the symmetry property \eqref{F - parity even x1 to x2}, hence, the ansatz for the tensor $F$ is a linear combination of these tensor structures:
\begin{equation}
F_{a_{1} a_{2} b_{1} b_{2} c_{1} c_{2} }(X) = \frac{1}{X^{3}} \, t_{a_{1} a_{2} b_{1} b_{2} c_{1} c_{2} }(\hat{X}) \, , \hspace{5mm} t_{a_{1} a_{2} b_{1} b_{2} c_{1} c_{2} }(\hat{X}) = \sum_{i=1}^{8} k_{i} \, t^{i}_{a_{1} a_{2} b_{1} b_{2} c_{1} c_{2} }(\hat{X}) \, , \label{FFF - parity even expansion}
\end{equation}
\vspace{-2mm}
where we have used the homogeneity constraint \eqref{FFF - homogeneity constraint}. 
It now remains to impose the differential constraint \eqref{newnew2}, which results in the following relations
\begin{subequations}
	\begin{gather}
	k_{3} = - 2 k_{1} - k_{2} \, , \hspace{10mm} k_{5} = k_{4} \, , \hspace{10mm} k_{6} = 15 k_{1} + 5 k_{2} - 5 k_{4} \, , \\
	k_{7} = - 7k_{1} - k_{2} + 3 k_{4} \, , \hspace{10mm} k_{8} = 28 k_{1} + 14 k_{2} - 7k_{4} \, .
	\end{gather}
\end{subequations}
Hence, we see that the differential constraint immediately fixes the parity-even sector down to three independent coefficients. It is at this step where the linear dependence of the first five tensor structures can be noticed, as the $k_{1}$ dependence can be removed by shifting the variables as follows: $k_{2} \rightarrow k_{2} - k_{1}$, $k_{3} \rightarrow k_{3} - k_{1}$, $ k_{4} \rightarrow k_{4} + 2 k_{1} $, $ k_{5} \rightarrow k_{5} + 2 k_{1} $. Alternatively it may be shown that the following linear dependence relation 
%(Schouten identity) 
holds
\begin{align}
& t^{1}_{a_{1} a_{2} b_{1} b_{2} c_{1} c_{2} }(\hat{X}) - t^{2}_{a_{1} a_{2} b_{1} b_{2} c_{1} c_{2} }(\hat{X}) - t^{3}_{a_{1} a_{2} b_{1} b_{2} c_{1} c_{2} }(\hat{X}) \nonumber \\
& \hspace{15mm} + 2 \, t^{4}_{a_{1} a_{2} b_{1} b_{2} c_{1} c_{2} }(\hat{X}) + 2 \, t^{5}_{a_{1} a_{2} b_{1} b_{2} c_{1} c_{2} }(\hat{X}) = 0 \,. 
\label{e2}
\end{align}
It is now clear that the $k_{1}$ term is redundant, hence, it can be completely removed from our analysis. This reduces our system of equations to
\begin{subequations}
	\begin{gather}
	k_{3} = - k_{2} \, , \hspace{10mm} k_{5} = k_{4} \, , \hspace{10mm} k_{6} = 5 k_{2} - 5 k_{4} \, , \\
	k_{7} = - k_{2} + 3 k_{4} \, , \hspace{10mm} k_{8} = 14 k_{2} - 7k_{4} \, .
	\end{gather}
\end{subequations}
Therefore, the parity-even sector of the three-point function is fixed at this stage up to two independent coefficients, $k_{2}$ and $k_{4}$, and the explicit solution for $F$ is
\begin{align}
F_{a_{1} a_{2} b_{1} b_{2} c_{1} c_{2} }(X) &= \frac{k_{2}}{X^{3}} \, \Big\{ t^{2}_{a_{1} a_{2} b_{1} b_{2} c_{1} c_{2} }(\hat{X}) - t^{3}_{a_{1} a_{2} b_{1} b_{2} c_{1} c_{2} }(\hat{X}) + 5 \, t^{6}_{a_{1} a_{2} b_{1} b_{2} c_{1} c_{2} }(\hat{X})\nonumber \\
& \hspace{20mm} - t^{7}_{a_{1} a_{2} b_{1} b_{2} c_{1} c_{2} }(\hat{X}) + 14 \, t^{8}_{a_{1} a_{2} b_{1} b_{2} c_{1} c_{2} }(\hat{X}) \Big\} \nonumber \\
& + \frac{k_{4}}{X^{3}} \, \Big\{ t^{4}_{a_{1} a_{2} b_{1} b_{2} c_{1} c_{2} }(\hat{X})  + t^{5}_{a_{1} a_{2} b_{1} b_{2} c_{1} c_{2} }(\hat{X}) - 5 \, t^{6}_{a_{1} a_{2} b_{1} b_{2} c_{1} c_{2} }(\hat{X}) \nonumber \\
& \hspace{20mm} + 3 \, t^{7}_{a_{1} a_{2} b_{1} b_{2} c_{1} c_{2} }(\hat{X}) - 7 \, t^{8}_{a_{1} a_{2} b_{1} b_{2} c_{1} c_{2} }(\hat{X}) \Big\} \, . \label{FFF - F solution}
\end{align}
The tensor $G$ is then determined in terms of $F$ using \eqref{F&G constraints}. However, we have not yet imposed the condition~\eqref{G - x1 to x2}. Since $G$ is an odd function of $X$ by virtue of \eqref{F&G constraints}, the constraint \eqref{G - x1 to x2} implies
\begin{equation}
G_{a_{1} a_{2} b_{1} b_{2} c_{1} c_{2} }(X) = - G_{b_{1} b_{2} a_{1} a_{2} c_{1} c_{2} }(X) \, . \label{G - parity even x1 to x2}
\end{equation}
After some calculations one can show that this results in an additional relation between the coefficients $k_2$ and $k_4$:
\be 
k_2=-2k_4\,. 
\label{newnew3}
\ee
Thus, the conservation equations and the proper transformation under the $z_1 \leftrightarrow z_{2}$ exchange fix the parity-even sector up to a single 
overall coefficient. 
Note that so far we have not imposed the $z_{1} \leftrightarrow z_{3}$ point-switch identity. 
It will be imposed later.

%We expect that these last two coefficients will be related to eachother after imposing the $z_{1} \leftrightarrow z_{3}$ point-switch identity, i.e supersymmetry imposes additional restrictions on the allowed structures in the correlation function.

\subsubsection{Parity-odd sector}

Let us now construct the parity-odd sector of the correlation function, where we begin by assuming that the tensor $\tilde{F}$ is an odd function of $X$. Due to \eqref{F - x1 to x2}, this implies that $\tilde{F}$ must satisfy
\begin{equation}
\tilde{F}_{a_{1} a_{2} b_{1} b_{2} c_{1} c_{2} }(X) = - \tilde{F}_{b_{1} b_{2} a_{1} a_{2} c_{1} c_{2} }(X) \, . \label{F - parity odd x1 to x2}
\end{equation}
Now let us construct an explicit solution for the tensor $F$; it must be an odd function of $X$, and each term must contain at most one instance of the Levi-Civita tensor (as products of the latter may be expressed in terms of the metric). We may decompose $\tilde{F}$ as follows:
\begin{align}
\tilde{F}_{a_{1} a_{2} b_{1} b_{2} c_{1} c_{2} }(X) &= \frac{1}{X^{3}} \big\{ \e_{a_{1} b_{1}}{}^{m} P^{1}_{m, a_{2} b_{2} c_{1} c_{2}} (\hat{X}) + \e_{a_{1} b_{2}}{}^{m} P^{2}_{m, a_{2} b_{1} c_{1} c_{2}}(\hat{X}) \nonumber \\
& \hspace{15mm} + \e_{a_{2} b_{1}}{}^{m} P^{3}_{m, a_{1} b_{2} c_{1} c_{2}}(\hat{X}) + \e_{a_{2} b_{2}}{}^{m} P^{4}_{m, a_{1} b_{1} c_{1} c_{2}}(\hat{X}) \big\} \, , \label{FFF - parity odd expansion}
\end{align}
where each $P^{i}$ must have the symmetry property $P^{i}_{m, a_{1} a_{2} b_{1} b_{2}}(X) = P^{i}_{m, (a_{1} a_{2}) (b_{1} b_{2})}(X)$. 
Requiring that the expansion \eqref{FFF - parity odd expansion} is consistent with the properties of pairwise index symmetry and \eqref{F - parity odd x1 to x2} implies that the $P^{i}$ must be identical. Hence we need to find a general expansion for a tensor $P_{m, a_{1} a_{2} b_{1} b_{2}}$ which is homogeneous degree 0 and is composed solely of $\hat{X}$ and the metric tensor. Using \textit{Mathematica} we can generate an ansatz consistent with the symmetry properties:
\begin{align}
P_{m, a_{1} a_{2} b_{1} b_{2}}(\hat{X}) &=
c_{1} \hat{X}^{a_{1}{}} \hat{X}^{a_{2}{}} \hat{X}^{b_{1}{}} \hat{X}^{b_{2}{}}
\hat{X}^{m} + c_{2} \hat{X}^{b_{1}{}}\hat{X}^{b_{2}{}} \hat{X}^{m}
\eta^{a_{1}{}a_{2}{}} + c_{3} \hat{X}^{a_{1}{}} \hat{X}^{a_{2}{}} \hat{X}^{m}
\eta^{b_{1}{}b_{2}{}} \nonumber \\ 
&  + c_{4} \big\{ \hat{X}^{a_{2}{}} \hat{X}^{b_{1}{}} \hat{X}^{b_{2}{}} \
\eta^{a_{1}{}m} + \hat{X}^{a_{1}{}} \hat{X}^{b_{1}{}} \hat{X}^{b_{2}{}} \
\eta^{a_{2}{}m} \big\} \nonumber \\ 
& + c_{5} \big\{ \hat{X}^{a_{1}{}} \hat{X}^{a_{2}{}} \hat{X}^{b_{2}{}} \
\eta^{b_{1}{}m} + \hat{X}^{a_{1}{}} \hat{X}^{a_{2}{}} \hat{X}^{b_{1}{}} \
\eta^{b_{2}{}m} \big\} \nonumber \\ 
& + c_{6} \big\{ \hat{X}^{a_{2}{}} \hat{X}^{b_{2}{}} \hat{X}^{m}
\eta^{a_{1}{}b_{1}{}} + \hat{X}^{a_{2}{}} \hat{X}^{b_{1}{}} \hat{X}^{m}
\eta^{a_{1}{}b_{2}{}} \nonumber \\
& \hspace{10mm} + \hat{X}^{a_{1}{}} \hat{X}^{b_{2}{}} \hat{X}^{m} \eta^{a_{2}{}b_{1}{}} + \hat{X}^{a_{1}{}} \hat{X}^{b_{1}{}} \hat{X}^{m} \
\eta^{a_{2}{}b_{2}{}} \big\} \nonumber \\ 
& + c_{7} \hat{X}^{m} \eta^{a_{1}{}a_{2}{}} \eta^{b_{1}{}b_{2}{}} + c_{8} \big\{ \hat{X}^{m} \eta^{a_{1}{}b_{2}{}} \eta^{a_{2}{}b_{1}{}} + \hat{X}^{m} \eta^{a_{1}{}b_{1}{}} \eta^{a_{2}{}b_{2}{}} \big\} \nonumber \\ 
& + c_{9} \big\{ \hat{X}^{b_{2}{}} \eta^{a_{1}{}m} \
\eta^{a_{2}{}b_{1}{}} + \hat{X}^{b_{1}{}} \eta^{a_{1}{}m} \
\eta^{a_{2}{}b_{2}{}} \nonumber \\
& \hspace{10mm} + \hat{X}^{b_{2}{}} \eta^{a_{1}{}b_{1}{}} \eta^{a_{2}{}m} + \hat{X}^{b_{1}{}} \eta^{a_{1}{}b_{2}{}} \eta^{a_{2}{}m} \big\} \nonumber \\ 
& + c_{10} \big\{ \hat{X}^{a_{2}{}} \eta^{a_{1}{}m} \
\eta^{b_{1}{}b_{2}{}} + \hat{X}^{a_{1}{}} \eta^{a_{2}{}m} \
\eta^{b_{1}{}b_{2}{}} \big\} \nonumber \\
& + c_{11} \big\{ \hat{X}^{b_{2}{}} \eta^{a_{1}{}a_{2}{}} \
\eta^{b_{1}{}m} + \hat{X}^{b_{1}{}} \eta^{a_{1}{}a_{2}{}} \
\eta^{b_{2}{}m} \big\} \nonumber \\ 
& + c_{12} \big\{ \hat{X}^{a_{2}{}} \eta^{a_{1}{}b_{2}{}} \
\eta^{b_{1}{}m} + \hat{X}^{a_{1}{}} \eta^{a_{2}{}b_{2}{}} \
\eta^{b_{1}{}m} \nonumber \\
& \hspace{10mm} + \hat{X}^{a_{2}{}} \eta^{a_{1}{}b_{1}{}} \
\eta^{b_{2}{}m} + \hat{X}^{a_{1}{}} \eta^{a_{2}{}b_{1}{}} \
\eta^{b_{2}{}m} \big\} \, .
\end{align}
Only 9 of these structures contribute when substituted into \eqref{FFF - parity odd expansion}, in particular the terms with $c_{9}$, $c_{4}$ and $c_{10}$ may be neglected. Imposing tracelessness on each pair of indices is tantamount to the following constraints on the coefficients
\begin{subequations}
	\begin{gather}
	c_{5} = c_{6} \, , \hspace{5mm} c_{12} = c_{8} \, , \hspace{5mm} c_{1} = - 6 c_{6} - 3 c_{3} \, , \\
	c_{7} = - \tfrac{2}{3} c_{8} - \tfrac{2}{3} c_{11} - \tfrac{1}{3} c_{2} \, .
	\end{gather}
\end{subequations}
It remains to impose the differential constraint for $\tilde{F}$ in \eqref{F&G constraints}, from which we find the additional relations
\begin{align}
c_{8} = -\tfrac{1}{4} c_{6} \, , \hspace{5mm} c_{3} = 0 \, , \hspace{5mm} c_{11} = \tfrac{1}{2} c_{6} \, , \hspace{5mm} c_{2} = - 2 c_{6} \, .
\end{align}
Hence, the solution for $\tilde{F}$ is fixed up to a single coefficient, 
$b = c_{6}$.\footnote{To account for linear dependence of the tensor structures, each constraint is checked by computationally analysing every element of the tensor for an arbitrary building block vector $X = ( X_{\textcolor{red}{0}}, X_{\textcolor{red}{1}}, X_{\textcolor{red}{2}} )$.} 
The solution for $\tilde{F}$ becomes
\begin{align}
\tilde{F}_{a_{1} a_{2} b_{1} b_{2} c_{1} c_{2} }(X) &= \frac{b}{X^{3}} \big\{ \e_{a_{1} b_{1}}{}^{m} P_{m, a_{2} b_{2} c_{1} c_{2}} (\hat{X}) + \e_{a_{1} b_{2}}{}^{m} P_{m, a_{2} b_{1} c_{1} c_{2}}(\hat{X}) \nonumber \\
& \hspace{10mm} + \e_{a_{2} b_{1}}{}^{m} P_{m, a_{1} b_{2} c_{1} c_{2}}(\hat{X}) + \e_{a_{2} b_{2}}{}^{m} P_{m, a_{1} b_{1} c_{1} c_{2}}(\hat{X}) \big\} \, , \label{F parity odd solution}
\end{align}
where the explicit solution for $P$ is
\begin{align}
P_{m, a_{1} a_{2} b_{1} b_{2}}(\hat{X}) &=
- 6 \hat{X}_{a_{1}{}} \hat{X}_{a_{2}{}} \hat{X}_{b_{1}{}} \hat{X}_{b_{2}{}} \hat{X}_{m} - 2 \hat{X}_{b_{1}{}} \hat{X}_{b_{2}{}} \hat{X}_{m} \eta_{a_{1}{}a_{2}{}} + \hat{X}_{a_{2}{}} \hat{X}_{b_{2}{}} \hat{X}_{m} \eta_{a_{1}{}b_{1}{}} \nonumber \\
& \hspace{5mm} + \hat{X}_{a_{2}{}} \hat{X}_{b_{1}{}} \hat{X}_{m} \eta_{a_{1}{}b_{2}{}} + \hat{X}_{a_{1}{}} \hat{X}_{b_{2}{}} \hat{X}_{m} \eta_{a_{2}{}b_{1}{}} + \hat{X}_{a_{1}{}} \hat{X}_{b_{1}{}} \hat{X}_{m} \eta_{a_{2}{}b_{2}{}} \nonumber \\
& \hspace{5mm} + \hat{X}_{a_{1}{}} \hat{X}_{a_{2}{}} \hat{X}_{b_{1}{}} \eta_{b_{2}{}m} + \hat{X}_{a_{1}{}} \hat{X}_{a_{2}{}} \hat{X}_{b_{2}{}} \eta_{b_{1}{}m} - \tfrac{1}{4} \hat{X}_{m} \eta_{a_{1}{}b_{1}{}} \eta_{a_{2}{}b_{2}{}} \nonumber \\
& \hspace{5mm} + \tfrac{1}{2} \hat{X}_{m} \eta_{a_{1}{}a_{2}{}} \eta_{b_{1}{}b_{2}{}} - \tfrac{1}{4} \hat{X}_{m} \eta_{a_{1}{}b_{2}{}} \eta_{a_{2}{}b_{1}{}} + \tfrac{1}{2} \hat{X}_{b_{2}{}} \eta_{a_{1}{}a_{2}{}} \eta_{b_{1}{}m} \nonumber \\
& \hspace{5mm} - \tfrac{1}{4} \hat{X}_{a_{2}{}} \eta_{a_{1}{}b_{2}{}} \eta_{b_{1}{}m} - \tfrac{1}{4} \hat{X}_{a_{1}{}} \eta_{a_{2}{}b_{2}{}} \eta_{b_{1}{}m} + \tfrac{1}{2} \hat{X}_{b_{1}{}} \eta_{a_{1}{}a_{2}{}} \eta_{b_{2}{}m} \nonumber \\
& \hspace{5mm} - \tfrac{1}{4} \hat{X}_{a_{2}{}} \eta_{a_{1}{}b_{1}{}} \eta_{b_{2}{}m} - \tfrac{1}{4} \hat{X}_{a_{1}{}} \eta_{a_{2}{}b_{1}{}} \eta_{b_{2}{}m} \, . \label{P solution}
\end{align}
The tensor $\tilde{G}$ is found using eq.~\eqref{F&G constraints}. However, we still need to impose the symmetry property~\eqref{G - x1 to x2}. Since $\tilde{G}$ is an even function of $X$, \eqref{G - x1 to x2} implies
\begin{equation}
\tilde{G}_{a_{1} a_{2} b_{1} b_{2} c_{1} c_{2} }(X) = \tilde{G}_{b_{1} b_{2} a_{1} a_{2} c_{1} c_{2} }(X) \, . \label{G - parity odd x1 to x2}
\end{equation}
After some calculations we find that eq.~\eqref{G - parity odd x1 to x2} is satisfied automatically and does not result in any restrictions on $b$. 
Thus, the conservation equations and the proper transformation under the $z_1 \leftrightarrow z_{2}$ exchange fix the parity-odd sector up to a single 
overall coefficient. 

%%%%%%%%%%%%%%%%%%%%%%%%%%%%%%%%%%%%%%%%%%%%%%%%%%%%%%%%%%%%%%%%%%%%%%%%%%%%

\subsection{Point-switch identity}\label{subsection4.2}

%%%%%%%%%%%%%%%%%%%%%%%%%%%%%%%%%%%%%%%%%%%%%%%%%%%%%%%%%%%%%%%%%%%%%%%%%%

The last constraint to be imposed on the correlation function $\langle \cF \cF \cF \rangle$ is invariance under the permutation of points $z_{1}$ and $z_{3}$, i.e we must have
\begin{equation}
\langle \cF_{\a(4)}(z_{1}) \, \cF_{\b(4)}(z_{2}) \, \cF_{\g(4)}(z_{3}) \rangle = \langle  \cF_{\g(4)}(z_{3}) \, \cF_{\b(4)}(z_{2}) \, \cF_{\a(4)}(z_{1}) \rangle \, .
\label{a1}
\end{equation}
This results in the following constraint on $\cH$
\begin{align}
\cH_{\a(4) \b(4) \g(4) }( \boldsymbol{X}_{3} , \Q_{3} ) &=  \frac{1}{\boldsymbol{x}_{13}^{6} \boldsymbol{X}_{3}^{6}} \, \prod_{i=1}^{4} \hat{\boldsymbol{x}}_{13 \, \a_{i}}{}^{\a'_{i}} \hat{\boldsymbol{x}}_{13 \, \g_{i}}{}^{\g'_{i}} \hat{\boldsymbol{x}}_{13}{}^{ \b'_{i} \d_{i} } \hat{\boldsymbol{X}}_{3 \, \d_{i} \b_{i}} \nonumber \\
& \hspace{10mm} \times \cH_{ \g'(4) \b'(4) \a'(4) }( -\boldsymbol{X}_{1}^{\text{T}} , -\Q_{1} )\, . \label{FFF - z1 to z3}
\end{align}
It is clear that direct calculation of \eqref{FFF - z1 to z3} is inefficient due to i) the large number of tensor structures in the solution for $\cH$, and ii) the linear dependence between the structures. Therefore, we will need to consider some alternative approaches, which will be explored in the next subsections. 

The superfield correlator $\langle \cF \cF \cF \rangle$ contains only two independent component correlators
\begin{align}
&\langle J_{a_{1} a_{2}}(x_{1}) \, J_{b_{1} b_{2}}(x_{2}) \, J_{c_{1} c_{2}}(x_{3}) \rangle \, , \hspace{10mm} \langle Q_{a_{1} a_{2}, \a}(x_{1}) \, J_{b_{1} b_{2}}(x_{2}) \, Q_{c_{1} c_{2}, \g}(x_{3}) \rangle \, .  \label{FFF - component correlators}
\end{align}
These may be obtained by bar-projection of the three-point function $\langle \cF \cF \cF \rangle$ as follows\footnote{To express each of these correlators in the form \eqref{FFF - component correlators}, we combine symmetric pairs of spinor indices into a vector index as in \eqref{H - vector notation} and use eq.~\eqref{a00}.}
\begin{subequations}
	\begin{align}
	\langle J_{\a(4)}(x_{1}) \, J_{\b(4)}(x_{2}) \, J_{\g(4)}(x_{3}) \rangle &= \langle \cF_{\a(4)}(z_{1}) \, \cF_{\b(4)}(z_{2}) \, \cF_{\g(4)}(z_{3}) \rangle \big| \, , \\[2mm]
	\langle Q_{\a(4), \a}(x_{1}) \, J_{\b(4)}(x_{2}) \, Q_{\g(4), \g}(x_{3}) \rangle &= 
	D_{(3) \g} D_{(1) \a} \langle \cF_{\a(4)}(z_{1}) \, \cF_{\b(4)}(z_{2}) \, \cF_{\g(4)}(z_{3}) \rangle \big| \, .
	\end{align}
\end{subequations}
All correlators involving the components $S_{\a(3)}$ and $B_{\a(4)}$ in eq.~\eqref{eeee} either vanish or are expressed in terms of~\eqref{FFF - component correlators}
by virtue of~\eqref{F multiplet - component constraints}.
From eq.~\eqref{a1} it follows that the component correlators~\eqref{FFF - component correlators} satisfy the following point-switch identities
\begin{subequations}
	\label{a2}
	\begin{align}
	&\langle J_{a_{1} a_{2}}(x_{1}) \, J_{b_{1} b_{2}}(x_{2}) \, J_{c_{1} c_{2}}(x_{3}) \rangle = 
	\langle J_{c_{1} c_{2}}(x_{3}) \, J_{b_{1} b_{2}}(x_{2}) \, J_{a_{1} a_{2}}(x_{1}) \rangle \, , \label{a2.1} \\[2mm]
	& \langle Q_{a_{1} a_{2}, \a}(x_{1}) \, J_{b_{1} b_{2}}(x_{2}) \, Q_{c_{1} c_{2}, \g}(x_{3}) \rangle=-
	\langle Q_{c_{1} c_{2}, \g}(x_{3}) \, J_{b_{1} b_{2}}(x_{2}) \, Q_{a_{1} a_{2}, \a}(x_{1}) \rangle \, . \label{a2.2}
	\end{align}
\end{subequations}
These relations will be studied analytically (though with extensive use of \textit{Mathematica}) in subsections \ref{subsubsection4.2.1} and \ref{subsubsection4.2.2}. 
However, proving eqs.~\eqref{a2} is not sufficient to prove eq.~\eqref{a1}. The reason is that we cannot use
eqs.~\eqref{F multiplet - component constraints} because we have not yet proven that the conservation law on the third point is satisfied. 
In fact, it will follow once we prove eq.~\eqref{a1}. Hence, to prove eq.~\eqref{a1} at the component level we must consider 
all component correlators obtained from~\eqref{a1} by the action of the superspace covariant derivatives followed by bar-projection. 
This is, clearly, impractical. Therefore, our approach will be to study eq.~\eqref{a1} at higher orders in $\theta_i$ numerically, which we do in \ref{subsubsection4.2.3}.
For this we will keep $\theta_i$ arbitrary but use various numeric values for the space-time points $x_1, x_2, x_3$.
Then the components of $\langle \cF_{\a(4)}(z_{1}) \, \cF_{\b(4)}(z_{2}) \, \cF_{\g(4)}(z_{3}) \rangle$ will be
polynomials in $\theta_i$ with numeric coefficients. Since these polynomials are quite complicated we are confident in our results despite the proof not being fully analytic. 

%%%%%%%%%%%%%%%%%%%%%%%%%%%%%%%%%%%%%%%%%%%%%%%%%%%%%%%%%%%%%%%%%%%%%%%%%

\subsubsection{Component correlator \texorpdfstring{$\langle J J J \, \rangle$}{< J J J >} }\label{subsubsection4.2.1}

%%%%%%%%%%%%%%%%%%%%%%%%%%%%%%%%%%%%%%%%%%%%%%%%%%%%%%%%%%%%%%%%%%%%%%%

The computation of the component correlator $\langle J J J \, \rangle$ is relatively straightforward, explicitly we have
\begin{align}
\langle J_{\a(4)}(x_{1}) \, J_{\b(4)}(x_{2}) \, J_{\g(4)}(x_{3}) \rangle &= \langle \cF_{\a(4)}(z_{1}) \, \cF_{\b(4)}(z_{2}) \, \cF_{\g(4)}(z_{3}) \rangle \big| \nonumber \\ 
& = \frac{ \prod_{i=1}^{4} \hat{\boldsymbol{x}}_{13 \, \a_{i}}{}^{\a'_{i}} \, \hat{\boldsymbol{x}}_{23 \, \b_{i}}{}^{\b'_{i}} }{(\boldsymbol{x}_{13}^{2})^{3} (\boldsymbol{x}_{23}^{2})^{3}} \, \cH_{\a'(4) \b'(4) \g(4) }( \boldsymbol{X}_{3}, \Q_{3} ) \big| \, . \nonumber
\end{align}
Since bar-projections of any objects involving $\Q$ vanish, combined with the result
\begin{equation}
\cH_{ \a'(4) \b'(4) \g(4) }( \boldsymbol{X}_{3}, \Q_{3} ) \big| = F_{ \a'(4) \b'(4) \g(4) }( X_{12} ) \, ,
\end{equation}
we obtain
\begin{align}
\langle J_{\a(4)}(x_{1}) \, J_{\b(4)}(x_{2}) \, J_{\g(4)}(x_{3}) \rangle & = \frac{ \prod_{i=1}^{4} \hat{x}_{13 \, \a_{i}}{}^{\a'_{i}} \, \hat{x}_{23 \, \b_{i}}{}^{\b'_{i}} }{(x_{13}^{2})^{3} (x_{23}^{2})^{3}} \, F_{\a'(4) \b'(4) \g(4) }( X_{12} ) \, .
\label{e3}
\end{align}
If we convert this result into vector notation by combining pairs of spinor indices, and apply the identity
\begin{equation}
I_{a_{1} a'_{1}}(x) = -\frac{1}{2} (\g_{a_{1}})^{\a_{1} \a_{2}} (\g_{a'_{1}})^{\a'_{1} \a'_{2}} \, \hat{x}_{ \a_{1} \a'_{1} } \hat{x}_{ \a_{2} \a'_{2} } \, ,
\end{equation}
we obtain the result
\begin{equation}
\langle J_{a_{1} a_{2}}(x_{1}) \, J_{b_{1} b_{2}}(x_{2}) \, J_{c_{1} c_{2}}(x_{3}) \rangle = \frac{\cI_{a_{1} a_{2}, a'_{1} a'_{2}}(x_{13}) \, \cI_{b_{1} b_{2}, b'_{1} b'_{2}}(x_{23})}{(x_{13}^{2})^{3} (x_{23}^{2})^{3} } \, F_{a'_{1} a'_{2} b'_{1} b'_{2} c_{1} c_{2} }( X_{12} ) \, .
\end{equation}
If we now use \eqref{FFF - parity even expansion}, then this component correlator can be put in the covariant canonical form
\begin{equation}
\langle J_{a_{1} a_{2}}(x_{1}) \, J_{b_{1} b_{2}}(x_{2}) \, J_{c_{1} c_{2}}(x_{3}) \rangle = \frac{\cI_{a_{1} a_{2}, a'_{1} a'_{2}}(x_{13}) \, \cI_{b_{1} b_{2}, b'_{1} b'_{2}}(x_{23})}{ x_{13}^{3} x_{23}^{3} x_{12}^{3} } \, t_{a'_{1} a'_{2} b'_{1} b'_{2} c_{1} c_{2} }( X_{12} ) \, ,
\end{equation}
It then follows that the
constrained tensors $t_{a'_{1} a'_{2} b'_{1} b'_{2} c_{1} c_{2} }( X_{12} )$
appearing in $\langle J J J \rangle$ satisfy all the same properties as those present in the energy-momentum tensor three-point function 
$\langle T T T \rangle$, so we can simply use the known results. 

It was shown in~\cite{Osborn:1993cr} 
that the parity-even contribution to the three-point function $\langle T T T \rangle$ in general 
dimensions is fixed up to three independent coefficients. 
However, in three-dimensional theories there is linear dependence between the tensor structures due to the identity~\eqref{e2}. 
This reduces the number of independent structures down to two. The solution for $t_{a_{1} a_{2} b_{1} b_{2} c_{1} c_{2} }( X )$
found in~\cite{Osborn:1993cr} is the same as given in our eqs.~\eqref{FFF - parity even expansion}, \eqref{FFF - F solution}. 
It was also shown in~\cite{Osborn:1993cr} that this solution satisfies the $z_1 \leftrightarrow z_{3}$ point-switch identity. 
Since the solution for the correlator $\langle J J J \rangle$ is identical to that of $\langle T T T \rangle$ 
it follows that the three-point function $\langle J J J \rangle$ in~\eqref{e3} with $F_{\a(4) \b(4) \g(4) }$ defined in~\eqref{FFF - F solution} 
is compatible with the point-switch identity~\eqref{a2.1} for arbitrary $k_2$ and $k_4$. In this case there is a further relation between $k_2$ and
$k_4$ in eq.~\eqref{newnew3} however it does not affect the $z_1 \leftrightarrow z_{3}$ point-switch identity.

The parity odd sector of the energy-momentum tensor three-point function was obtained in~\cite{Giombi:2011rz}. It was shown that it is fixed up to one independent 
structure given in eqs.~\eqref{F parity odd solution}, \eqref{P solution}.\footnote{We use a different approach and notation than the authors in~\cite{Giombi:2011rz} however our results agree.} 
Hence, eqs.~\eqref{F parity odd solution}, \eqref{P solution} are also compatible 
with the point-switch identity~\eqref{a2.1}. In the remaining subsections we will consider the relation~\eqref{a1} at higher orders in $\theta_i$. 

%%%%%%%%%%%%%%%%%%%%%%%%%%%%%%%%%%%%%%%%%%%%%%%%%%%%%%%%%%%%%%%%%%%%%%%%%%%%%

\subsubsection{Component correlator \texorpdfstring{$\langle Q J Q \rangle$}{< Q J Q >}}\label{subsubsection4.2.2}

%%%%%%%%%%%%%%%%%%%%%%%%%%%%%%%%%%%%%%%%%%%%%%%%%%%%%%%%%%%%%%%%%%%%%%%%%%%%%%

The correlator  $\langle Q J Q \rangle$ can be computed as follows
\begin{align}
\langle Q_{\a(4), \a}(x_{1}) \, J_{\b(4)}(x_{2}) \, Q_{\g(4), \g}(x_{3}) \rangle &= D_{(3) \g} D_{(1) \a} \langle \cF_{\a(4)}(z_{1}) \, \cF_{\b(4)}(z_{2}) \, \cF_{\g(4)}(z_{3}) \rangle \big| \nonumber \\ 
& = D_{(3) \g} D_{(1) \a} \Big\{ \tfrac{ \prod_{i=1}^{4} \hat{\boldsymbol{x}}_{13 \, \a_{i}}{}^{\a'_{i}} \, \hat{\boldsymbol{x}}_{23 \, \b_{i}}{}^{\b'_{i}} }{(\boldsymbol{x}_{13}^{2})^{3} (\boldsymbol{x}_{23}^{2})^{3}} \, \cH_{ \a'(4) \b'(4) \g(4) }( \boldsymbol{X}_{3}, \Q_{3} ) \Big\} \Big|  \nonumber \\
&= A + B\,.
\end{align}
After evaluating the derivatives, one finds that the calculation is broken up into two relevant parts: the $A$ contribution is due to the derivatives hitting the prefactor,
\begin{equation}
A = D_{(3) \g} D_{(1) \a} \Bigg\{ \frac{ \prod_{i=1}^{4} \hat{\boldsymbol{x}}_{13 \, \a_{i}}{}^{\a'_{i}} \, \hat{\boldsymbol{x}}_{23 \, \b_{i}}{}^{\b'_{i}} }{(\boldsymbol{x}_{13}^{2})^{3} (\boldsymbol{x}_{23}^{2})^{3}} \Bigg\} \, \cH_{ \a'(4) \b'(4) \g(4) }( \boldsymbol{X}_{3}, \Q_{3} ) \bigg| \, ,
\end{equation}
while the $B$ contribution arises due to the derivatives hitting $\cH$, 
\begin{align}
B &= \frac{ \prod_{i=1}^{4} \hat{\boldsymbol{x}}_{13 \, \a_{i}}{}^{\a'_{i}} \, \hat{\boldsymbol{x}}_{23 \, \b_{i}}{}^{\b'_{i}} }{(\boldsymbol{x}_{13}^{2})^{3} (\boldsymbol{x}_{23}^{2})^{3}} \, D_{(3) \g} D_{(1) \a} \Big\{ \cH_{ \a'(4) \b'(4) \g(4) }( \boldsymbol{X}_{3}, \Q_{3} ) \Big\} \Big| \, .
\end{align}
Other combinations in which either derivative acts on the prefactor and $\cH$ result in terms that are at least linear in $\q_{1}, \q_{3}$ or $\Q_{3}$, which vanish upon bar projection, so they may be neglected. The $A$ contribution can be written in the form
\begin{align}
A &= \frac{1}{(x_{13}^{2})^{7/2} (x_{23}^{2})^{3}} \, \hat{x}_{13 \, \a}{}^{\a'} \hat{x}_{13 \, \a_{1}}{}^{\a'_{1}} \hat{x}_{13 \, \a_{2}}{}^{\a'_{2}} \hat{x}_{13 \, \a_{3}}{}^{\a'_{3}} \hat{x}_{13 \, \a_{4}}{}^{\a'_{4}} \nonumber \\
& \hspace{35mm} \times \hat{x}_{23 \, \b_{1}}{}^{\b'_{1}} \hat{x}_{23 \, \b_{2}}{}^{\b'_{2}} \hat{x}_{23 \, \b_{3}}{}^{\b'_{3}} \hat{x}_{23 \, \b_{4}}{}^{\b'_{4}} \, \cT^{A}_{\a', \a'(4) \b'(4) \g, \g(4)}(X_{12}) \, ,
\end{align}
with $\cT^{A}$ defined as
\begin{equation}
\cT^{A}_{\a, \a(4) \, \b(4) \, \g, \g(4) }(X) = - 10 \text{i} \, \ve_{\g (\a}  F_{ \a_{1} \a_{2} \a_{3} \a_{4} ) \b(4) \g(4) }(X) \, .
\end{equation}
Similarly if we evaluate the $B$ contribution, we find it can be written in the form
\begin{align}
B &= \frac{1}{(x_{13}^{2})^{7/2} (x_{23}^{2})^{3}} \, \hat{x}_{13 \, \a}{}^{\a'} \hat{x}_{13 \, \a_{1}}{}^{\a'_{1}} \hat{x}_{13 \, \a_{2}}{}^{\a'_{2}} \hat{x}_{13 \, \a_{3}}{}^{\a'_{3}} \hat{x}_{13 \, \a_{4}}{}^{\a'_{4}} \nonumber \\
& \hspace{35mm} \times \hat{x}_{23 \, \b_{1}}{}^{\b'_{1}} \hat{x}_{23 \, \b_{2}}{}^{\b'_{2}} \hat{x}_{23 \, \b_{3}}{}^{\b'_{3}} \hat{x}_{23 \, \b_{4}}{}^{\b'_{4}} \, \cT^{B}_{\a', \a'(4) \b'(4) \g , \g(4)}(X_{12}) \, ,
\end{align}
with $\cT^{B}$ given by the expression
\begin{align}
\cT^{B}_{\a, \a(4) \, \b(4) \, \g, \g(4) }(X) &= -\text{i} \, (\g^{m})_{\a \s} X^{\s}{}_{\g} \, \pa_{m} F_{ \a(4) \b(4) \g(4) }(X) - 2 X_{\a \g} G_{ \a(4) \b(4) \g(4) }(X) \, .
\end{align}
Hence, we see that the correlation function $\langle Q J Q \rangle$ may be written in the following covariant canonical form
\begin{align}
\langle Q_{\a(4), \a}(x_{1}) \, J_{\b(4)}(x_{2}) \, Q_{\g(4), \g}(x_{3}) \rangle &= \frac{1}{(x_{13}^{2})^{7/2} (x_{23}^{2})^{3}} \, \hat{x}_{13 \, \a}{}^{\a'} \hat{x}_{13 \, \a_{1}}{}^{\a'_{1}} \hat{x}_{13 \, \a_{2}}{}^{\a'_{2}} \hat{x}_{13 \, \a_{3}}{}^{\a'_{3}} \hat{x}_{13 \, \a_{4}}{}^{\a'_{4}} \nonumber \\
& \times \hat{x}_{23 \, \b_{1}}{}^{\b'_{1}} \hat{x}_{23 \, \b_{2}}{}^{\b'_{2}} \hat{x}_{23 \, \b_{3}}{}^{\b'_{3}} \hat{x}_{23 \, \b_{4}}{}^{\b'_{4}} \, \cT_{\a', \a'(4) \b'(4) \g, \g(4)}(X_{12}) \, ,
\end{align}
with
\begin{equation}
\cT_{\a, \a(4) \b(4) \g, \g(4)}(X) = \cT^{A}_{\a, \a(4) \b(4) \g, \g(4)}(X) + \cT^{B}_{\a, \a(4) \b(4) \g, \g(4)}(X) \, .
\end{equation}
Additional details regarding this calculation are contained in appendix \ref{AppB}. It is worth commenting that it is not immediately obvious that $\cT$ is totally symmetric in the $\a$ indices; indeed it may shown by direct calculations that this symmetry is manifest by virtue of \eqref{H - differential constraint} (and by extension \eqref{F&G constraints}).\footnote{Recall that in \eqref{F multiplet - component constraints} it was shown that the component field $Q$ is totally symmetric after imposing conservation of $\cF$. Since we have already imposed conservation of $\langle \cF \cF \cF \rangle$ at $z_{1}$, the fact that $\langle Q J Q \rangle$ is totally symmetric in $\a$ is implicit.}
To make subsequent calculations more tractable, we convert this entire expression into vector notation. The component three-point function may then be written in the following form
\begin{align}
\langle Q_{a_{1} a_{2}, \a}(x_{1}) \, J_{b_{1} b_{2} }(x_{2}) \, Q_{c_{1} c_{2}, \g}(x_{3}) \rangle &=  \frac{\hat{x}_{13}^{m}}{(x_{13}^{2})^{7/2} (x_{23}^{2})^{3}} \, \cI_{a_{1} a_{2}, a'_{1} a'_{2}}(x_{13}) \, \cI_{b_{1} b_{2}, b'_{1} b'_{2}}(x_{23}) \nonumber \\ 
& \hspace{30mm} \times \cT_{m, a'_{1} a'_{2} b'_{1} b'_{2} c_{1} c_{2}, \a \g}(X_{12}) \,. \label{QJQ - covariant representation}
\end{align}
It is convenient to decompose the tensor $\cT$ into the symmetric and antisymmetric parts
\begin{align}
\cT_{m, a_{1} a_{2} b_{1} b_{2} c_{1} c_{2}, \a \g}(X) = \ve_{\a \g} A_{m, a_{1} a_{2} b_{1} b_{2} c_{1} c_{2}}(X) 
+ (\g_{n})_{\a \g} S^{n}{}_{ m, a_{1} a_{2} b_{1} b_{2} c_{1} c_{2}}(X) \, .
\end{align}
We find the following expressions for  the tensors $A$ and $S$:
\begin{align}
A_{m, a_{1} a_{2} b_{1} b_{2} c_{1} c_{2}}(X) &= \text{i} \e_{m}{}^{p q} X_{q} \pa_{p} F_{ a_{1} a_{2} b_{1} b_{2} c_{1} c_{2} }(X) + 2 X_{m}  G_{ a_{1} a_{2} b_{1} b_{2} c_{1} c_{2} }(X) \nonumber \\
& \hspace{40mm} - 2\text{i} \, \P_{m, a_{1} a_{2} m_{1} m_{2}} F^{ m_{1} m_{2} }{}_{b_{1} b_{2} c_{1} c_{2} }(X) \, , \label{QJQ - A term}\\[2mm]
S^{n}{}_{m, a_{1} a_{2} b_{1} b_{2} c_{1} c_{2}}(X) &= \text{i} \mathfrak{D}^{n}{}_{m} F_{ a_{1} a_{2} b_{1} b_{2} c_{1} c_{2} }(X) + 2 \e^{n}{}_{m p} X^{p} G_{ a_{1} a_{2} b_{1} b_{2} c_{1} c_{2} }(X) \nonumber \\
& \hspace{40mm} - 2\text{i} \, \X^{n}{}_{m, a_{1} a_{2} m_{1} m_{2}} F^{m_{1} m_{2}}{}_{ b_{1} b_{2} c_{1} c_{2} }(X) \, . \label{QJQ - S term}
\end{align}
The differential operator $\mathfrak{D}$ and the constant ``projection" tensors $\P$, $\X$ naturally arise when expressing $\langle Q J Q \rangle$ in the covariant form \eqref{QJQ - covariant representation}. They have the following definitions
\begin{align}
\mathfrak{D}_{n m} &= X_{n} \pa_{m} - X_{m} \pa_{n} + \eta_{n m} X^{p} \pa_{p} - \eta_{n m} \, , \\[2mm]
\P_{m, a_{1} a_{2}, b_{1} b_{2}} &= \tfrac{1}{2} \epsilon_{a_{2}{}b_{2}{}n} \eta_{a_{1}{}b_{1}{}} + \tfrac{1}{2} \epsilon_{a_{2}{}b_{1}{}n} \eta_{a_{1}{}b_{2}{}} + \tfrac{1}{2} \epsilon_{a_{1}{}b_{2}{}n} \eta_{a_{2}{}b_{1}{}} + \tfrac{1}{2} \epsilon_{a_{1}{}b_{1}{}n} \eta_{a_{2}{}b_{2}{}} \, , \\[2mm]
\X_{n m, a_{1} a_{2} b_{1} b_{2}} &= \tfrac{1}{2} \eta_{a_{1}{}n} \eta_{a_{2}{}b_{2}{}} \eta_{b_{1}{}m} + \tfrac{1}{2} \eta_{a_{1}{}b_{2}{}} \eta_{a_{2}{}n} \eta_{b_{1}{}m} - \tfrac{1}{2} \eta_{a_{1}{}m} \eta_{a_{2}{}b_{2}{}} \eta_{b_{1}{}n} - \tfrac{1}{2} \eta_{a_{1}{}b_{2}{}} \eta_{a_{2}{}m} \eta_{b_{1}{}n} \nonumber \\
& \hspace{2mm} + \tfrac{1}{2} \eta_{a_{1}{}n} \eta_{a_{2}{}b_{1}{}} \eta_{b_{2}{}m} + \tfrac{1}{2} \eta_{a_{1}{}b_{1}{}} \eta_{a_{2}{}n} \eta_{b_{2}{}m} - \tfrac{1}{2} \eta_{a_{1}{}m} \eta_{a_{2}{}b_{1}{}} \eta_{b_{2}{}n} -  \tfrac{1}{2} \eta_{a_{1}{}b_{1}{}} \eta_{a_{2}{}m} \eta_{b_{2}{}n} \nonumber \\
& \hspace{2mm} - \eta_{a_{1}{}b_{2}{}} \eta_{a_{2}{}b_{1}{}} \eta_{mn} -  \eta_{a_{1}{}b_{1}{}} \eta_{a_{2}{}b_{2}{}} \eta_{mn} + \tfrac{2}{3} \eta_{a_{1}{}a_{2}{}} \eta_{b_{1}{}b_{2}{}} \eta_{mn} \, .
\end{align}
The point switch identity on $\langle Q J Q \rangle$, eq.~\eqref{a2.2}
%
%\begin{equation}
%	\langle Q_{a_{1} a_{2}, \a}(x_{1}) \, J_{b_{1} b_{2} }(x_{2}) \, Q_{c_{1} c_{2}, \g}(x_{3}) \rangle = - \langle Q_{c_{1} c_{2}, \g}(x_{3}) \, J_{b_{1} b_{2} }(x_{2}) \, Q_{a_{1} a_{2}, \a}(x_{1}) \rangle \, . \label{QJQ - z1 to z3}
%\end{equation}
%
%This will fix the remaining coefficients of the correlation function $\langle \cF \cF \cF \rangle$. In particular \eqref{QJQ - z1 to z3} implies the following constraints on $A$ and $S$
%
can be written in terms of the following two equations involving only vector indices:
\begin{subequations}
	\begin{align}
	\cI_{b_{1} b_{2}}{}^{ b'_{1} b'_{2}}(X) \, A_{m, a_{1} a_{2} b'_{1} b'_{2} c_{1} c_{2}}(X) + A_{m, c_{1} c_{2} b_{1} b_{2} a_{1} a_{2}}(-X) &= 0 \, , \label{QJQ - x1 to x3 A term}\\[2mm]
	\cI_{b_{1} b_{2}}{}^{ b'_{1} b'_{2}}(X) \, S^{n}{}_{m, a_{1} a_{2} b'_{1} b'_{2} c_{1} c_{2}}(X) - S^{n}{}_{m, c_{1} c_{2} b_{1} b_{2} a_{1} a_{2}}(-X) &= 0 \, . \label{QJQ - x1 to x3 S term}
	\end{align}
\end{subequations}
To recall, here the tensors $A$ and $S$ are given by eqs.~\eqref{QJQ - A term}, \eqref{QJQ - S term}, where the tensor $F$ is given by eq.~\eqref{FFF - F solution} 
in the parity-even case and in eqs.~\eqref{F parity odd solution}, \eqref{P solution} in the parity-odd case, and the tensor $G$ in both cases is obtained from $F$ 
using eq.~\eqref{F&G constraints}.

Now the task is to substitute $F$ and $G$ into eqs.~\eqref{QJQ - A term}, \eqref{QJQ - S term} in the parity-even and in the parity-odd cases separately to determine
if there are additional, different from eq.~\eqref{newnew3} constrains on the coefficients $k_2, k_4$ and $b$ in eqs.~\eqref{FFF - F solution} and~\eqref{F parity odd solution}. 
Since $A$ and $S$ have rather complicated definitions, it is futile to attempt to impose them by hand, however computation of these 
identities is possible in \textit{Mathematica} using the \textit{xAct} package \cite{MARTINGARCIA2008597}. The package allows for symbolic manipulation of tensors using index notation, and contains a suite of ``canonicalisation" functions which can essentially manipulate tensor structures down to their simplest form. In this way the computations are completely symbolic and are exactly the same as if they were done ``by hand". Once a given tensor is canonicalised, we can then convert the expression into an array using in-built functions. 

\noindent \textbf{Parity-even sector:}
Evaluating \eqref{QJQ - x1 to x3 A term} using definitions \eqref{QJQ - A term} and the solution \eqref{FFF - F solution} results in  $\approx400$ terms after canonicalisation. 
On the other hand, \eqref{QJQ - x1 to x3 S term} results in $\approx800$. The tensor structures in each identity should cancel amongst each other for some relation between the coefficients $k_{2}$ and $k_{4}$. However if we naively just collect all of the tensor structures one would find that $k_{2} = k_{4} = 0$, as there is a hidden linear dependence 
between the terms.  
A way around this is to go into a coordinate basis and check every component of the LHS of \eqref{QJQ - x1 to x3 A term} and \eqref{QJQ - x1 to x3 S term}. If we carry out this computation, the identities are satisfied for the choice $k_{2} = - 2 k_{4}$. Hence,  we do not get any new relations in the parity-even 
sector and it is still fixed up to an overall coefficient. 

\noindent \textbf{Parity-odd sector:}
We now carry out an identical analysis for the parity-odd solution \eqref{F parity odd solution}, which turns out to be more computationally intensive. In this case there are $\approx800$ tensor structures after canonicalisation of the LHS of \eqref{QJQ - x1 to x3 A term}, while there are $\approx1600$ for \eqref{QJQ - x1 to x3 S term}. 
If one goes into a coordinate basis the identities are satisfied for an arbitrary choice of the coefficient $b$. Hence, the parity-odd sector is also fixed up to a single tensor structure.

%%%%%%%%%%%%%%%%%%%%%%%%%%%%%%%%%%%%%%%%%%%%%%%%%%%%%%%%%%%%%%%%%%%%%%%%%%%%%%%%%%%%%%%%%%

\subsubsection{Numerical analysis}\label{subsubsection4.2.3}

%%%%%%%%%%%%%%%%%%%%%%%%%%%%%%%%%%%%%%%%%%%%%%%%%%%%%%%%%%%%%%%%%%%%%%%%%%%%%%%%%%%

To supplement the results above, we will carry out a numerical analysis of the point switch identity by substituting in various configurations of points. 
To do this, first we convert the ansatz \eqref{FFF - ansatz} into vector notation. This can be done by introducing the following $\cN = 1$ object
\begin{align}
I_{a b}(\boldsymbol{x}_{12}) &= - \frac{1}{2} (\g_{a})^{\a_{1} \a_{2}} (\g_{b})^{\a'_{1} \a'_{2}} (\hat{\boldsymbol{x}}_{12})_{ \a_{1} \a'_{1}} 
(\hat{\boldsymbol{x}}_{12})_{ \a_{2} \a'_{2}}
\nonumber  \\[2mm]
&= I_{a b}(y_{12}) - \text{i} \epsilon_{a b m} \, \hat{y}_{12}^{m} \frac{\theta^{2}}{y_{12}} \, .
\end{align}
To recall, $\boldsymbol{x}_{12}$ is given in eq.~\eqref{Two-point building blocks 1 - properties 1}, the vector $y_{12}$ is 
given in~\eqref{newnew1} and $\hat{y}_{12}^{m}= y_{12}^m/{y_{12}}$.
$I_{a b}(\boldsymbol{x}_{12})$  may be thought of as the supersymmetric generalisation of \eqref{Conformal inversion tensors}. It obeys some useful properties such as
\begin{equation}
I_{a}{}^{m}(\boldsymbol{x}_{12}) \, I_{m b}(-\boldsymbol{x}_{12}) = \eta_{a b} \, , \hspace{5mm}
I_{a}{}^{m}(\boldsymbol{x}_{12}) \, I_{m b}(\boldsymbol{x}_{12}) = \eta_{a b} - 2 \text{i} \epsilon_{a b m} \, \hat{y}_{12}^{m} \frac{\theta^{2}}{y_{12}} \, .
\end{equation}
Using this new object, the ansatz \eqref{FFF - ansatz} can be written in the form
\begin{align}
\langle \cF_{a_{1} a_{2}}(z_{1}) \, \cF_{b_{1} b_{2}}(z_{2}) \, \cF_{c_{1} c_{2}}(z_{3}) \rangle &= \frac{ I_{a_{1} a'_{1}}(\boldsymbol{x}_{13}) \, I_{a_{2} a'_{2}}(\boldsymbol{x}_{13}) \, I_{b_{1} b'_{1}}(\boldsymbol{x}_{23}) \, I_{b_{2} b'_{2}}(\boldsymbol{x}_{23}) }{(\boldsymbol{x}_{13}^{2})^{3} (\boldsymbol{x}_{23}^{2})^{3}} \nonumber \\
& \hspace{15mm} \times \, \cH_{a'_{1} a'_{2} b'_{1} b'_{2} c_{1} c_{2} }( \boldsymbol{X}_{3}, \Q_{3} ) \, . \label{FFF - vector ansatz}
\end{align}
Now to check the point switch identity, we will introduce null vectors $\l_{1}, \l_{2}, \l_{3}$, and contract them with the ansatz to obtain:
\begin{align}
\langle \cF(z_{1}) \, \cF(z_{2}) \, \cF(z_{3}) \rangle &= \langle \cF_{a_{1} a_{2}}(z_{1}) \, \cF_{b_{1} b_{2}}(z_{2}) \, \cF_{c_{1} c_{2}}(z_{3}) \rangle \, \l_{1}^{a_{1}} \,  \l_{1}^{a_{2}} \, \l_{2}^{b_{1}} \, \l_{2}^{b_{2}} \, \l_{3}^{c_{1}} \, \l_{3}^{c_{2}} \, . \label{FFF - polynomial}
\end{align}
Essentially our approach is to pick a configuration of points $x_{1}, x_{2}, x_{3}$, and null vectors $\l_{1}, \l_{2}, \l_{3}$, then expand out \eqref{FFF - polynomial} to all orders and combinations of the fermionic superspace coordinates $\theta_{1}, \theta_{2}, \theta_{3}$. This simplifies the point switch identity
\begin{align}
\langle \cF(z_{1}) \, \cF(z_{2}) \, \cF(z_{3}) \rangle = \langle \cF(z_{3}) \, \cF(z_{2}) \, \cF(z_{1}) \rangle \, , \label{FFF - point switch polynomial}
\end{align}
to a polynomial expression in the fermionic coordinates. We then check whether the point switch identity is satisfied for both the parity-even and parity-odd solutions that we found in Subsection \ref{subsection4.1}. To carry out these computations we must make use of the following expansions for the fermionic two-point (three-point) functions, which follow from the definitions \eqref{Two-point building blocks 1 - properties 1}, \eqref{Three-point building blocks 1b}:
\begin{subequations}
	\begin{align}
	\theta_{13}^{2} &= \theta_{1}^{2} + \theta_{3}^{2} - 2 \, \theta_{1} \cdot \theta_{3} \, ,  \\[2mm]
	\Theta_{3}^{2} &= \frac{\theta_{13}^{2}}{y_{13}^{2}} + \frac{\theta_{23}^{2}}{y_{23}^{2}} + \frac{2}{y_{13}^{2} y_{23}^{2}} \, y_{13}^{m} y_{23}^{n} \big\{ \eta_{m n} \theta_{13} \cdot \theta_{23} - \epsilon_{m n c} \, (\theta_{13} \cdot \g \cdot \theta_{23})^{c} \big\} \, ,
	\end{align}
\end{subequations}
where we have used the notation
\begin{equation}
\theta_{i} \cdot \theta_{j} = \theta_{i}^{\a} \theta_{j \, \a} \, , \hspace{5mm} ( \theta_{ij} \cdot \g \cdot \theta_{jk} )^{a} = (\g^{a})_{\a \b} \theta_{ij}^{\a} \theta_{jk}^{\b} \, .
\end{equation}
Expansions for the other building blocks are obtained by cyclic permutations of superspace points. Hence we see that the resulting polynomial from \eqref{FFF - polynomial} will be a function of $\theta_{i}^{2}$, $\theta_{i} \cdot \theta_{j}$, $( \theta_{ij} \cdot \g \cdot \theta_{jk} )^{a}$, and combinations/products of these objects.\footnote{Not all these objects 
	are linearly independent since $\theta_i$ are Grassmann odd but one can choose a convenient basis.}
All the $\theta$ expansions and numerical calculations are done computationally. We performed a numeric analysis for various 
configurations of points and null vectors and always found the same result. 
Below we present one example when the polynomials are relatively simple.

Let us  pick the following points and null vectors
\begin{subequations}
	\begin{align}
	x_{1} &= (0,-1,0) \, , &  x_{2} &= (0,1,0) \, , &  x_{3} &= (0,0,1) \, , \\
	\l_{1} &= (1,0,1) \, , &  \l_{2} &= (1,1,0) \, , & \l_{3} &= (1,-1,0) \, .
	\end{align}
\end{subequations}
We now substitute the above values into \eqref{FFF - point switch polynomial}. For the parity-even solution (denoted by subscript $E$) we obtain
\begin{align}
&\langle \cF(z_{1}) \, \cF(z_{2}) \, \cF(z_{3}) \rangle_{E} - \langle \cF(z_{3}) \, \cF(z_{2}) \, \cF(z_{1}) \rangle_{E} = (\tfrac{k_{2}}{128} + \tfrac{k_{4}}{64} ) \, \theta_{1}^{2} \theta_{2}^{2} + (\tfrac{k_{2}}{128} + \tfrac{k_{4}}{64} ) \, \theta_{1}^{2} \theta_{3}^{2}  \nonumber \\
& + (\tfrac{k_{2}}{128} + \tfrac{k_{4}}{64} ) \, \theta_{2}^{2} \theta_{3}^{2} + \text{i} (\tfrac{k_{2}}{64} + \tfrac{k_{4}}{32} ) \, \theta_{1} \cdot \theta_{3} + \text{i} (\tfrac{k_{2}}{64} + \tfrac{k_{4}}{32} ) \, \theta_{2} \cdot \theta_{3} - \text{i} (\tfrac{k_{2}}{64} + \tfrac{k_{4}}{32} ) \, ( \theta_{13} \cdot \g \cdot \theta_{23} )^{1} \nonumber \\
& + (\tfrac{k_{2}}{32} + \tfrac{k_{4}}{16} ) \, \theta_{1} \cdot \theta_{3} \, ( \theta_{13} \cdot \g \cdot \theta_{23} )^{1} - (\tfrac{k_{2}}{16} + \tfrac{k_{4}}{8} ) \, \theta_{2} \cdot \theta_{3} \, ( \theta_{13} \cdot \g \cdot \theta_{23} )^{1} \nonumber \\
& + \theta_{1}^{2} \big\{ - \tfrac{\text{i} k_{2}}{128} - \tfrac{\text{i} k_{4}}{64} - ( \tfrac{k_{2}}{64} + \tfrac{ k_{4}}{32} ) \,\theta_{2} \cdot \theta_{3} - ( \tfrac{k_{2}}{64} + \tfrac{ k_{4}}{32} ) \, ( \theta_{13} \cdot \g \cdot \theta_{23} )^{1} \big\} \nonumber \\
& + \theta_{2}^{2} \big\{ - \tfrac{\text{i} k_{2}}{128} - \tfrac{\text{i} k_{4}}{64} - ( \tfrac{k_{2}}{64} + \tfrac{ k_{4}}{32} ) \,\theta_{1} \cdot \theta_{3} + ( \tfrac{k_{2}}{32} + \tfrac{ k_{4}}{16} ) \, ( \theta_{13} \cdot \g \cdot \theta_{23} )^{1} \big\} \nonumber \\
& + \theta_{3}^{2} \big\{ - \tfrac{\text{i} k_{2}}{64} - \tfrac{\text{i} k_{4}}{32} - ( \tfrac{k_{2}}{64} + \tfrac{ k_{4}}{32} ) \,\theta_{1} \cdot \theta_{2} + ( \tfrac{k_{2}}{64} + \tfrac{ k_{4}}{32} ) \, ( \theta_{13} \cdot \g \cdot \theta_{23} )^{1} \big\} \, .
\end{align}
Clearly, this expression vanishes at each order of $\theta$ for the choice $k_{2} = - 2 k_{4}$ which is the same condition as found previously in eq.~\eqref{newnew3}. 
Hence, the numerical evaluations agree with our previous calculations and does not give any new relations. 

Next we perform the same calculation for the parity odd solution (denoted by $O$). Explicit evaluation of $\langle \cF(z_{1}) \, \cF(z_{2}) \, \cF(z_{3}) \rangle_{O}$ yields the following polynomial
\begin{align}
\langle \cF(z_{1}) \, \cF(z_{2}) \, \cF(z_{3}) \rangle_{O} &= - \tfrac{15}{64} b \, \theta_{1}^{2} \theta_{2}^{2} - \tfrac{15}{64} b \, \theta_{1}^{2} \theta_{3}^{2} - \tfrac{15}{64} b \, \theta_{2}^{2} \theta_{3}^{2} \nonumber \\
& + \tfrac{3 \text{i}}{16} b \, \theta_{1} \cdot \theta_{3} - \tfrac{3 \text{i}}{32} b \, \theta_{2} \cdot \theta_{3} - \tfrac{9 \text{i}}{32} b \, ( \theta_{13} \cdot \g \cdot \theta_{23} )^{1} \nonumber \\
& - \tfrac{3}{16} b \, \theta_{1} \cdot \theta_{3} \, ( \theta_{13} \cdot \g \cdot \theta_{23} )^{1} + \tfrac{3}{4} b \, \theta_{2} \cdot \theta_{3} \, ( \theta_{13} \cdot \g \cdot \theta_{23} )^{1} \nonumber \\
& + \theta_{1}^{2} \big\{ - \tfrac{3 \text{i}}{32} b + \tfrac{15}{32} b \,\theta_{2} \cdot \theta_{3} + \tfrac{3}{32} b \, ( \theta_{13} \cdot \g \cdot \theta_{23} )^{1} \big\} \nonumber \\
& + \theta_{2}^{2} \big\{ \tfrac{3 \text{i}}{64} b + \tfrac{15}{32} b \,\theta_{1} \cdot \theta_{3} - \tfrac{3}{8} b \, ( \theta_{13} \cdot \g \cdot \theta_{23} )^{1} \big\} \nonumber \\
& + \theta_{3}^{2} \big\{ - \tfrac{3 \text{i}}{64} b + \tfrac{15}{32} b \,\theta_{1} \cdot \theta_{2} - \tfrac{9}{32} b \, ( \theta_{13} \cdot \g \cdot \theta_{23} )^{1} \big\} \, .
\end{align}
When we similarly compute $\langle \cF(z_{3}) \, \cF(z_{2}) \, \cF(z_{1}) \rangle_{O}$ we find the same result. 
Hence, we observe cancellation at every order and therefore the odd solution also satisfies the point switch identity for an arbitrary coefficient $b$. 

Note that the polynomials in both parity-even and parity-odd cases are quite non-trivial even for a simple choice of the points and null vectors.  
We performed a similar numeric analysis for various other choices and obtained the same result as above. 
However, in all other cases the polynomials are quite large so we will not present them here. The complexity of the polynomials makes any 
accidental cancellations highly unlikely. Hence, we are  confident that the point-switch identity is satisfied for $k_{2} = - 2 k_{4}$ and arbitrary $b$.

%\noindent \textbf{Configurations 2 \& 3:}
%For other sets of points the outputs become quite cumbersome to present, so here we will simply state the explicit values of points that we tried. Setting $x_{i}$, $\l_{i}$ to the following values
% 
%\begin{subequations}
%	\begin{align}
%		x_{1} &= (-1,2,1) \, , & \hspace{5mm} x_{2} &= (0,1,-2) \, , & \hspace{5mm} x_{3} &= (1,3,1) \, , \\
%		\l_{1} &= (-1,1,0) \, , & \hspace{5mm} \l_{2} &= (1,0,1) \, , & \hspace{5mm} \l_{3} &= (5,-3,4) \, ,
%	\end{align}
%\end{subequations}
%
%we found that the point switch was satisfied for both the parity even solution (for the choice $k_{2} = - 2k_{4}$) and parity odd solution. The same conclusion was found for following set of points and null vectors
% 
%\begin{subequations}
%	\begin{align}
%		x_{1} &= (1,6,2) \, , & \hspace{5mm} x_{2} &= (-1,3,-4) \, , & \hspace{5mm} x_{3} &= (2,7,0) \, , \\
%		\l_{1} &= (-1,0,1) \, , & \hspace{5mm} \l_{2} &= (-2,2,0) \, , & \hspace{5mm} \l_{3} &= (5,0,5) \, .
%	\end{align}
%\end{subequations}
%

%%%%%%%%%%%%%%%%%%%%%%%%%%%%%%%%%%%%%%%%%%%%%%%%%%%%%%%%%%%%%%%%%%%%%%%%%%%%%%%%%%%%%

\subsection{Summary of results}\label{subsection4.3}

%%%%%%%%%%%%%%%%%%%%%%%%%%%%%%%%%%%%%%%%%%%%%%%%%%%%%%%%%%%%%%%%%%%%%%%%%%%%%%%%%%%%

Since our analysis is rather technical and involves analytic and numeric computations of the superfield and component expressions, we will collect all the pieces together and summarise our results. 
%In subsection \ref{subsection4.1} we carried out a superfield analysis of the three-point function $\langle \cF \cF \cF \rangle$, in particular we showed that the parity even sector of the correlation function is fixed up to two independent coefficients after imposing conservation and invariance under permutation of points $z_{1}$ and $z_{2}$. 
%In a similar manner the parity odd sector is found to be fixed up to a single tensor structure. 
%Next we analyse the $z_{1} \leftrightarrow z_{3}$ point-switch identity. In subsection \ref{subsection4.2} 
%we compute the two independent component correlators of conserved currents
%contained within $\langle \cF \cF \cF \rangle$. We found that the point-switch identity for the $\langle Q J Q \rangle$ component correlator 
%further fixes the parity-even sector up to a single tensor structure, while the parity-odd solution satisfied 
%the point-switch identity automatically. The calculations are totally analytic, and were carried out using the \textit{xAct} package \cite{MARTINGARCIA2008597}. 
%Next we supplemented this analytic calculation by a numerical check of the $\langle \cF \cF \cF \rangle$ point-switch identity for various point configurations. 
%We obtained no further relations between the coefficients. 
We found that the correlation function $\langle \cF \cF \cF \rangle$ contains two independent tensor structures after imposing all of the constraints; 
one of them is parity-even, while the other is parity-odd. 
In particular, we found that the parity-odd contribution 
is present unlike in all cases of three-point functions involving the supercurrent and flavour current multiplets~\cite{Buchbinder:2015qsa, Buchbinder:2021gwu}.

The correlation function found above has the following structure
\begin{equation}
\langle \cF_{\a(4)}(z_{1}) \, \cF_{\b(4)}(z_{2}) \, \cF_{\g(4)}(z_{3}) \rangle = \frac{ \prod_{i=1}^{4} \hat{\boldsymbol{x}}_{13 \, \a_{i}}{}^{\a'_{i}} \, \hat{\boldsymbol{x}}_{23 \, \b_{i}}{}^{\b'_{i}} }{(\boldsymbol{x}_{13}^{2})^{3} (\boldsymbol{x}_{23}^{2})^{3}} \, \cH_{\a'(4) \b'(4) \g(4) }( \boldsymbol{X}_{3}, \Q_{3} ) \, ,
\end{equation}
where $\cH$ can also be written  as follows
\begin{align}
\cH_{\a_{1} \a_{2} \a_{3} \a_{4} \b_{1} \b_{2} \b_{3} \b_{4} \g_{1} \g_{2} \g_{3} \g_{4} }( \boldsymbol{X} , \Q ) &= (\g^{a_{1}})_{\a_{1} \a_{2}} (\g^{a_{2}})_{\a_{3} \a_{4}} (\g^{b_{1}})_{\b_{1} \b_{2}} (\g^{b_{2}})_{\b_{3} \b_{4}} \nonumber \\
& \hspace{10mm} \times (\g^{c_{1}})_{\g_{1} \g_{2}} (\g^{c_{2}})_{\g_{3} \g_{4}} \cH_{a_{1} a_{2} b_{1} b_{2} c_{1} c_{2} }( \boldsymbol{X} , \Q ) \, .
\end{align}
The tensor $\cH$ in vector notation then may be split into parity-even and parity-odd sectors
\begin{equation}
\cH_{a_{1} a_{2} b_{1} b_{2} c_{1} c_{2} }(\boldsymbol{X}, \Q) = \cH_{a_{1} a_{2} b_{1} b_{2} c_{1} c_{2} }(\boldsymbol{X}, \Q)_{E} + \cH_{a_{1} a_{2} b_{1} b_{2} c_{1} c_{2} }(\boldsymbol{X}, \Q)_{O} \, ,
\end{equation}
where each solution admits the following expansion
\begin{subequations}
	\begin{align}
	\cH_{a_{1} a_{2} b_{1} b_{2} c_{1} c_{2} }(\boldsymbol{X}, \Q)_E  &= F_{a_{1} a_{2} b_{1} b_{2} c_{1} c_{2} }(X) + \Q^{2} G_{a_{1} a_{2} b_{1} b_{2} c_{1} c_{2} }(X) \, , \\
	\cH_{a_{1} a_{2} b_{1} b_{2} c_{1} c_{2} }(\boldsymbol{X}, \Q)_O  &= \tilde{F}_{a_{1} a_{2} b_{1} b_{2} c_{1} c_{2} }(X) + \Q^{2} 
	\tilde{G}_{a_{1} a_{2} b_{1} b_{2} c_{1} c_{2} }(X) \, , 
	\end{align}
\end{subequations}
with $G$, $\tilde{G}$ determined in terms of $F$, $\tilde{F}$ by the equations~\eqref{F&G constraints},
After imposing all the constraints, we find that the solution for the tensor $F$ in the even sector is
\begin{align}
F_{a_{1} a_{2} b_{1} b_{2} c_{1} c_{2} }(X) &= \frac{a}{X^{3}} \, \Big\{ -2 \, t^{2}_{a_{1} a_{2} b_{1} b_{2} c_{1} c_{2} }(\hat{X}) + 2 \, t^{3}_{a_{1} a_{2} b_{1} b_{2} c_{1} c_{2} }(\hat{X}) + t^{4}_{a_{1} a_{2} b_{1} b_{2} c_{1} c_{2} }(\hat{X}) \nonumber \\
& \hspace{20mm} + t^{5}_{a_{1} a_{2} b_{1} b_{2} c_{1} c_{2} }(\hat{X}) - 15 \, t^{6}_{a_{1} a_{2} b_{1} b_{2} c_{1} c_{2} }(\hat{X}) \nonumber \\
& \hspace{25mm} + 5 \, t^{7}_{a_{1} a_{2} b_{1} b_{2} c_{1} c_{2} }(\hat{X}) - 35 \, t^{8}_{a_{1} a_{2} b_{1} b_{2} c_{1} c_{2} }(\hat{X})  \Big\} \, ,
\end{align}
where we have relabeled $k_{4} \rightarrow a$ and $t^i (\hat{X})$ are given by eqs.~\eqref{newnew10}, \eqref{newnew11}.

On the other hand, we find the solution in the odd sector to be
\begin{align}
\tilde{F}_{a_{1} a_{2} b_{1} b_{2} c_{1} c_{2} }(X) &= \frac{b}{X^{3}} \big\{ \e_{a_{1} b_{1}}{}^{m} P_{m, a_{2} b_{2} c_{1} c_{2}} (\hat{X}) + \e_{a_{1} b_{2}}{}^{m} P_{m, a_{2} b_{1} c_{1} c_{2}}(\hat{X}) \nonumber \\
& \hspace{15mm} + \e_{a_{2} b_{1}}{}^{m} P_{m, a_{1} b_{2} c_{1} c_{2}}(\hat{X}) + \e_{a_{2} b_{2}}{}^{m} P_{m, a_{1} b_{1} c_{1} c_{2}}(\hat{X}) \big\} \, ,
\end{align}
with $P$ defined as in \eqref{P solution}.

%%%%%%%%%%%%%%%%%%%%%%%%%%%%%%%%%%%%%%%%%%%%%%%%%%%%%%%%%%%%%%%%%%%%%%%%%

\section{Mixed correlators}\label{section5}

In this section we will discuss some basic examples of three-point functions of $\cF$ with other fields such as a scalar superfield $\cO$ of dimension $\D$, 
and the non-abelian flavour current superfield $L^{\abar}_{\a}$ of dimension $3/2$. The calculations are straightforward compared to the $\langle \cF \cF \cF \rangle$ case, 
so we will not require computational methods here. These three-point functions were also previously studied in~\cite{Nizami:2013tpa}, however, our method
is different and more explicit. 

%%%%%%%%%%%%%%%%%%%%%%%%%%%%%%%%%%%%%%%%%%%%%%%%%%%%%%%%%%%%%%%%%%%%%%%%%

%%%%%%%%%%%%%%%%%%%%%%%%%%%%%%%%%%%%%%%%%%%%%%%%%%%%%%%%%%%%%%%%%%%%%%%%%
\subsection{Correlation function \texorpdfstring{$\langle \cO \cF \cO \rangle$}{< O F O >}}\label{subsection5.1}

Let us now compute the correlation function $\langle \cO \cF \cO \rangle$, which admits the general ansatz
\begin{equation}
\langle \cO(z_{1}) \, \cF_{\a(4)}(z_{2}) \, \cO(z_{3}) \rangle = \frac{\prod_{i=1}^{4} \boldsymbol{x}_{23 \, \a_{i}}{}^{\a'_{i}} }{(\boldsymbol{x}_{13}^{2})^{\D} (\boldsymbol{x}_{23}^{2})^{3} } \, \cH_{\a'(4)}( \boldsymbol{X}_{3}, \Q_{3} ) \, .
\end{equation}
As usual, the tensor $\cH$ is required to satisfy covariant constraints arising from conservation equations and point-switch identities. They are summarised below:
\begin{enumerate}
	\item[\textbf{(i)}] \textbf{Homogeneity constraint}
	
	Covariance of the correlation function under scale transformations of superspace results in the following constraint on $\cH$
	\begin{align}
	\cH_{\a(4)}( \l^{2} \boldsymbol{X} , \l \Q ) = (\l^{2})^{-3} \cH_{\a(4) }( \boldsymbol{X} , \Q ) \, , \label{OFO - homogeneity constraint}
	\end{align}
	which implies that $\cH$ is a homogeneous tensor field of degree $-3$.
	
	\item[\textbf{(ii)}] \textbf{Differential constraints}
	
	The conservation equation \eqref{F - conservation equation} implies that the correlation function must satisfy the following constraint
	\begin{equation}
	D_{(2)}^{\s} \langle \cO(z_{1}) \, \cF_{\s \a(3)}(z_{2}) \, \cO(z_{3}) \rangle = 0 \, . \label{OFO - differential constraint}
	\end{equation}
	Application of the identities \eqref{Three-point building blocks 1c - differential identities 3} results in the following differential constraint on $\cH$
	\begin{equation}
	\cQ^{\s} \cH_{\s \a(3) }( \boldsymbol{X} , \Q ) = 0 \, .
	\end{equation}
	
	\item[\textbf{(iii)}] \textbf{Point-switch identity}
	
	Invariance under permutation of the superspace points $z_{1}$ and $z_{3}$ results in the following constraint on the correlation function
	\begin{equation}
	\langle \cO(z_{1}) \, \cF_{\a(4)}(z_{2}) \, \cO(z_{3}) \rangle = \langle \cO(z_{3}) \, \cF_{\a(4)}(z_{2})  \, \cO(z_{1}) \rangle \, ,
	\end{equation}
	which results in the following constraint on $\cH$
	\begin{equation}
	\cH_{\a(4)}( \boldsymbol{X}_{3} , \Q_{3} ) = \frac{\prod_{i=1}^{4} \hat{\boldsymbol{x}}_{13}^{\a'_{i} \d_{i} } \hat{\boldsymbol{X}}_{3 \, \d_{i} \a_{i}} }{\boldsymbol{x}_{13}^{6} \boldsymbol{X}_{3}^{6}} \, \cH_{\a'(4) }( -\boldsymbol{X}_{1}^{\text{T}} , -\Q_{1} ) \, . \label{OFO - z1 to z3}
	\end{equation}
\end{enumerate}

Now we must construct an explicit solution; analogous to the $\langle \cF \cF\cF \rangle$ case, we combine symmetric pairs of spinor indices into vector ones as follows:
\begin{equation}
\cH_{\a_{1} \a_{2} \a_{3} \a_{4} }( \boldsymbol{X}, \Q ) = (\g^{a_{1}} )_{\a_{1}\a_{2} } (\g^{a_{2}} )_{\a_{3} \a_{4} } \cH_{ a_{1} a_{2} }( \boldsymbol{X}, \Q) \, ,
\end{equation}
where it is required that $\cH$ in vector notation is both symmetric and traceless. It has the expansion
\begin{equation}
\cH_{a_{1} a_{2}}( \boldsymbol{X}, \Q ) = F_{a_{1} a_{2} }(X) + \Q^{2} G_{a_{1} a_{2}}(X) \, .
\end{equation}
The component fields $F$ and $G$ are both required to be symmetric and traceless. If we now impose \eqref{OFO - differential constraint}, we obtain the constraints
\begin{equation}
\pa^{a_{1}} F_{a_{1} a_{2}} = 0 \, , \hspace{10mm} G_{a_{1} a_{2}} = \tfrac{\text{i}}{2} \e_{(a_{1}}{}^{m n} \pa_{n} F_{a_{2}) m} \, . \label{OFO - differential constraints 2}
\end{equation}
Therefore we need only solve for the field $F$. A general expansion consistent with the tensor symmetries and homogeneity is
\begin{equation}
F_{a_{1} a_{2} } = \frac{c}{X^{3}} \, \bigg\{ \eta_{a_{1} a_{2}} - \frac{3 X_{a_{1}} X_{a_{2}} }{X^{2}} \bigg\} \, .
\end{equation}
Note that no parity violating structures are permitted as there is simply not enough indices on the tensor $F$ to allow for such contributions. Substituting this solution into \eqref{OFO - differential constraints 2} shows that it is satisfied for any value of $c$, while $G=0$. The final solution for the tensor $\cH$ in spinor notation is
\begin{align}
\cH_{\a_{1} \a_{2} \a_{3} \a_{4}}(\boldsymbol{X}, \Q) &= c \, \bigg\{ - \frac{1}{\boldsymbol{X}^{3} } \big( \ve_{\a_{1} \a_{3}} \ve_{\a_{2} \a_{4}} + \ve_{\a_{1} \a_{4}} \ve_{\a_{2} \a_{3}} \big) - \frac{3}{\boldsymbol{X}^{5}} \boldsymbol{X}_{\a_{1} \a_{2}} \boldsymbol{X}_{\a_{3} \a_{4}} \nonumber \\[2mm]
& \hspace{20mm} - \frac{3 \text{i}}{2} \big( \ve_{\a_{1} \a_{2}} \boldsymbol{X}_{\a_{3}\a_{4}} + \ve_{\a_{3} \a_{4}} \boldsymbol{X}_{\a_{1}\a_{2}} \big) \frac{\Q^{2}}{\boldsymbol{X}^{5}} \, \bigg\} \, .
\end{align}
Indeed, substitution of this solution into \eqref{OFO - z1 to z3} demonstrates that it is compatible with the point-switch identity. Hence this correlation function is determined up to a single parity-even tensor structure.  A similar result was obtained in~\cite{Nizami:2013tpa}.

%%%%%%%%%%%%%%%%%%%%%%%%%%%%%%%%%%%%%%%%%%%%%%%%%%%%%%%%%%%%%%%%%%%%%%%%%

\subsection{Correlation function \texorpdfstring{$\langle \cF L L \rangle$}{< F L L >}}\label{subsection5.2}

%%%%%%%%%%%%%%%%%%%%%%%%%%%%%%%%%%%%%%%%%%%%%%%%%%%%%%%%%%%%%%%%%%%%%%%%%

In this subsection we will compute the correlation function $\langle \cF L L \rangle$, where $L$ is the non-abelian flavour current superfield of dimension $3/2$ which obeys the conservation equation
\begin{equation}
D^{\a} L^{\abar}_{\a} = 0 \, . \label{L - conservation equation}
\end{equation}
The correlation function admits the general ansatz
\begin{equation}
\langle \cF_{\a(4)}(z_{1}) L^{\abar}_{\b}(z_{2}) L^{\bbar}_{\g}(z_{3}) \rangle = \d^{\abar \bbar} \, \frac{ \big( \prod_{i=1}^{4} \hat{\boldsymbol{x}}_{13 \, \a_{i}}{}^{\a'_{i} } \big) \, \hat{\boldsymbol{x}}_{23 \, \b}{}^{\b'}  }{(\boldsymbol{x}_{13}^{2})^{3} (\boldsymbol{x}_{23}^{2})^{3/2} } \, \cH_{\a'(4) \b' \g}( \boldsymbol{X}_{3}, \Q_{3} ) \, .
\end{equation}
The constraints on this three-point function are summarised below:
\begin{enumerate}
	\item[\textbf{(i)}] \textbf{Homogeneity constraint}
	
	Covariance under scale transformations of superspace results in the following constraint on $\cH$
	\begin{align}
	\cH_{\a(4) \b \g}( \l^{2} \boldsymbol{X} , \l \Q ) = (\l^{2})^{-3} \cH_{\a(4) \b \g }( \boldsymbol{X} , \Q ) \, , \label{FLL - homogeneity constraint}
	\end{align}
	which implies that $\cH$ is a homogeneous tensor field of degree $-3$.
	
	\item[\textbf{(ii)}] \textbf{Differential constraints}
	
	The conservation equations \eqref{F - conservation equation}, \eqref{L - conservation equation} imply the following constraints
	\begin{subequations}
		\begin{align}
		D_{(1)}^{\s} \langle \cF_{ \s \a(3)}(z_{1}) \, L^{\abar}_{\a}(z_{2}) \, L^{\bbar}_{\g}(z_{3}) \rangle &= 0 \, , \label{HLL - differential constraint 1a} \\[2mm] 
		D_{(2)}^{\b} \langle \cF_{\a(4)}(z_{1}) \, L^{\abar}_{\b}(z_{2}) \, L^{\bbar}_{\g}(z_{3}) \rangle &= 0 \, . \label{HLL - differential constraint 1b}
		\end{align}
	\end{subequations}
	Application of the identities \eqref{Three-point building blocks 1c - differential identities 3} then gives
	\begin{subequations}
		\begin{align}
		\cD^{\s} \cH_{\s \a(3) \b \g}( \boldsymbol{X} , \Q ) = 0 \, , \label{HLL - differential constraint 2a} \\[2mm]
		\cQ^{\b} \cH_{\a(4) \b \g}( \boldsymbol{X} , \Q ) = 0 \, . \label{HLL - differential constraint 2b}
		\end{align}
	\end{subequations}
	
	\item[\textbf{(iii)}] \textbf{Point-switch identity}
	
	Invariance under permutation of the superspace points $z_{2}$ and $z_{3}$ is equivalent to the condition
	\begin{equation}
	\langle \cF_{\a(4)}(z_{1}) \, L^{\abar}_{\b}(z_{2}) \, L^{\bbar}_{\g}(z_{3}) \rangle = - \langle \cF_{\a(4)}(z_{1}) \, L^{\bbar}_{\g}(z_{3}) \, L^{\abar}_{\b}(z_{2}) \rangle \, ,
	\end{equation}
	which results in the following constraint on $\cH$
	\begin{equation}
	\cH_{\a(4) \b \g}( \boldsymbol{X}_{3} , \Q_{3} ) = \frac{ \hat{\boldsymbol{x}}_{23 \, \b}{}^{\b'} \hat{\boldsymbol{x}}_{23 \, \g}{}^{\g'} \prod_{i=1}^{4} \hat{\boldsymbol{x}}_{23}^{\a'_{i} \d_{i} } \hat{\boldsymbol{X}}_{3 \, \a_{i} \d_{i} } }{\boldsymbol{x}_{23}^{6} \boldsymbol{X}_{3}^{6}} \, \cH_{\a'(4) \g' \b' }( -\boldsymbol{X}_{2}^{\text{T}} , -\Q_{2} ) \, . \label{HLL - z2 to z3}
	\end{equation}
\end{enumerate}

As before we combine symmetric pairs of spinor indices into vector ones as follows:
\begin{equation}
\cH_{\a_{1} \a_{2} \a_{3} \a_{4} \b \g }( \boldsymbol{X}, \Q ) = (\g^{a_{1}} )_{\a_{1} \a_{2} } (\g^{a_{2}} )_{\a_{3} \a_{4} } \cH_{ a_{1} a_{2}, \b \g }( \boldsymbol{X}, \Q) \, .
\end{equation}
The above decomposition holds provided that $\cH_{a_{1} a_{2}, \, \b \g} $ is symmetric and traceless in $a_{1}, a_{2}$.\footnote{In the RHS we require that the antisymmetric part in $\a_{2}$, $\a_{3}$ vanishes. Using eq. \eqref{Gamma product}, it can be seen that $\cH$ must be symmetric and traceless in $a_{1}$ and $a_{2}$.} We then expand this in irreducible components as follows:
\begin{equation}
\cH_{a_{1} a_{2} , \, \b \g}( \boldsymbol{X}, \Q ) = \ve_{\b \g} A_{a_{1} a_{2} }( \boldsymbol{X} , \Q) + (\g^{c})_{\b \g} S_{a_{1} a_{2}, c}(\boldsymbol{X}, \Q) \, ,
\end{equation}
with
\begin{subequations}
	\begin{align}
	A_{a_{1} a_{2}}( \boldsymbol{X}, \Q) &= A^{1}_{a_{1} a_{2}}(X) + \Q^{2} A^{2}_{a_{1} a_{2}}(X) \, , \\[2mm]
	S_{a_{1} a_{2}, c}( \boldsymbol{X}, \Q) &= S^{1}_{a_{1} a_{2}, c}(X) + \Q^{2} S^{2}_{a_{1} a_{2}, c}(X) \, .
	\end{align}
\end{subequations}
Here the $A^{i}$ and $S^{i}$ are both symmetric and traceless in $a_{1}, a_{2}$. Imposing the differential relation \eqref{HLL - differential constraint 2a} results in the following constraints on the tensors $A^{i}$ and $S^{i}$
\begin{subequations}
	\begin{align} 
	\pa^{m} A^{1}_{m a_{2}}(X) &= 0 \, , & A^{2}_{a_{1} a_{2}}(X) &= - \tfrac{\text{i}}{2} \e_{(a_{1}}{}^{m n} \pa_{m} A^{1}_{a_{2}) n}(X) \, , \label{HLL - differential constraint 3a - 1} \\[2mm]
	\pa^{m} S^{1}_{m a_{2},c}(X) &= 0 \, , & S^{2}_{a_{1} a_{2},c}(X) &= - \tfrac{\text{i}}{2} \e_{(a_{1}}{}^{m n} \pa_{m} S^{1}_{a_{2}) n,c}(X) \, , \label{HLL - differential constraint 3a - 2}
	\end{align}
\end{subequations}
while \eqref{HLL - differential constraint 2b} gives the additional relations
\begin{subequations}
	\begin{align}
	A^{2}_{a_{1} a_{2}}(X) & = \tfrac{\text{i}}{2} \pa^{m} S^{1}_{a_{1} a_{2}, m}(X) \, , \label{HLL - differential constraint 3b - 1} \\[2mm]
	S^{2}_{a_{1} a_{2}, c}(X) &= - \tfrac{\text{i}}{2} \big\{ \pa_{c} A^{1}_{a_{1} a_{2} }(X) + \e_{c}{}^{m n} \pa_{m} S^{1}_{ a_{1} a_{2}, n}(X) \big\} \, . \label{HLL - differential constraint 3b - 2}
	\end{align}
\end{subequations}
Hence, we may treat $A^{1}$ and $S^{1}$ as independent. The only solution for the tensor $A^{1}$ compatible with the symmetries is
\begin{align}
A^{1}_{a_{1} a_{2}}(X) &= \frac{c}{X^{3}} \bigg\{ \eta_{a_{1} a_{2}} - \frac{3 X_{a_{1}} X_{a_{2} }}{X^{2}} \bigg\} \, ,
\end{align}
while for $S$ we have the general ansatz
\begin{align}
S^{1}_{a_{1} a_{2},c}(X) &= k_{1} \frac{X_{a_{1}} X_{a_{2}} X_{c}}{X^{6}} + k_{2} \, \bigg\{ \frac{\e_{a_{1} c}{}^{m} X_{m} X_{a_{2}}}{X^{5}} +  \frac{\e_{a_{2} c}{}^{m} X_{m} X_{a_{1}}}{X^{5}} \bigg\} \nonumber \\ 
& \hspace{10mm} + k_{3} \, \bigg\{ \frac{ \eta_{c a_{1} } X_{a_{2}}}{X^{4}} +  \frac{ \eta_{c a_{2} } X_{a_{1}}  }{X^{4}} \bigg\} + k_{4} \frac{\eta_{a_{1} a_{2}} X_{c} }{X^{4}} \, .
\end{align}
Imposing tracelessness in $a_{1}, a_{2}$ on this expansion results in the constraint
\begin{equation}
k_{3} = - \tfrac{3}{2} k_{4} - \tfrac{1}{2} k_{1} \, .
\end{equation}
The solution then becomes
\begin{align}
S^{1}_{a_{1} a_{2},c}(X) &= k_{1} \, \bigg\{ \frac{X_{a_{1}} X_{a_{2}} X_{c}}{X^{6}} - \frac{1}{2} \frac{ \eta_{c a_{2}} X_{a_{1}}}{X^{4}} - \frac{1}{2} \frac{ \eta_{c a_{1}} X_{a_{2}}}{X^{4}} \bigg\} \nonumber \\
& \hspace{5mm} + k_{2} \, \bigg\{ \frac{\e_{a_{1} c}{}^{m} X_{m} X_{a_{2}}}{X^{5}} +  \frac{\e_{a_{2} c}{}^{m} X_{m} X_{a_{1}}}{X^{5}} \bigg\} \nonumber \\ 
& \hspace{5mm} + k_{4} \, \bigg\{ \frac{\eta_{a_{1} a_{2}} X_{c} }{X^{4}} - \frac{3}{2} \frac{\eta_{c a_{2}} X_{a_{1}}}{X^{4}} - \frac{3}{2} \frac{\eta_{c a_{1}} X_{a_{2}}}{X^{4}} \bigg\} \, .
\end{align}
It remains to impose the differential constraints. In particular, the equations \eqref{HLL - differential constraint 3a - 1}, \eqref{HLL - differential constraint 3a - 2} result in
\begin{equation}
k_{1} = k_{4} = 0 \, ,
\end{equation}
while $A^{2}$ vanishes. After making the replacement $k_{2} \rightarrow \tilde{c}$, the solutions for the tensors $A^{i}$, $S^{i}$ now become
\begin{subequations}
	\begin{align}
	A^{1}_{a_{1} a_{2}}(X) &= \frac{c}{X^{3}} \, \bigg\{ \eta_{a_{1} a_{2}} - \frac{3 X_{a_{1}} X_{a_{2} }}{X^{2}} \bigg\} \, , \hspace{10mm} A^{2}_{a_{1} a_{2}}(X) = 0 \, , \label{FLL - solutions a} \\[2mm]
	S^{1}_{a_{1} a_{2},c}(X) &= \tilde{c} \, \bigg\{ \frac{\e_{a_{1} c}{}^{m} X_{m} X_{a_{2}}}{X^{5}} + \frac{\e_{a_{2} c}{}^{m} X_{m} X_{a_{1}}}{X^{5}} \bigg\} \, ,  \label{FLL - solutions b} \\[2mm]
	S^{2}_{a_{1} a_{2},c}(X) &= \tilde{c} \, \bigg\{ - \frac{5 \text{i}}{2} \frac{X_{a_{1}} X_{a_{2}} X_{c}}{X^{7}} + \frac{\text{i}}{2} \frac{ \eta_{c a_{2}} X_{a_{1}}}{X^{5}} + \frac{\text{i}}{2} \frac{ \eta_{c a_{1}} X_{a_{2}}}{X^{5}} + \frac{\text{i}}{2} \frac{ \eta_{a_{1} a_{2}} X_{c}}{X^{5}} \bigg\} \, .  \label{FLL - solutions c}
	\end{align}
\end{subequations}
These solutions are consistent with the remaining constraints \eqref{HLL - differential constraint 3b - 1}, \eqref{HLL - differential constraint 3b - 2} for the choice $\tilde{c} = -3 c$. It can also be shown by direct substitution that this solution is consistent with the point-switch identity \eqref{HLL - z2 to z3}. 
Hence this correlator is determined up to a single tensor structure.

Let us comment on the absence of parity-odd contributions.\footnote{In our formalism, the presence of the antisymmetric $\e$ tensor in the tensor $\cH$ does not necessarily imply it is parity-odd. Instead one must count the overall number of $\g$-matrices contained in both $\cH$ and the prefactor after performing superspace reduction, 
	then make use of identities such as $\e_{m n p} = -\tfrac{1}{2} \text{Tr}( \g_{m} \g_{n} \g_{p} )$. This approach was applied to the study of mixed correlators of the supercurrent and flavour current multiplets \cite{Buchbinder:2021gwu}.} They could only potentially come from the following terms contained in $S^1$
\begin{equation}
S^{1}_{({\rm odd}) a_{1} a_{2},c}(X) = k_{1} \frac{X_{a_{1}} X_{a_{2}} X_{c}}{X^{6}}  +
k_{3} \, \bigg\{ \frac{ \eta_{c a_{1} } X_{a_{2}}}{X^{4}} +  \frac{ \eta_{c a_{2} } X_{a_{1}}  }{X^{4}} \bigg\} + k_{4} \frac{\eta_{a_{1} a_{2}} X_{c} }{X^{4}} \, ,
\end{equation}
which are odd under $X^m \to - X^m$. However, this expression cannot be at the same time traceless and transverse for any choice of the coefficient $k_1, k_3, k_4$,
which can be easily checked. 

This result is contrary to the computation carried out using the polarisation spinor formalism in~\cite{Nizami:2013tpa}, where it was shown that a parity violating contribution can exist. 
A direct comparison with the results obtained in~\cite{Nizami:2013tpa} is difficult as our approach and notation are quite different. 
Our formalism, however, has the benefit that it is analytic and rather explicit.\footnote{The corresponding 
	result in \cite{Nizami:2013tpa} is listed in Table 2 with few details provided. To our knowledge it is based mostly on numerical methods, whereas our result is obtained analytically.}
As a consistency check, in Appendix \ref{AppC} we reformulate this problem and use the $\langle L L \cF \rangle$ ansatz. The evaluation procedure is slightly different 
but the same conclusion is obtained.

%%%%%%%%%%%%%%%%%%%%%%%%%%%%%%%%%%%%%%%%%%%%%%%%%%%%%%%%%%%%%%%%%%%%%%%%%
\section{Conclusion}\label{section6}
%%%%%%%%%%%%%%%%%%%%%%%%%%%%%%%%%%%%%%%%%%%%%%%%%%%%%%%%%%%%%%%%%%%%%%%%%

In this paper we analysed various correlation functions involving a conserved superspin-2 current  multiplet $\cF_{\a(4)}$. 
The case of $\langle \cF \cF \cF \rangle$ is particularly challenging due to the proliferation of tensor structures in the solution; indeed we found that it could 
only be studied efficiently using computational methods. We obtained that the three-point function $\langle \cF \cF \cF \rangle$ contains one parity-even and one parity-odd structure.

The appearance of a single parity-even structure can be understood intuitively and is somewhat expected. Indeed, the superfield $\cF_{\a(4)}$ contains 
a conserved spin-2 current $J_{\a(4)}$ as the lowest component which, though being different from the energy-momentum tensor, satisfies the same conservation equation. 
Its three-point function has two parity-even structures which can be attributed to contributions from a free boson and a free fermion. 
Since supersymmetry relates bosons and fermions, it is reasonable to expect that these structures become related, giving rise to a single independent contribution. 
On the other hand, the existence of the parity-odd structure in $\langle \cF \cF \cF \rangle$ is rather non-trivial because, as was pointed out in the introduction, 
there is an apparent tension between parity-odd structures and supersymmetry:
all three-point functions involving the energy-momentum tensor and  vector currents admit parity-odd structures in the non-supersymmetric case~\cite{Giombi:2011rz}
but not in the supersymmetric one~\cite{Buchbinder:2015qsa, Buchbinder:2021gwu}.

Let us now clarify a possibly confusing point. The three-point function of the energy-momentum tensor $T$ does not allow parity-odd structures
in the supersymmetric case, whereas the three-point function of the similar spin-2 current $J$ does. This might look paradoxical because 
$T$ and $J$ have the same symmetry properties and satisfy the same conservation equation. However, it is important to remember that $T$ 
and $J$ belong to different supermultiplets and, hence, transform differently under supersymmetry. Therefore, restrictions on their correlation functions 
due to supersymmetry are different. 

A natural extension of our results is to study the three-point functions of higher-spin current multiplets of (arbitrary) higher (super)spin. 
For non-supersymmetric conformal field theories, the three-point functions of bosonic higher spin currents were found in~\cite{Zhiboedov:2012bm, Stanev:2013eha, Elkhidir:2014woa}.
In four-dimensional supersymmetric conformal field theories correlation functions of higher-spin spinor currents were recently studied in~\cite{Buchbinder:2021kjk} 
(see also~\cite{Buchbinder:2021izb}). Deriving explicit solutions becomes increasingly difficult for fields with higher-spins. 
It is possible that other approaches, for example, based on supersymmetric generalisations of the embedding formalism~\cite{Weinberg:2010fx, Goldberger:2011yp, Goldberger:2012xb, Maio:2012tx} 
or of the spinor-helicity formalism~\cite{Jain:2021wyn, Jain:2021vrv, Jain:2021gwa}, can be more efficient. It would be interesting to explore them as well. 

Another natural question is to find explicit realisations of superconformal field theories possessing a conserved superspin--2 current multiplet. 
Since this multiplet also contains a higher-spin current one should expect that these theories possess infinitely many conserved higher-(super)spin 
currents~\cite{Boulanger:2013zza}.

%%%%%%%%%%%%%%%%%%%%%%%%%%%%%%%%%%%%%%%%%%%%%%%%%%%%%%%%%%%%%%%%%%%%%%%%%

\section*{Acknowledgements}
The authors would like to thank Daniel Hutchings, Sergei Kuzenko, and Kai Turner for valuable discussions. 
The work of E.I.B. is supported in part by the Australian Research Council, project No. DP200101944.
The work of B.S. is supported by the \textit{Bruce and Betty Green Postgraduate Research Scholarship} under the Australian Government Research Training Program.

%%%%%%%%%%%%%%%%%%%%%%%%%%%%%%%%%%%%%%%%%%%%%%%%%%%%%%%%%%%%%%%%%%%%%%%%%

%%%%%%%%%%%%%%%%%%%%%%%%%%%%%%%%%%%%%%%%%%%%%%%%%%%%%%%%%%%%%%%%%%%%%%%%%

\appendix

\section{3D conventions and notation}\label{AppA}

For the Minkowski metric we use the ``mostly plus'' convention: $\eta_{mn} = \text{diag}(-1,1,1)$. Spinor indices are then raised and lowered with the $\text{SL}(2,\mathbb{R})$ invariant anti-symmetric $\varepsilon$-tensor
\begin{align}
\ve_{\a \b} = 
\begingroup
\setlength\arraycolsep{4pt}
\begin{pmatrix}
\, 0 & -1 \, \\
\, 1 & 0 \,
\end{pmatrix}
\endgroup 
\, , & \hspace{5mm}
%%%%%%%%%%%%%%%%%%%%%
\ve^{\a \b} =
\begingroup
\setlength\arraycolsep{4pt}
\begin{pmatrix}
\, 0 & 1 \, \\
\, -1 & 0 \,
\end{pmatrix}
\endgroup 
\, , \hspace{5mm}
\ve_{\a \g} \ve^{\g \b} = \d_{\a}{}^{\b} \, , \\[4mm]
& \hspace{-8mm} \f_{\a} = \ve_{\a \b} \, \f^{\b} \, , \hspace{10mm} \f^{\a} = \ve^{\a \b} \, \f_{\b} \, .
\end{align}
The $\g$-matrices are chosen to be real, and are expressed in terms of the Pauli matrices $\s$ as follows:
\begin{subequations}
	\begin{align}
	(\g_{0})_{\a}{}^{\b} = - \text{i} \s_{2} = 
	\begingroup
	\setlength\arraycolsep{4pt}
	\begin{pmatrix}
	\, 0 & -1 \, \\
	\, 1 & 0 \,
	\end{pmatrix}
	\endgroup 
	\, , & \hspace{8mm}
	%%%%%%%%%%%%%%%%%%%%%
	(\g_{1})_{\a}{}^{\b} = \s_{3} = 
	\begingroup
	\setlength\arraycolsep{4pt}
	\begin{pmatrix}
	\, 1 & 0 \, \\
	\, 0 & -1 \,
	\end{pmatrix}
	\endgroup 
	\, , \\[3mm]
	%%%%%%%%%%%%%%%%%%%%%
	(\g_{2})_{\a}{}^{\b} = - \s_{1} &= 
	\begingroup
	\setlength\arraycolsep{4pt}
	\begin{pmatrix}
	\, 0 & -1 \, \\
	\, -1 & 0 \,
	\end{pmatrix}
	\endgroup 
	\, ,
	\end{align}
\end{subequations}
\begin{equation}
(\g_{m})_{\a \b} = \ve_{\b \d} (\g_{m})_{\a}{}^{\d} \, , \hspace{10mm} (\g_{m})^{\a \b} = \ve^{\a \d} (\g_{m})_{\d}{}^{\b} \, .
\end{equation}
The $\g$-matrices are traceless and symmetric
\begin{equation}
(\g_{m})^{\a}{}_{\a} = 0 \, , \hspace{10mm} (\g_{m})_{\a \b} = (\g_{m})_{\b \a} \, ,
\end{equation} 
and also satisfy the Clifford algebra
\begin{equation}
\g_{m} \g_{n} + \g_{n} \g_{m} = 2 \eta_{mn} \, .
\end{equation}
Products of $\g$-matrices are then
\begin{subequations}
	\begin{align}
	(\g_{m})_{\a}{}^{\r} (\g_{n})_{\r}{}^{\b} &= \eta_{mn} \d_{\a}{}^{\b} + \e_{mnp} (\g^{p})_{\a}{}^{\b} \, , \label{Gamma product} \\[2mm]
	(\g_{m})_{\a}{}^{\r} (\g_{n})_{\r}{}^{\s} (\g_{p})_{\s}{}^{\b} &= \eta_{mn} (\g_{p})_{\a}{}^{\b} - \eta_{mp} (\g_{n})_{\a}{}^{\b} + \eta_{np} (\g_{m})_{\a}{}^{\b} + \e_{mnp} \d_{\a}{}^{\b} \, ,
	\end{align}
\end{subequations}
where we have introduced the 3D Levi-Civita tensor $\e$, with $\e^{012} = - \e_{012} = 1$. It satisfies the following identities:
\begin{subequations}
	\begin{align}
	\e_{mnp} \e_{m' n' p'} &= - \eta_{mm'} ( \eta_{nn'} \eta_{pp'} - \eta_{np'} \eta_{pn'} ) - ( n' \leftrightarrow m' ) - ( m' \leftrightarrow p' ) \, , \\
	\e_{mnp} \e^{m}{}_{ n' p'} &= - \eta_{nn'} \eta_{pp'} + \eta_{n p'} \eta_{p n'} \, , \\
	\e_{mnp} \e^{mn}{}_{ p'} &= - 2 \eta_{ p p'} \, , \\
	\e_{mnp} \e^{mnp} &= -6 \, . 
	\end{align}
\end{subequations}
We also have the orthogonality and completeness relations for the $\g$-matrices
\begin{equation}
(\g^{m})_{\a \b} (\g_{m})^{\r \s} = - \d_{\a}{}^{\r} \d_{\b}{}^{\s}  - \d_{\a}{}^{\s}  \d_{\b}{}^{\r} \, , \hspace{5mm} (\g_{m})_{\a \b} (\g_{n})^{\a \b} = -2 \eta_{mn} \, .
\end{equation}
Finally, the $\g$-matrices are used to swap from vector to spinor indices. For example, given some three-vector $x_{m}$, it can be expressed equivalently in terms of a symmetric second-rank spinor $x_{\a \b}$ as follows:
\begin{gather}
x^{\a \b} = (\g^{m})^{\a \b} x_{m}  \, , \hspace{5mm} x_{m} = - \frac{1}{2} (\g_{m})^{\a \b} x_{\a \b} \, , \\[2mm]
\det (x_{\a \b}) = \frac{1}{2} x^{\a \b} x_{\a \b} = - x^{m} x_{m} = -x^{2} \, .
\end{gather}
The same conventions are also adopted for the spacetime partial derivatives $\partial_{m}$
\begin{gather}
\partial^{\a \b} = \partial^{m} (\g_{m})^{\a \b} \, , \hspace{5mm} \partial_{m} = - \frac{1}{2} (\g_{m})^{\a \b} \partial_{\a \b} \, , \\[2mm]
\partial_{m} x^{n} = \d_{m}^{n} \, , \hspace{5mm} \partial_{\a \b} x^{\r \s} = - \d_{\a}{}^{\r} \d_{\b}{}^{\s}  - \d_{\a}{}^{\s}  \d_{\b}{}^{\r} \, , \\[2mm]
\x^{m} \partial_{m} = - \frac{1}{2} \x^{\a \b} \partial_{\a \b} \, .
\end{gather}
We also define the supersymmetry generators $Q^{I}_{\a}$
\begin{equation}
Q_{\a} = \text{i} \frac{\partial}{\partial \q^{\a}} + (\g^{m})_{\a \b} \q^{\b} \frac{\partial}{\partial x^{m}} \, , \label{Supercharges}
\end{equation}
and the covariant spinor derivatives
\begin{equation}
D_{\a} = \frac{\partial}{\partial \q^{\a}} + \text{i} (\g^{m})_{\a \b} \q^{\b} \frac{\partial}{\partial x^{m}} \, , \label{Covariant spinor derivatives}
\end{equation}
which anti-commute with the supersymmetry generators, $\{ Q_{\a} , D_{\b}\} = 0$, and obey the standard anti-commutation relations
\begin{equation}
\big\{ D_{\a} , D_{\b} \big\} = 2 \text{i} \, (\g^{m})_{\a \b} \partial_{m} \, .
\end{equation}

%%%%%%%%%%%%%%%%%%%%%%%%%%%%%%%%%%%%%%%%%%%%%%%%%%%%%%%%%%%%%%%%%%%%%%%%%
%\section{Component analysis}
%\label{AppB}

%In this appendix we will provide some additional details regarding the component reduction from $\langle \cF \cF \cF \rangle$ to $\langle Q J Q \rangle$.

\section{Component reduction: \texorpdfstring{$\langle \cF \cF \cF \rangle$ $\rightarrow$ $\langle Q J Q \rangle$}{< F F F > to < Q J Q >}}\label{AppB}

In this appendix we will provide some additional details regarding the component reduction from $\langle \cF \cF \cF \rangle$ to $\langle Q J Q \rangle$.
We recall from Subsection \ref{subsection4.2} that the component correlation function $\langle Q J Q \rangle$ is obtained from $\langle \cF \cF \cF \rangle$ as follows:
\begin{align}
\langle Q_{\a(4), \a}(x_{1}) \, J_{\b(4)}(x_{2}) \, Q_{\g(4), \g}(x_{3}) \rangle &= D_{(3) \g} D_{(1) \a} \langle \cF_{\a(4)}(z_{1}) \, \cF_{\b(4)}(z_{2}) \, \cF_{\g(4)}(z_{3}) \rangle \big| \nonumber \\ 
& = D_{(3) \g} D_{(1) \a} \Big\{ \tfrac{ \prod_{i=1}^{4} \hat{\boldsymbol{x}}_{13 \, \a_{i}}{}^{\a'_{i}} \, \hat{\boldsymbol{x}}_{23 \, \b_{i}}{}^{\b'_{i}} }{(\boldsymbol{x}_{13}^{2})^{3} (\boldsymbol{x}_{23}^{2})^{3}} \, \cH_{ \a'(4) \b'(4) \g(4) }( \boldsymbol{X}_{3}, \Q_{3} ) \Big\} \Big|  \nonumber \\
&= A + B \, .
\end{align}
The calculation is broken up into two relevant parts: the $A$ contribution is due to the derivatives hitting the prefactor,
\begin{equation}
A = D_{(3) \g} D_{(1) \a} \Bigg\{ \frac{ \prod_{i=1}^{4} \hat{\boldsymbol{x}}_{13 \, \a_{i}}{}^{\a'_{i}} \, \hat{\boldsymbol{x}}_{23 \, \b_{i}}{}^{\b'_{i}} }{(\boldsymbol{x}_{13}^{2})^{3} (\boldsymbol{x}_{23}^{2})^{3}} \Bigg\} \, \cH_{ \a'(4) \b'(4) \g(4) }( \boldsymbol{X}_{3}, \Q_{3} ) \bigg| \, ,
\end{equation}
while the $B$ contribution arises due to the derivatives hitting $\cH$, 
\begin{align}
B &= \frac{ \prod_{i=1}^{4} \hat{\boldsymbol{x}}_{13 \, \a_{i}}{}^{\a'_{i}} \, \hat{\boldsymbol{x}}_{23 \, \b_{i}}{}^{\b'_{i}} }{(\boldsymbol{x}_{13}^{2})^{3} (\boldsymbol{x}_{23}^{2})^{3}} \, D_{(3) \g} D_{(1) \a} \Big\{ \cH_{ \a'(4) \b'(4) \g(4) }( \boldsymbol{X}_{3}, \Q_{3} ) \Big\} \Big| \, .
\end{align}
Let us start with the $A$ term. After distributing the derivatives we obtain
\begin{align}
D_{(3) \g} D_{(1) \a} \Bigg\{ \frac{ \prod_{i=1}^{4} \hat{\boldsymbol{x}}_{13 \, \a_{i}}{}^{\a'_{i}} \, \hat{\boldsymbol{x}}_{23 \, \b_{i}}{}^{\b'_{i}} }{(\boldsymbol{x}_{13}^{2})^{3} (\boldsymbol{x}_{23}^{2})^{3}} \Bigg\} &= D_{(3) \g} D_{(1) \a} \Bigg\{ \frac{\hat{\boldsymbol{x}}_{13 \, \a_{1}}{}^{\a'_{1}} \hat{\boldsymbol{x}}_{13 \, \a_{2}}{}^{\a'_{2}} \hat{\boldsymbol{x}}_{13 \, \a_{3}}{}^{\a'_{3}} \hat{\boldsymbol{x}}_{13 \, \a_{4}}{}^{\a'_{4}} }{(\boldsymbol{x}_{13}^{2})^{3}} \Bigg\} \nonumber \\
& \hspace{10mm} \times \frac{\hat{\boldsymbol{x}}_{23 \, \b_{1}}{}^{\b'_{1}} \hat{\boldsymbol{x}}_{23 \, \b_{2}}{}^{\b'_{2}} \hat{\boldsymbol{x}}_{23 \, \b_{3}}{}^{\b'_{3}} \hat{\boldsymbol{x}}_{23 \, \b_{4}}{}^{\b'_{4}} }{(\boldsymbol{x}_{23}^{2})^{3}} \, ,
\end{align}
where we have used the fact that $D_{(3)}$ hitting the objects $\boldsymbol{x}_{23}$ result in $\q$ linear terms, hence they do not contribute. We then find
\begin{align*}
& D_{(3) \g} D_{(1) \a} \Bigg\{ \frac{\hat{\boldsymbol{x}}_{13 \, \a_{1}}{}^{\a'_{1}} \hat{\boldsymbol{x}}_{13 \, \a_{2}}{}^{\a'_{2}} \hat{\boldsymbol{x}}_{13 \, \a_{3}}{}^{\a'_{3}} \hat{\boldsymbol{x}}_{13 \, \a_{4}}{}^{\a'_{4}} }{(x_{13}^{2})^{3}} \Bigg\} \Bigg| \\[2mm]
& = \frac{2 \text{i}}{ (\boldsymbol{x}_{13}^{2})^{5} } \Big\{ \ve_{\a \a_{1}} \d_{\g}{}^{\a'_{1}} x_{13 \, \a_{2}}{}^{\a'_{2}} x_{13 \, \a_{3}}{}^{\a'_{3}} x_{13 \, \a_{4}}{}^{\a'_{4}}
+ \ve_{\a \a_{2}} \d_{\g}{}^{\a'_{2}} x_{13 \, \a_{1}}{}^{\a'_{1}} x_{13 \, \a_{3}}{}^{\a'_{3}} x_{13 \, \a_{4}}{}^{\a'_{4}} \\
& \hspace{30mm} + \ve_{\a \a_{3}} \d_{\g}{}^{\a'_{3}} x_{13 \, \a_{1}}{}^{\a'_{1}} x_{13 \, \a_{2}}{}^{\a'_{2}} x_{13 \, \a_{4}}{}^{\a'_{4}} 
+ \ve_{\a \a_{4}} \d_{\g}{}^{\a'_{4}} x_{13 \, \a_{1}}{}^{\a'_{1}} x_{13 \, \a_{2}}{}^{\a'_{2}} x_{13 \, \a_{3}}{}^{\a'_{3}} \Big\} \\
& \hspace{40mm} + \frac{10 \text{i}}{ (x_{13}^{2})^{6} } \, x_{13 \, \a \g} x_{13 \, \a_{1}}{}^{\a'_{1}} x_{13 \, \a_{2}}{}^{\a'_{2}} x_{13 \, \a_{3}}{}^{\a'_{3}} x_{13 \, \a_{4}}{}^{\a'_{4}} \, .
\end{align*}
Finally, after repeated application of the identity 
\begin{equation}
\ve_{\a \a_{1}} \ve_{\g \a'_{1}} = \frac{x_{13 \a \a'_{1}} x_{13 \g \a_{1}} - x_{13 \a \g} x_{ 13 \a_{1} \a'_{1}} }{x_{13}^{2}} \, ,
\end{equation}
we obtain the result
\begin{align*}
& D_{(3) \g} D_{(1) \a} \Bigg\{ \frac{\hat{\boldsymbol{x}}_{13 \, \a_{1}}{}^{\a'_{1}} \hat{\boldsymbol{x}}_{13 \, \a_{2}}{}^{\a'_{2}} \hat{\boldsymbol{x}}_{13 \, \a_{3}}{}^{\a'_{3}} \hat{\boldsymbol{x}}_{13 \, \a_{4}}{}^{\a'_{4}} }{(\boldsymbol{x}_{13}^{2})^{3}} \Bigg\} \Bigg| \\[2mm]
& = \frac{2 \text{i}}{ (x_{13}^{2})^{6} } \Big\{ x_{13 \a}{}^{\a'_{1}} x_{13 \g \a_{1}} x_{13 \, \a_{2}}{}^{\a'_{2}} x_{13 \, \a_{3}}{}^{\a'_{3}} x_{13 \, \a_{4}}{}^{\a'_{4}}
+ x_{13 \a}{}^{\a'_{2}} x_{13 \g \a_{2}} x_{13 \, \a_{1}}{}^{\a'_{1}} x_{13 \, \a_{3}}{}^{\a'_{3}} x_{13 \, \a_{4}}{}^{\a'_{4}} \\
& \hspace{25mm} + x_{13 \a}{}^{\a'_{3}} x_{13 \g \a_{3}} x_{13 \, \a_{1}}{}^{\a'_{1}} x_{13 \, \a_{2}}{}^{\a'_{2}} x_{13 \, \a_{4}}{}^{\a'_{4}} 
+ x_{13 \a}{}^{\a'_{4}} x_{13 \g \a_{4}} x_{13 \, \a_{1}}{}^{\a'_{1}} x_{13 \, \a_{2}}{}^{\a'_{2}} x_{13 \, \a_{3}}{}^{\a'_{3}}  \\
& \hspace{35mm} + x_{13 \, \a \g} x_{13 \, \a_{1}}{}^{\a'_{1}} x_{13 \, \a_{2}}{}^{\a'_{2}} x_{13 \, \a_{3}}{}^{\a'_{3}} x_{13 \, \a_{4}}{}^{\a'_{4}} \Big\} \, .
\end{align*}
After some additional minor manipulations we obtain
\begin{align}
A &= \frac{1}{(x_{13}^{2})^{7/2} (x_{23}^{2})^{3}} \, \hat{x}_{13 \, \a}{}^{\a'} \hat{x}_{13 \, \a_{1}}{}^{\a'_{1}} \hat{x}_{13 \, \a_{2}}{}^{\a'_{2}} \hat{x}_{13 \, \a_{3}}{}^{\a'_{3}} \hat{x}_{13 \, \a_{4}}{}^{\a'_{4}} \nonumber \\
& \hspace{40mm} \times \hat{x}_{23 \, \b_{1}}{}^{\b'_{1}} \hat{x}_{23 \, \a_{2}}{}^{\b'_{2}} \hat{x}_{23 \, \b_{3}}{}^{\b'_{3}} \hat{x}_{23 \, \b_{4}}{}^{\b'_{4}} \, \cT^{A}_{\a' \a'(4) \b'(4) \g \g(4)}(X_{12}) \, .
\end{align}
Now consider the $B$ term, in particular we need to evaluate
\begin{align}
D_{(3) \g} D_{(1) \a} \cH_{ \a'(4) \b'(4) \g(4) }( \boldsymbol{X}_{3}, \Q_{3} ) \, .
\end{align}
Using the identities \eqref{Three-point building blocks 1c - differential identities 3} we obtain
\begin{equation}
D_{(3) \g} D_{(1) \a} \cH_{ \a'(4) \b'(4) \g(4) }( \boldsymbol{X}_{3}, \Q_{3} ) = D_{(3) \g} \bigg\{ - \frac{\boldsymbol{x}_{13 \a}{}^{\a'}}{\boldsymbol{x}_{13}^{2}} \cD_{(3) \a'} \cH_{ \a'(4) \b'(4) \g(4) }( \boldsymbol{X}_{3}, \Q_{3} ) \bigg\} \, .
\end{equation}
Evaluating the derivative within the brackets gives
\begin{align}
\cD_{\a'} \cH_{ \a'(4) \b'(4) \g(4) }( \boldsymbol{X}, \Q ) &= \text{i} (\g^{m})_{\a' \d} \Q^{\d} \pa_{m} F_{ \a'(4) \b'(4) \g(4) }(X) \nonumber \\
& \hspace{25mm} + 2 \Q_{\a'} G_{ \a'(4) \b'(4) \g(4) }(X) \, .
\end{align}
Now in order to compute 
\begin{equation}
D_{(3) \g} D_{(1) \a} \cH_{ \a'(4) \b'(4) \g(4) }( \boldsymbol{X}_{3}, \Q_{3} ) \big| \, ,
\end{equation}
we note that contributions in which the spinor derivative acts on $\boldsymbol{x}_{13}$ or $X_{12}$ produce terms that are linear in $\q$, so they may be neglected as they vanish after bar projection. On the other hand the following identity holds
\begin{equation}
D_{(3) \a'} \Q_{3}^{\d} \big| = X_{12 \a'}{}^{\d} \, .
\end{equation}
Hence we obtain
\begin{align}
D_{(3) \g} D_{(1) \a} \cH_{ \a'(4) \b'(4) \g(4) }( \boldsymbol{X}_{3}, \Q_{3} ) \big| &= - \frac{x_{13 \a}{}^{\a'}}{x_{13}^{2}} \Big\{ \text{i} (\g^{m})_{\a' \d} X_{12 \, \g}^{\d} \pa_{m} F_{ \a'(4) \b'(4) \g(4) }(X_{12}) \nonumber \\
& \hspace{25mm} + 2 X_{12 \, \a' \g} G_{ \a'(4) \b'(4) \g(4) }(X_{12}) \Big\} \, .
\end{align}
Therefore the $B$ contribution may be expressed in the form
\begin{align}
B &= \frac{1}{(x_{13}^{2})^{7/2} (x_{23}^{2})^{3}} \, \hat{x}_{13 \, \a}{}^{\a'} \hat{x}_{13 \, \a_{1}}{}^{\a'_{1}} \hat{x}_{13 \, \a_{2}}{}^{\a'_{2}} \hat{x}_{13 \, \a_{3}}{}^{\a'_{3}} \hat{x}_{13 \, \a_{4}}{}^{\a'_{4}} \nonumber \\
& \hspace{35mm} \times \hat{x}_{23 \, \b_{1}}{}^{\b'_{1}} \hat{x}_{23 \, \b_{2}}{}^{\b'_{2}} \hat{x}_{23 \, \b_{3}}{}^{\b'_{3}} \hat{x}_{23 \, \b_{4}}{}^{\b'_{4}} \, \cT^{B}_{\a', \a'(4) \b'(4) \g , \g(4)}(X_{12}) \, ,
\end{align}
with $\cT^{B}$ given by the expression
\begin{align}
\cT^{B}_{\a, \a(4) \, \b(4) \, \g, \g(4) }(X) &= -\text{i} \, (\g^{m})_{\a \s} X^{\s}{}_{\g} \, \pa_{m} F_{ \a(4) \b(4) \g(4) }(X) - 2 X_{\a \g} G_{ \a(4) \b(4) \g(4) }(X) \, .
\end{align}
Combining both the $A$ and $B$ terms we obtain the component correlation function 
\begin{align}
\langle Q_{\a(4), \a}(x_{1}) \, J_{\b(4)}(x_{2}) \, Q_{\g(4), \g}(x_{3}) \rangle &= \frac{1}{(x_{13}^{2})^{7/2} (x_{23}^{2})^{3}} \, \hat{x}_{13 \, \a}{}^{\a'} \hat{x}_{13 \, \a_{1}}{}^{\a'_{1}} \hat{x}_{13 \, \a_{2}}{}^{\a'_{2}} \hat{x}_{13 \, \a_{3}}{}^{\a'_{3}} \hat{x}_{13 \, \a_{4}}{}^{\a'_{4}} \nonumber \\
& \times \hat{x}_{23 \, \b_{1}}{}^{\b'_{1}} \hat{x}_{23 \, \b_{2}}{}^{\b'_{2}} \hat{x}_{23 \, \b_{3}}{}^{\b'_{3}} \hat{x}_{23 \, \b_{4}}{}^{\b'_{4}} \, \cT_{\a', \a'(4) \b'(4) \g, \g(4)}(X_{12}) \, ,
\end{align}
with
\begin{equation}
\cT_{\a, \a(4) \b(4) \g, \g(4)}(X) = \cT^{A}_{\a, \a(4) \b(4) \g, \g(4)}(X) + \cT^{B}_{\a, \a(4) \b(4) \g, \g(4)}(X) \, .
\end{equation}
%

%%%%%%%%%%%%%%%%%%%%%%%%%%%%%%%%%%%%%%%%%%%%%%%%%%%%%%%%%%%%%%%%%%%%%%%%%

%%%%%%%%%%%%%%%%%%%%%%%%%%%%%%%%%%%%%%%%%%%%%%%%%%%%%%%%%%%%%%%%%%%%%%%%%
\section{Consistency checks}\label{AppC}

\subsection{Correlator \texorpdfstring{$\langle L L \cO \rangle$}{< L L O >}}\label{AppC1}

In this sub-appendix we derive the general form of the correlation function $\langle L L \cO \rangle$. We also demonstrate that our solution is consistent with the results of \cite{Nizami:2013tpa} in terms of the number of independent tensor structures. The ansatz for $\langle L L \cO \rangle$ is
\begin{equation}
\langle L^{\abar}_{\a}(z_{1}) \, L^{\bbar}_{\b}(z_{2}) \, \cO(z_{3}) \rangle = \d^{\abar \bbar} \, \frac{ \hat{\boldsymbol{x}}_{13 \, \a}{}^{\a'} \, \hat{\boldsymbol{x}}_{23 \, \b}{}^{\b'}  }{(\boldsymbol{x}_{13}^{2})^{3/2} (\boldsymbol{x}_{23}^{2})^{3/2} } \, \cH_{\a' \b'}( \boldsymbol{X}_{3}, \Q_{3} ) \, . \label{LLO - ansatz}
\end{equation}
The constraints on this three-point function are summarised below:
\begin{enumerate}
	\item[\textbf{(i)}] \textbf{Homogeneity constraint}
	
	Covariance under scale transformations of superspace results in the following constraint on $\cH$
	\begin{align}
	\cH_{\a \b}( \l^{2} \boldsymbol{X} , \l \Q ) = (\l^{2})^{-\t} \cH_{\a \b}( \boldsymbol{X} , \Q ) \, , \label{LLO - homogeneity constraint}
	\end{align}
	which implies that $\cH$ is homogeneous degree $\t = 3 - \D$.
	
	\item[\textbf{(ii)}] \textbf{Differential constraints}
	
	The conservation equations \eqref{F - conservation equation} imply the following constraints
	\begin{equation}
	D_{(1)}^{\a} \langle L^{\abar}_{\a}(z_{1}) \, L^{\bbar}_{\b}(z_{2}) \, \cO(z_{3}) \rangle = 0 \, , \label{LLO - differential constraint 1}
	\end{equation}
	Application of the identities \eqref{Three-point building blocks 1c - differential identities 3} to \eqref{LLO - differential constraint 1} gives
	\begin{align}
	\cD^{\a} \cH_{\a \b}( \boldsymbol{X} , \Q ) = 0 \, . \label{LLO - differential constraint 1a}
	\end{align}

	\item[\textbf{(iii)}] \textbf{Point-switch identity}
	
	Invariance under permutation of the superspace points $z_{1}$ and $z_{2}$ is equivalent to the condition
	\begin{equation}
	\langle L^{\abar}_{\a}(z_{1}) \, L^{\bbar}_{\b}(z_{2}) \, \cO(z_{3}) \rangle = - \langle L^{\bbar}_{\b}(z_{2}) \, L^{\abar}_{\a}(z_{1}) \, \cO(z_{3}) \rangle \, ,
	\end{equation}
	which results in the following constraint on $\cH$
	\begin{equation}
	\cH_{\a \b}( \boldsymbol{X} , \Q ) = - \cH_{\a \b}( -\boldsymbol{X}^{\text{T}}, -\Q )  \, . \label{LLO - z1 to z2}
	\end{equation}
\end{enumerate}

An irreducible expansion for $\cH$ is
\begin{equation}
\cH_{ \a \b }( \boldsymbol{X}, \Q ) = \ve_{\a \b} A( \boldsymbol{X} , \Q) + (\g^{a})_{\a \b} S_{a}(\boldsymbol{X}, \Q) \, ,
\end{equation}
with
\begin{subequations}
	\begin{align}
	A( \boldsymbol{X}, \Q) &= A^{1}(X) + \Q^{2} A^{2}(X) \, , \\[2mm]
	S_{a}( \boldsymbol{X}, \Q) &= S^{1}_{a}(X) + \Q^{2} S^{2}_{a}(X) \, .
	\end{align}
\end{subequations}
The point switch identity \eqref{LLO - z1 to z2} implies
\begin{subequations}
	\begin{align}
	A^{1}(X) &= A^{1}(-X) \, , &  A^{2}(X) &= A^{2}(-X) \, , \label{LLO - z1 to z2 2a} \\[2mm]
	S^{1}_{a}(X) &= - S^{1}_{a}(-X) \, , & S^{2}_{a}(X) &= - S^{2}_{a}(-X) \, . \label{LLO - z1 to z2 2b}
	\end{align}
\end{subequations}
Imposing the differential relation \eqref{LLO - differential constraint 1a} results in the following constraints on the tensors $A^{i}$ and $S^{i}$
\begin{subequations}
	\begin{align}
	A^{2}(X) & = \tfrac{\text{i}}{2} \pa^{m} S^{1}_{m}(X) \, , \label{LLO - differential constraint 2a} \\[2mm]
	S^{2}_{a}(X) &= - \tfrac{\text{i}}{2} \big\{ \pa_{a} A^{1}(X) + \e_{a}{}^{m n} \pa_{m} S^{1}_{n}(X) \big\} \, . \label{LLO - differential constraint 2b}
	\end{align}
\end{subequations}
Hence we may treat $A^{1}$ and $S^{1}$ as independent. Explicit solutions for the tensors $A^{1}$ and $S^{1}$ are
\begin{equation}
A^{1}(X) = \frac{a}{X^{\t}} \, , \hspace{10mm} S^{1}_{a}(X) = b \, \frac{X_{a}}{X^{\t+1}} \, .
\end{equation}
These solutions are trivially compatible with \eqref{LLO - z1 to z2 2a}, \eqref{LLO - z1 to z2 2b}. Using \eqref{LLO - differential constraint 2a}, \eqref{LLO - differential constraint 2b} we obtain expressions for $A^{2}$ and $S^{2}$
\begin{equation}
A^{2}(X) = \frac{\text{i} b}{2} (2 - \t) \frac{1}{X^{\t+1}} \, , \hspace{10mm} S^{2}_{a}(X) = \frac{\text{i} a}{2} \t \frac{X_{a}}{X^{\t+2}} \, .
\end{equation}
Following \cite{Nizami:2013tpa} we set $\D=1$ $(\t = 2)$ and obtain the set of solutions
\begin{subequations}
	\begin{align}
	A^{1}(X) &= \frac{a}{X^{2}} \, , &  A^{2}(X) &= 0 \, , \\[2mm]
	S^{1}_{a}(X) &= b \frac{X_{a}}{X^{3}} \, , &  S^{2}_{a}(X) &= \text{i} a  \frac{X_{a}}{X^{4}} \, .
	\end{align}
\end{subequations}
The solution for $\cH$ in spinor notation is then
\begin{equation}
\cH_{\a \b}(\boldsymbol{X}, \Q) = a \, \bigg\{ \frac{\ve_{\a \b}}{\boldsymbol{X}^{2}} + \frac{ \text{i} \boldsymbol{X}_{\a \b} \Q^{2}}{\boldsymbol{X}^{4}}\bigg\} + b \, \bigg\{ \frac{\boldsymbol{X}_{\a \b}}{\boldsymbol{X}^{3}} + \frac{\text{i}}{2} \frac{\ve_{\a \b} \Q^{2}}{\boldsymbol{X}^{3}} \bigg\} \, ,
\end{equation}
which clearly contains both parity even and parity odd contributions. Our notation is quite different so it is difficult to make a direct comparison, however we agree on the number of independent tensor structures. 

\subsection{Correlator \texorpdfstring{$\langle \cF L L \rangle$}{< F L L >} -- alternative ansatz}\label{AppC2}

In this subsection we investigate an alternative formulation of the correlation function $\langle \cF L L \rangle$, it serves as a consistency check of our result in Subsection \ref{subsection5.2}. The starting point is the alternative ansatz $\langle L L \cO \rangle$:
\begin{equation}
\langle L^{\abar}_{\b}(z_{1}) L^{\bbar}_{\g}(z_{2}) \cF_{\a(4)}(z_{3}) \rangle = \d^{\abar \bbar} \, \frac{ \hat{\boldsymbol{x}}_{13 \, \b}{}^{\b'} \, \hat{\boldsymbol{x}}_{23 \, \g}{}^{\g'}  }{(\boldsymbol{x}_{13}^{2})^{3/2} (\boldsymbol{x}_{23}^{2})^{3/2} } \, \cH_{\b' \g' \a(4) }( \boldsymbol{X}_{3}, \Q_{3} ) \, . \label{LLF - ansatz}
\end{equation}
The constraints on this three-point function are summarised below:
\begin{enumerate}
	\item[\textbf{(i)}] \textbf{Homogeneity constraint}
	
	Covariance under scale transformations of superspace results in the following constraint on $\cH$
	\begin{align}
	\cH_{\b \g \a(4)}( \l^{2} \boldsymbol{X} , \l \Q ) = \cH_{\b \g \a(4)}( \boldsymbol{X} , \Q ) \, , \label{LLF - homogeneity constraint}
	\end{align}
	which implies that $\cH$ is homogeneous degree $0$.
	
	\item[\textbf{(ii)}] \textbf{Differential constraints}
	
	The conservation equations \eqref{F - conservation equation} imply the following constraints
	\begin{subequations}
		\begin{align}
		D_{(1)}^{\b} \langle L^{\abar}_{\b}(z_{1}) \, L^{\bbar}_{\g}(z_{2}) \, \cF_{\a(4)}(z_{3}) \rangle &= 0 \, , \label{LLH - differential constraint 1a} \\[2mm] 
		D_{(3)}^{\s} \langle L^{\abar}_{\b}(z_{1}) \, L^{\bbar}_{\g}(z_{2}) \, \cF_{\s \a(3)}(z_{3}) \rangle &= 0 \, . \label{LLH - differential constraint 1b}
		\end{align}
	\end{subequations}
	Application of the identities \eqref{Three-point building blocks 1c - differential identities 3} to \eqref{LLH - differential constraint 1a} gives
	\begin{align}
	\cD^{\b} \cH_{\b \g \a(4)}( \boldsymbol{X} , \Q ) = 0 \, . \label{LLH - differential constraint 2a}
	\end{align}
	Imposing \eqref{HLL - differential constraint 1b} is rather non-trivial, it will be handled later in this section.
	
	\item[\textbf{(iii)}] \textbf{Point-switch identity}
	
	Invariance under permutation of the superspace points $z_{1}$ and $z_{2}$ is equivalent to the condition
	\begin{equation}
	\langle L^{\abar}_{\b}(z_{1}) \, L^{\bbar}_{\g}(z_{2}) \, \cF_{\a(4)}(z_{3}) \rangle = - \langle  L^{\bbar}_{\g}(z_{2}) \, L^{\abar}_{\b}(z_{1}) \, \cF_{\a(4)}(z_{3}) \rangle \, ,
	\end{equation}
	which results in the following constraint on $\cH$
	\begin{equation}
	\cH_{\b \g \a(4)}( \boldsymbol{X} , \Q ) = - \cH_{\g \b \a(4)}( -\boldsymbol{X} ^{\text{T}}, -\Q )  \, . \label{LLH - z1 to z2}
	\end{equation}
\end{enumerate}

First we combine symmetric pairs of spinor indices into vector ones as follows:
\begin{equation}
\cH_{ \b \g \a_{1} \a_{2} \a_{3} \a_{4} }( \boldsymbol{X}, \Q ) = (\g^{a_{1}} )_{\a_{1} \a_{2} } (\g^{a_{2}} )_{\a_{3} \a_{4} } \cH_{ \b \g , a_{1} a_{2} }( \boldsymbol{X}, \Q) \, ,
\end{equation}
where it is required that $\cH_{a_{1} a_{2}, \, \b \g} $ is symmetric and traceless in $a_{1}, a_{2}$. We then expand this in irreducible components as follows:
\begin{equation}
\cH_{ \b \g , a_{1} a_{2} }( \boldsymbol{X}, \Q ) = \ve_{\b \g} A_{a_{1} a_{2} }( \boldsymbol{X} , \Q) + (\g^{c})_{\b \g} S_{a_{1} a_{2}, c}(\boldsymbol{X}, \Q) \, ,
\end{equation}
with
\begin{subequations}
	\begin{align}
	A_{a_{1} a_{2}}( \boldsymbol{X}, \Q) &= A^{1}_{a_{1} a_{2}}(X) + \Q^{2} A^{2}_{a_{1} a_{2}}(X) \, , \\[2mm]
	S_{a_{1} a_{2}, c}( \boldsymbol{X}, \Q) &= S^{1}_{a_{1} a_{2}, c}(X) + \Q^{2} S^{2}_{a_{1} a_{2}, c}(X) \, .
	\end{align}
\end{subequations}
Here the $A^{i}$ and $S^{i}$ are both symmetric and traceless in $a_{1}, a_{2}$. The point switch identity \eqref{LLH - z1 to z2} implies
\begin{subequations}
	\begin{align}
	A^{1}_{a_{1} a_{2}}(X) &= A^{1}_{a_{1} a_{2}}(-X) \, , \hspace{16mm} A^{2}_{a_{1} a_{2}}(X) = A^{2}_{a_{1} a_{2}}(-X) \, , \label{LLH - z1 to z2 2a} \\[2mm]
	S^{1}_{a_{1} a_{2}, c}(X) &= - S^{1}_{a_{1} a_{2}, c}(-X) \, , \hspace{10mm} S^{2}_{a_{1} a_{2}, c}(X) = - S^{2}_{a_{1} a_{2}, c}(-X) \, . \label{LLH - z1 to z2 2b}
	\end{align}
\end{subequations}
Imposing the differential relation \eqref{LLH - differential constraint 2a} results in the following constraints on the tensors $A^{i}$ and $S^{i}$
\begin{subequations}
	\begin{align}
	A^{2}_{a_{1} a_{2}}(X) & = \tfrac{\text{i}}{2} \pa^{m} S^{1}_{a_{1} a_{2}, m}(X) \, , \label{LLH - differential constraint 3b - 1} \\[2mm]
	S^{2}_{a_{1} a_{2}, c}(X) &= - \tfrac{\text{i}}{2} \big\{ \pa_{c} A^{1}_{a_{1} a_{2} }(X) + \e_{c}{}^{m n} \pa_{m} S^{1}_{ a_{1} a_{2}, n}(X) \big\} \,. \label{LLH - differential constraint 3b - 2}
	\end{align}
\end{subequations}
Hence, we may treat $A^{1}$ and $S^{1}$ as independent. The only solution for the tensor $A^{1}$ compatible with the symmetries is
\begin{align}
A^{1}_{a_{1} a_{2}}(X) &= c \, \bigg\{ \eta_{a_{1} a_{2}} - \frac{3 X_{a_{1}} X_{a_{2} }}{X^{2}} \bigg\} \, ,
\end{align}
while for $S$ we have the general ansatz
\begin{align}
S^{1}_{a_{1} a_{2},c}(X) &= k_{1} \frac{X_{a_{1}} X_{a_{2}} X_{c}}{X^{3}} + k_{2} \, \bigg\{ \frac{\e_{a_{1} c}{}^{m} X_{m} X_{a_{2}}}{X^{2}} +  \frac{\e_{a_{2} c}{}^{m} X_{m} X_{a_{1}}}{X^{2}} \bigg\} \nonumber \\ 
& \hspace{10mm} + k_{3} \, \bigg\{ \frac{ \eta_{c a_{1} } X_{a_{2}}}{X} +  \frac{ \eta_{c a_{2} } X_{a_{1}}  }{X} \bigg\} + k_{4} \frac{\eta_{a_{1} a_{2}} X_{c} }{X} \, .
\end{align}
Imposing tracelessness on $a_{1}, a_{2}$ on this expansion, along with the conditions \eqref{LLH - z1 to z2 2a}, \eqref{LLH - z1 to z2 2b} results in the constraints
\begin{equation}
k_{3} = - \tfrac{3}{2} k_{4} - \tfrac{1}{2} k_{1} \, , \hspace{5mm} k_{2} = 0 \, .
\end{equation}
The solution then becomes
\begin{align}
S^{1}_{a_{1} a_{2},c}(X) &= k_{1} \, \bigg\{ \frac{X_{a_{1}} X_{a_{2}} X_{c}}{X^{3}} - \frac{1}{2} \frac{ \eta_{c a_{2}} X_{a_{1}}}{X} - \frac{1}{2} \frac{ \eta_{c a_{1}} X_{a_{2}}}{X} \bigg\} \nonumber \\ 
& \hspace{5mm} + k_{4} \, \bigg\{ \frac{\eta_{a_{1} a_{2}} X_{c} }{X} - \frac{3}{2} \frac{\eta_{c a_{2}} X_{a_{1}}}{X} - \frac{3}{2} \frac{\eta_{c a_{1}} X_{a_{2}}}{X} \bigg\} \, .
\end{align}
The expressions for $A^{2}$ and $S^{2}$ are
\begin{subequations}
	\begin{align}
	A^{2}_{a_{1} a_{2}}(X) &= \frac{\text{i} (k_{1} + k_{4})}{2} \bigg\{ \frac{\eta_{a_{1} a_{2}}}{X} - \frac{3 X_{a_{1}} X_{a_{2} }}{X^{3}} \bigg\} \, , \\[2mm]
	S^{2}_{a_{1} a_{2},c}(X) &= - 3 \text{i} c \, \bigg\{ \frac{X_{a_{1}} X_{a_{2}} X_{c}}{X^{4}} - \frac{1}{2} \frac{ \eta_{c a_{2}} X_{a_{1}}}{X^{2}} - \frac{1}{2} \frac{ \eta_{c a_{1}} X_{a_{2}}}{X^{2}} \bigg\} \nonumber \\ 
	& \hspace{10mm} + \frac{ \text{i} k_{1} }{4} \, \bigg\{ \frac{\e_{a_{1} c}{}^{m} X_{m} X_{a_{2}}}{X^{3}} +  \frac{\e_{a_{2} c}{}^{m} X_{m} X_{a_{1}}}{X^{3}} \bigg\} \nonumber \\
	& \hspace{15mm} - \frac{ \text{3i} k_{4} }{4} \, \bigg\{ \frac{\e_{a_{1} c}{}^{m} X_{m} X_{a_{2}}}{X^{3}} +  \frac{\e_{a_{2} c}{}^{m} X_{m} X_{a_{1}}}{X^{3}} \bigg\} \, .
	\end{align}
\end{subequations}
Imposing the symmetry conditions \eqref{LLH - z1 to z2 2a}, \eqref{LLH - z1 to z2 2b} on these solutions results in the constraint $k_{1} = 3 k_{4} $. After making the replacement $k_{4} \rightarrow \tilde{c}$, the solutions for the tensors $A^{i}$ and $S^{i}$ become
\begin{subequations}
	\begin{align}
	A^{1}_{a_{1} a_{2}}(X) &= c \, \bigg\{ \eta_{a_{1} a_{2}} - \frac{3 X_{a_{1}} X_{a_{2} }}{X^{2}} \bigg\} \, , \\[2mm]
	A^{2}_{a_{1} a_{2}}(X) &= 2 \text{i} \tilde{c} \, \bigg\{ \frac{\eta_{a_{1} a_{2}}}{X} - \frac{3 X_{a_{1}} X_{a_{2} }}{X^{3}} \bigg\} \, , \\[2mm]
	S^{1}_{a_{1} a_{2},c}(X) &= \tilde{c} \, \bigg\{ \frac{ 3 X_{a_{1}} X_{a_{2}} X_{c}}{X^{3}} + \frac{\eta_{a_{1} a_{2}} X_{c} }{X} - \frac{ 3 \eta_{c a_{2}} X_{a_{1}}}{X} - \frac{ 3 \eta_{c a_{1}} X_{a_{2}}}{X} \bigg\} \, , \\[2mm]
	S^{2}_{a_{1} a_{2},c}(X) &= - 3 \text{i} c \, \bigg\{ \frac{X_{a_{1}} X_{a_{2}} X_{c}}{X^{4}} - \frac{1}{2} \frac{ \eta_{c a_{2}} X_{a_{1}}}{X^{2}} - \frac{1}{2} \frac{ \eta_{c a_{1}} X_{a_{2}}}{X^{2}} \bigg\} \, ,
	\end{align}
\end{subequations}
where we note that the correlation function is presently determined up to two coefficients. However, it remains to impose the final relation \eqref{LLH - differential constraint 1b}. 
To provide a comparison with our results is Subsection \ref{subsection5.2} it is sufficient to 
analyse conservation on one of the component correlators to see whether this reduces the number of tensor structures. First let us define the component fields 
by bar-projection\footnote{There are three component fields contained within the flavour current multiplet, they are $\{ \psi_{\a}, V_{\a \b}, \chi_{\a} \}$. The superfield conservation equation \eqref{L - conservation equation} then implies that $V$ satisfies $\pa^{a} V_{a} = 0$, while $\chi$ is auxiliary. The calculations are similar to those in Section \ref{section3}. }
\begin{align}
\psi^{\abar}_{\a}(x) = L^{\abar}_{\a}(z) \big| \, , \hspace{10mm} J_{a_{1} a_{2}}(x) = \frac{1}{4} (\g_{a_{1}})^{\a_{1} \a_{2}} (\g_{a_{2}})^{\a_{3} \a_{4}} \cF_{\a_{1} \a_{2} \a_{3} \a_{4}}(z) \big| \, .
\end{align}
The three-point function $\langle \psi \psi J \rangle$ is then defined as follows
\begin{equation}
\langle \psi^{\abar}_{\b}(x_{1}) \, \psi^{\bbar}_{\g}(x_{2}) \, J_{a_{1} a_{2}}(x_{3}) \rangle = \langle L^{\abar}_{\b}(z_{1}) \, L^{\bbar}_{\g}(z_{2}) \,  \cF_{a_{1} a_{2}}(z_{3}) \rangle \big| \, .
\end{equation}
where bar-projection denotes switching off the fermionic variables at each superspace point. Using \eqref{LLF - ansatz}, this three-point function has the general form:
\begin{subequations}
	\begin{align}
	\langle \psi^{\abar}_{\b}(x_{1}) \, \psi^{\bbar}_{\g}(x_{2}) \, J_{a_{1} a_{2}}(x_{3}) \rangle &= \d^{\abar \bbar} \, \frac{ \hat{x}_{13 \, \b}{}^{\b'} \, \hat{x}_{23 \, \g}{}^{\g'}  }{(x_{13}^{2})^{3/2} (x_{23}^{2})^{3/2} } \, H_{\b' \g' , a_{1} a_{2} }( X_{12} ) \, , \\[2mm]
	H_{\b \g , a_{1} a_{2} }( X ) &= \ve_{\b \g} A^{1}_{a_{1} a_{2}}(X) + (\g^{c})_{\b \g} S^{1}_{a_{1} a_{2} , c}(X)  \, ,
	\end{align}
\end{subequations}
where $A^{1}$ and $S^{1}$ are the solutions given above. We will then impose conservation by transforming the three-point function such that it is represented in the following way
\begin{equation}
\langle \psi^{\abar}_{\b}(x_{1}) \, \psi^{\bbar}_{\g}(x_{2}) \, J_{a_{1} a_{2}}(x_{3}) \rangle \hspace{5mm} \Rightarrow \hspace{5mm} \langle J_{a_{1} a_{2}}(x_{3}) \, \psi^{\bbar}_{\g}(x_{2}) \, \psi^{\abar}_{\b}(x_{1}) \rangle \, .
\end{equation} 
Using the explicit expressions for $A^{1}$ and $S^{1}$, after some calculation it may be shown that
\begin{align}
\langle J_{a_{1} a_{2}}(x_{3}) \, \psi^{\bbar}_{\g}(x_{2}) \, \psi^{\abar}_{\b}(x_{1}) \rangle &= \d^{\abar \bbar} \, \frac{ \cI_{a_{1} a_{2}, a'_{1} a'_{2} }( \hat{x}_{31} ) \, \hat{x}_{21 \, \g}{}^{\g'}  }{(x_{31}^{2})^{3} (x_{21}^{2})^{3/2} } \, \tilde{H}_{a'_{1} a'_{2} , \g' \b }( X_{32} ) \, , \\[2mm]
\tilde{H}_{a_{1} a_{2}, \g \b }( X ) &= \ve_{\b \g} \tilde{A}^{1}_{a_{1} a_{2}}(X) + (\g^{c})_{\b \g} \tilde{S}^{1}_{a_{1} a_{2} , c}(X) \, ,
\end{align}
where $\tilde{H}$ is homogeneous degree $-3$ and the tensors $\tilde{A}^{1}$ and $\tilde{S}^{1}$ are found to be 
\begin{align}
\tilde{A}^{1}_{a_{1} a_{2}}(X) &= \frac{\tilde{c}}{X^{3}} \bigg\{ \eta_{a_{1} a_{2}} - \frac{ 3 X_{a_{1}} X_{a_{2}}}{X^{2}} \bigg\} \, , \\[2mm]
\tilde{S}^{1}_{a_{1} a_{2}, c}(X) &= c \, \bigg\{ \frac{3 X_{a_{1}} X_{a_{2}} X_{c} }{X^{6}} - \frac{\eta_{a_{1} a_{2}} X_{c} }{X^{4}} \bigg\} \nonumber \\
& \hspace{5mm} - 3 \tilde{c} \, \bigg\{ \frac{\e_{a_{1} c m} X^{m} X_{a_{2} }}{X^{5}} + \frac{\e_{a_{2} c m} X^{m} X_{a_{1} }}{X^{5}} \bigg\} \, .
\end{align}
At this stage we note that the $\tilde{c}$-terms exactly match the solutions \eqref{FLL - solutions a}, \eqref{FLL - solutions b}, 
however we have picked up an extra tensor structure (the $c$-terms). 
If we now relabel the points in the $ \langle J \psi \psi \rangle$ ansatz such that $x_{3} \rightarrow x_{1}$, $x_{1} \rightarrow x_{3}$, 
then the tensors $\tilde{A}^{1}$, $\tilde{S}^{1}$ must satisfy the constraints \eqref{HLL - differential constraint 3a - 1}, \eqref{HLL - differential constraint 3a - 2}. 
The solutions above are compatible with these constraints provided that $c=0$. Hence we have found that this correlator is fixed up to a 
single tensor structure and fully agrees with what we found in Subsection~\ref{subsection5.2}.

%%%%%%%%%%%%%%%%%%%%%%%%%%%%%%%%%%%%%%%%%%%%%%%%%%%%%%%%%%%%%%%%%%%%%%%%%
	
%\section{Computational results}\label{AppD}
%*Incomplete*
%
%Might paste snippets of code setup in here or will simply link to a github repository containing my code.

%\addcontentsline{toc}{section}{References}
%\bibliographystyle{bibstyle} %siam abbrv unsrt
%\bibliography{Masterbib}

\printbibliography[heading=bibintoc,title={References}]

%Print all references%
%\nocite{*}	

%\begin{thebibliography}{1}
%\input{mixedcor-final-v4.bbl}
%\end{thebibliography}

\end{document}